\documentclass[twocolumn,aps,prd,showpacs,nofootinbib,superscriptaddress,preprintnumbers,floatfix,10pt]{revtex4-1}

\usepackage{amsmath}
\usepackage{amsfonts}
\usepackage{amssymb}
\usepackage{epsfig}
\usepackage{graphicx}
\usepackage{color}
\usepackage{slashed}
\usepackage{epstopdf}
\usepackage{url}

\usepackage{hyperref}

\usepackage{feynmp-auto}
\usepackage{mathrsfs}

\usepackage{verbatim}
\usepackage{ulem}
\usepackage{slashed}
\usepackage{subfigure}
\usepackage[toc,page]{appendix}
\usepackage{mathtools}
\usepackage{siunitx}
\usepackage{bbm}
\usepackage{atbegshi,picture}
\usepackage{lipsum}

\newcommand{\Tr}{\text{Tr}}

\newcommand{\mmo}[1]{\begin{pmatrix} #1 \end{pmatrix}}
\numberwithin{equation}{section}
\newcommand{\muu}[1]{\begin{align} #1 \end{align}}

\newcommand{\abc}{\begin{enumerate}[label=\alph*)]}

\newcommand{\zogh}{Z_{\omega}}

\newcommand{\zdia}{Z_C}
\newcommand{\zadia}{Z_{\bar{C}}}

\AtBeginShipoutNext{\AtBeginShipoutUpperLeft{%
  \put(\dimexpr\paperwidth-1cm\relax,-1cm){\makebox[0pt][r]{\framebox{CHIBA-EP-224, \today}}}%
}}

\begin{document}


\begin{fmffile}{1stPart}

\fmfcmd{%
vardef cross_bar (expr p, len, ang) =
((-len/2,0)--(len/2,0))
rotated (ang + angle direction length(p)/2 of p)
shifted point length(p)/2 of p
enddef;
style_def crossedV expr p =
cdraw wiggly p;
ccutdraw cross_bar (p, 5mm, 45);
ccutdraw cross_bar (p, 5mm, -45)
enddef;}

\fmfcmd{%
vardef cross_bar (expr p, len, ang) =
((-len/2,0)--(len/2,0))
rotated (ang + angle direction length(p)/2 of p)
shifted point length(p)/2 of p
enddef;
style_def crossedX expr p =
cdraw curly p;
ccutdraw cross_bar (p, 5mm, 45);
ccutdraw cross_bar (p, 5mm, -45)
enddef;}


\preprint{}

\title {Composite operator and condensate in the $SU(N)$ Yang-Mills theory with $U(N-1)$ stability group} 

\author{Matthias Warschinke}
\email{m\_warschinke@chiba-u.jp}
\affiliation{Department of Physics, Graduate School of Science, Chiba University, Chiba 263-8522, Japan}

\author{Ryutaro Matsudo}
\email{afca3071@chiba-u.jp}
\affiliation{Department of Physics, Graduate School of Science and Engineering, Chiba University, Chiba 263-8522, Japan}
\author{Shogo Nishino}
\email{shogo.nishino@chiba-u.jp}
\affiliation{Department of Physics, Graduate School of Science, Chiba University, Chiba 263-8522, Japan}

\author{Toru Shinohara}
\email{sinohara@graduate.chiba-u.jp}
\affiliation{Department of Physics, Graduate School of Science, Chiba University, Chiba 263-8522, Japan}

\author{Kei-Ichi Kondo}
\email{kondok@faculty.chiba-u.jp}
\affiliation{Department of Physics, Graduate School of Science, Chiba University, Chiba 263-8522, Japan}
\affiliation{Department of Physics, Graduate School of Science and Engineering, Chiba University, Chiba 263-8522, Japan}

\begin{abstract}
Recently, some reformulations of the Yang-Mills theory inspired by the
Cho-Faddeev-Niemi decomposition have been developed in order to
understand confinement from the viewpoint of the dual superconductivity.
In this paper we focus on the reformulated $SU(N)$ Yang-Mills theory in
the minimal option with $U(N-1)$ stability group. Despite existing numerical simulations on the lattice we perform the
perturbative analysis to one-loop level as a first step towards the
non-perturbative analytical treatment.
First, we give the Feynman rules and calculate all renormalization
factors to obtain the standard renormalization group functions to one-loop level in light of the renormalizability of this theory. Then we introduce
a mixed gluon-ghost composite operator of mass dimension two and show
the BRST invariance and the multiplicative renormalizability.  Armed
with these results, we argue the existence of the mixed gluon-ghost
condensate by means of the so-called local composite operator formalism, which leads to various interesting implications for confinement as shown in preceding works.
\end{abstract}


\maketitle

\section{Introduction}
\label{intro}
The dual superconductivity picture \cite{nambu1974strings, Zichichi:1976uh, mandelstam1976ii, hooft1981topology} represents one of the most popular attempts to explain color confinement. In order for this mechanism to work, the existence of magnetic monopoles is crucial, raising the question of how to extract them from the underlying theory. Famous examples of such monopole configurations are the Dirac-monopole in the Abelian Maxwell theory \cite{Dirac60}, represented by a singular gauge field, or the 't Hooft-Polyakov monopole in non-Abelian gauge theories, which, however, relies on the presence of an adjoint scalar field \cite{hooft1974magnetic,Polyakov:1974ek}. Dealing with pure Yang-Mills theory, one has to find a way to define the monopole in the absence of any scalar field. For $SU(2)$, even in this case, a possibility to obtain the monopoles is given by performing a gauge-covariant decomposition of the gauge field, the Cho-Duan-Ge-Faddeev-Niemi-Shabanov decomposition \cite{duan19792,cho1980restricted, cho1981extended,cho1980extended,cho1980colored,faddeev1999partially, faddeev2007spin,faddeev1999partial,faddeev1999decomposing, bolokhov2004infrared, shabanov1999effective,1999PhLB..463..263S}, which also has been extended to the general $SU(N)$ case by Cho \cite{cho1980extended, cho1980colored,bae2001extended, cho2003color} and Faddeev-Niemi \cite{faddeev1999partial, faddeev1999decomposing, bolokhov2004infrared}. It relies on the introduction of the so-called adjoint color-field $\mathfrak{n}(x)$, which is used to define the decomposition of the gauge field $\mathcal{A}_\mu = \mathcal{V}_\mu + \mathcal{X}_\mu$ into the residual (or restricted) field $\mathcal{V}_\mu$ and the coset (or remaining) field $\mathcal{X}_\mu$.

Recently, this idea has been readdressed under a different viewpoint by regarding this decomposition merely as a non-linear change of variables. The resulting reformulation of the Yang-Mills theory has been first performed in the $SU(2)$ case \cite{kondo2006yang, *kondo2005brst} and later extended to the general $SU(N)$ case \cite{kondo2008reformulating}. It turned out, however, that for $N\geq 3$ the decomposition is no longer unique, as the gauge field is decomposed into the part lying in the stability group $H$ and its remainder $SU(N)/H$. But already in the $N=3$ case, for example, there are two options for the stability group,
\muu{
H^{max}=U(1)\times U(1)\ \ \ \text{or} \ \ \ \ H^{min}=U(2).
}
The first case is referred to as maximal option and involves the definition of two color-fields. In the second case, the minimal option, we only need one color-field to define the decomposition. Only for $SU(2)$ the choice is unique and the maximal and minimal options are equivalent.

In this paper we will consider the decomposition in the minimal option, that is decomposing $G=SU(N)$ into the stability group $H=U(N-1)$ and the coset $G/H=SU(N)/U(N-1)$. However, counting the degrees of freedom in the color-field extended Yang-Mills theory one finds that they exceed those of the original Yang-Mills theory. This is taken care of by means of the so-called reduction condition,
\muu{
[\mathfrak{n}, D_\mu[\mathcal{A}]D^\mu[\mathcal{A}]\mathfrak{n}]=0.\label{redcond}
}
Solving this differential equation gives $\mathfrak{n}$ as a functional of $\mathcal{A}_\mu$, eliminating the superfluous degrees of freedom. Even though this procedure is reminiscent of a ``gauge fixing'' from the extended Yang-Mills theory to a theory equipollent to the original $SU(N)$ Yang-Mills theory, it should be remarked that the idea behind it is conceptually different from the usual gauge fixing.

Another key aspect within this reformulation is that one could introduce a gauge invariant mass term for the homogeneously transforming coset field \cite{kondo2006yang},
\muu{
m_X^2 \Tr_{G/H} \left(\mathcal{X}_\mu \mathcal{X}^\mu \right).
}
This leads to the idea of a non-vanishing covariant coset gluon condensate that could lead to a dynamically generated mass term coming from the quartic self-interaction term. In fact, at least in the $SU(2)$ and $SU(3)$ case the ``abelian'' dominance implied by such a condensate has already been observed in lattice simulations in terms of an exponential falloff of the covariant coset field two-point function in the infrared, suggesting the dynamical generation of the gluon mass \cite{shibata2007compact, shibata2013non}. Moreover, in preceding works the assumption of a non-zero condensate lead to many more implications such as removal of the Nielsen-Olesen instability in the Savvidy vacuum \cite{kondo2006gauge, kondo2014stability}, the Faddeev-Niemi model (describing glueballs as knot-solitons) as a low-energy effective theory of this reformulated Yang-Mills theory \cite{kondo2004magnetic,*kondo2005magnetic,kondo2006glueball}, or quark confinement at low temperatures \cite{kondo2015confinement}. For a more detailed introduction of the reformulated Yang-Mills theory and its main features please see the review \cite{KONDO20151}.

The main goal of this paper is to complement the lattice simulations by an analytical study of the coset field mass generation. In order to do so we consider a slightly modified dimension-2 operator,
\muu{
\mathcal{O}=\frac{1}{2}\mathcal{X}_\mu^a \mathcal{X}^{\mu a} - i \xi C^a \bar{C}^a,
}
as suggested in  \cite{kondo2001vacuum}. Here, $\xi$ is the ``gauge fixing'' parameter corresponding to the reduction condition and the index $a$ runs over the coset space $SU(N)/U(N-1)$. The pure gluon condensate is recovered for $\xi=0$. It should be remarked that this operator has already been analyzed in different gauges, such as for a non-decomposed gauge group (and thus with the index running over the whole gauge group) and in the usual covariant gauge fixing \cite{kondo2002renormalizing}, within the Curci-Ferrari gauge \cite{curci1976class, curci1976slavnov}, or within the maximal Abelian gauge (MAG) \cite{hooft1981topology, kronfeld1987topology, kronfeld1987monopole}. In order to construct a well-defined effective action, a new method called local composite operator (LCO) formalism has been developed \cite{verschelde1995perturbative, knecht2001new} and used to show not only the existence of the ``full'' gluon condensate $A_\mu^A A^{\mu A}$ in linear covariant gauges \cite{verschelde2001non, dudal2004dynamical} but also the existence of the gluon-ghost condensate in both aforementioned gauges \cite{dudal2003anomalous, dudal2003gluon, dudal2004analytic}. We would like to readdress this issue within our novel decomposition.

The paper is organized as follows. In the second section we set up the Lagrangian, explaining the decomposition of the gauge field and the incorporation of the ``gauge fixing'' related to the reduction condition. We then briefly discuss the one-loop renormalization and calculate all renormalization group (RG) functions. The third section is dedicated to the proper introduction of the composite operator $\mathcal{O}=\frac{1}{2}\mathcal{X}_\mu^a \mathcal{X}^{\mu a} - i \xi C^a \bar{C}^a$. We prove its (on-shell) BRST invariance and the multiplicative renormalizability to one-loop level. In the fourth section we use the LCO formalism in order to deal with divergences quadratic in the source, coming from the composite operator source term $J\mathcal{O}$. In the last section, the one-loop effective potential for the composite operator is calculated and the existence of the condensate is discussed.

\section{Lagrangian in the minimal option}
\label{sec:reformulated}

Before discussing the decomposition we restrict ourselves to the case of a space-time independent color-field, thus discarding the monopole degrees of freedom. The analytical treatment of the dynamical color-field is a highly complicated task and plays only a minor role when investigating the coset field condensate. Therefore, considering a fixed color-field is sufficient for our purposes. In particular, we choose the color-field to be the last Cartan generator,
\begin{align}
\mathfrak{n}=T^\gamma,\  \gamma=N^2-1,\label{nchoice}
\end{align}
where the generators are normalized as $2\Tr(T^A T^B)=\delta^{AB}$. Even after the choice \eqref{nchoice}, the theory still has the local (and global) $U(N-1)$ symmetry, because the constant form of the color-field is maintained under the $U(N-1)$ gauge transformations, as it is supposed to transform in the adjoint way, $\mathfrak{n}(x)\to U(x) \mathfrak{n}(x) U^\dagger (x)$ for $U(x)\in SU(N)$. The gauge field is then decomposed into $\mathcal{A}_\mu=\mathcal{V}_\mu+\mathcal{X}_\mu$, where the residual field $\mathcal{V}_\mu$ takes value in $U(N-1)$ and the covariant coset field $\mathcal{X}_\mu$ takes value in the coset space $SU(N) /U(N-1)$. It is convenient to further decompose $U(N-1) \to SU(N-1) \times U(1)$, since $T^\gamma$ commutes with all other generators of $U(N-1)$. In a suitable basis we write 
\muu{
\mathcal{X}_\mu &= X_\mu^a T^a \in \mathfrak{su}(N)-\mathfrak{u}(N-1),\nonumber\\
\mathcal{V}_\mu &=V_\mu^J T^J = V_\mu^j T^j + V_\mu^\gamma T^\gamma \in \mathfrak{su}(N-1)+\mathfrak{u}(1),
}
where the generators obey the following commutator relations,
\muu{
[T^a,T^b]&=if^{abJ}T^J, \nonumber \\
[T^J,T^a]&=if^{Jab}T^b, \nonumber \\
[T^J,T^K]&=if^{JKL}T^L,\nonumber \\
}
and the last relation is further decomposed according to $U(N-1)=SU(N-1)\times U(1)$,
\muu{
[T^j,T^k]&=if^{jkl}T^l,\nonumber \\
[T^\gamma,T^j]&=0, \nonumber \\
[T^\gamma,T^\gamma]&=0.\nonumber\\
}
In the $SU(3)$ case for example the different indices take the values $a\in\{4,5,6,7\}$, $j\in\{1,2,3\}$ and $\gamma =8$. The $SU(2)$ case is special in the sense that the decomposition reads $SU(2) \to SU(2) / U(1) \times U(1)$ and therefore the residual field does not possess the $SU(N-1)$ part. We simply find $a\in\{1,2\}$ and $\gamma=3$. Using the fact that in this decomposition the only non-vanishing structure constants are $f^{abJ}$ and $f^{JKL}$ (note however that $f^{\gamma KL}=0$), the Yang-Mills Lagrangian is decomposed as, 
\muu{
\mathcal{L}_{YM}=-\frac{1}{4} F_{\mu \nu}^a F^{\mu \nu a} - \frac{1}{4}F_{\mu \nu}^J F^{\mu \nu J},
}
where 
\muu{
F_{\mu \nu}^a &= D_\mu^{ab} X_\nu^b - D^{ab}_{\nu} X_\mu^b,  \nonumber \\
F_{\mu \nu}^J &= \partial_\mu V_\nu^J - \partial_\nu V_\mu^J +g f^{Jab} X_\mu^a X_\nu^b+gf^{JKL}V_\mu^K V_\nu^L,
}
and the covariant derivative is defined with respect to the residual field, \muu{
D_\mu^{ab}=\delta^{ab}\partial_\mu - gf^{abJ} V_\mu^J.
}
Finally, we want to remark that due to the fixing of the color-field $\mathfrak{n}$, the originally gauge invariant mass term $m^2_X \Tr\left(\mathcal{X}_\mu \mathcal{X}^\mu \right)$ loses its gauge invariance. However, one is at least able to construct a BRST invariant composite operator containing the coset gluon condensate, cf. section \ref{sec:the composite}.

\subsection{BRST invariance and gauge fixing}
\label{sec:BRST}

In the following, we want to incorporate the reduction condition \eqref{redcond} by means of a ``gauge fixing''. First, let us recall the BRST transformation $\delta_B$, 
\muu{
\delta_B X_\mu^a &= D_\mu^{ab} \omega^{ b}+gf^{abJ}X_\mu^b C^J,  \nonumber\\
\delta_B V_\mu^J &= \partial_\mu C^J + gf^{JKL}V_\mu^K C^L + gf^{Jab}X_\mu^a \omega^b, \nonumber \\
\delta_B \omega^a &= -g f^{abJ} \omega^b C^J, \nonumber\\
\delta_B C^J &= -\frac{g}{2} f^{JKL} C^K C^L -\frac{g}{2}f^{Jab}\omega^a \omega^b, \nonumber \\
\delta_B \bar{\omega}^a &= i N^a, \ \ \ \delta_B \bar{C}^J=iN^J,\nonumber\\
\delta_B N^a &= \delta_B N^J =0, \label{BRST}
}
and the anti-BRST transformation $\bar{\delta}_B$,
\muu{
\bar{\delta}_B X_\mu^a &= D_\mu^{ab} \bar{\omega}^{ b}+gf^{abJ}X_\mu^b \bar{C}^J,  \nonumber\\
\bar{\delta}_B V_\mu^J &= \partial_\mu \bar{C}^J + gf^{JKL}V_\mu^K \bar{C}^L + gf^{Jab}X_\mu^a \bar{\omega}^b, \nonumber \\
\bar{\delta}_B \bar{\omega}^a &= -g f^{abJ} \bar{\omega}^b \bar{C}^J, \nonumber\\
\bar{\delta}_B \bar{C}^J &= -\frac{g}{2} f^{JKL} \bar{C}^K \bar{C}^L -\frac{g}{2}f^{Jab}\bar{\omega}^a \bar{\omega}^b, \nonumber \\
\bar{\delta}_B \omega^a &= i \bar{N}^a, \ \ \ \bar{\delta}_B C^J=i\bar{N}^J,\nonumber\\
\bar{\delta}_B \bar{N}^a &= \bar{\delta}_B \bar{N}^J =0, \label{antiBRST}
}
where we have introduced the Nakanishi-Lautrup field $\mathcal{N}=(N^a,N^j.N^\gamma)$, the ghosts $\mathcal{C}=(\omega^a,C^j,C^\gamma)$ and anti-ghosts $\bar{\mathcal{C}}=(\bar{\omega}^a,\bar{C}^j,\bar{C}^\gamma)$ according to the three parts $SU(N)/U(N-1)$, $SU(N-1)$ and $U(1)$, respectively. The quantity $\bar{\mathcal{N}}$ finally is defined as $\bar{\mathcal{N}}=g[\mathcal{C},\bar{\mathcal{C}}]-\mathcal{N}$. Both transformations are nilpotent, $\delta_B^2 = \bar{\delta}_B^2=0$ and satisfy $\{\delta_B, \bar{\delta}_B\}=0$. It is shown \cite{KONDO20151} that the reduction condition \eqref{redcond} can be cast into the form
\muu{
D_\mu^{ab} X^{\mu b}=0. 
} 
Then we can introduce the ``gauge fixing term'' as 
\muu{
\mathcal{L}_{GF}^{red}&=i\delta_B \bar{\delta}_B \left( \frac{1}{2} X_\mu^a X^{\mu a} -i\frac{\xi}{2}\omega^a \bar{\omega}^a \right)\nonumber \\
&=-i\delta_B \left(  \bar{\omega}^a \left[ D_\mu^{ab} X^{\mu b}  + \frac{\xi}{2}N^a \right]-i\frac{\xi g}{2}f^{abJ}\bar{\omega}^a \bar{\omega}^b C^J \right).
}
Beside its different interpretation this also differs from the standard gauge fixing procedure by the last term, generating the four-ghost interaction after performing the BRST transformation. This is necessary since we deal with a non-linear gauge fixing, in which case the four-ghost interaction preserves the renormalizability of the theory \cite{min1985renormalization}. Indeed, we find
\muu{
\mathcal{L}_{GF}^{red}&=\frac{\xi}{2} N^a N^a + i\bar{\omega}^a D^{\mu ab}[V]D_\mu^{bc}[V]\omega^c +N^a D_\mu^{ab}[V]X^{\mu b} \nonumber\\
&+igf^{abJ}\bar{\omega}^a (D^{\mu bc}[V] X_\mu^c ) C^J - i\xi g f^{aJb}  C^J \bar{\omega}^b N^a \nonumber \\
&+\frac{\xi g^2}{4} f^{abJ}f^{cdJ} \bar{\omega}^a \bar{\omega}^b \omega^c \omega^d +ig^2 f^{abJ}f^{cdJ}  X^{\mu a} X_\mu^c \bar{\omega}^b \omega^d \nonumber\\
&-\frac{\xi g^2}{4}f^{jkl} f^{ajb} \bar{\omega}^a \bar{\omega}^b C^k C^l, 
}
where the four-ghost interactions are obtained.

We are left with fixing the residual $U(N-1)$ symmetry. We choose the simple Lorenz gauge, where according to the decomposition $U(N-1)=SU(N-1) \times U(1)$ two different gauge fixing parameters are introduced,
\muu{
\mathcal{L}_{GF}^{res}&=-i\delta_B \left[\bar{C}^j \left(\partial_\mu V^{\mu j}+ \frac{\lambda}{2} N^j \right)\right]\nonumber\\
&\ \ \ \ \ \ \ \ -i\delta_B \left[\bar{C}^\gamma \left(\partial_\mu V^{\mu \gamma}+ \frac{\alpha}{2} N^\gamma \right)\right]\nonumber\\
&=\frac{\lambda}{2} N^j N^j +\frac{\alpha}{2}N^\gamma N^\gamma + N^J \partial_\mu V^{\mu J} + i \bar{C}^J \partial^2  C^J \nonumber\\
&+ i g f^{jkl}\bar{C}^j \partial_\mu \left(V^{\mu k} C^l \right) + i g f^{Jab} \bar{C}^J \partial_\mu \left(X^{\mu a} \omega^b \right).
}
Even though two different gauge fixing parameters are introduced, it can be shown that both the $SU(N-1)$ and the $U(1)$ part of the residual gauge fixing Lagrangian are independently invariant under the global $U(1)$, the global $SU(N-1)$ and the combined global $U(N-1)$ transformations, irrespective of the choice of $\lambda$ and $\alpha$. In particular, it is shown later that $\lambda$ and $\alpha$ receive different one-loop corrections, such that this distinction is actually necessary from the viewpoint of renormalization.

Finally, the Nakanishi-Lautrup field is integrated out, which casts the ``gauge fixing'' Lagrangian into the form

\muu{
\mathcal{L}_{GF}^{red}&=-\frac{1}{2\xi}(D_\mu^{ab}[V]X^{\mu b})^2 +i\bar{\omega}^a D^{\mu ab}[V]D_\mu^{bc}[V]\omega^c \nonumber \\
&+\frac{\xi g^2}{4} f^{abJ}f^{cdJ} \bar{\omega}^a \bar{\omega}^b \omega^c \omega^d +ig^2 f^{abJ}f^{cdJ}  X^{\mu a} X_\mu^c \bar{\omega}^b \omega^d, \label{LRED}\\
\mathcal{L}_{GF}^{res}&=-\frac{1}{2\lambda}\left( \partial_\mu V^{\mu j}\right)^2-\frac{1}{2\alpha}\left( \partial_\mu V^{\mu \gamma}\right)^2+i \bar{C}^J \partial^2 C^J\nonumber \\
&  +igf^{jkl}\bar{C}^j \partial_\mu (V^{\mu k}C^l)+igf^{Jab}\bar{C}^J \partial_\mu (X^{\mu a}\omega^b),\label{LRES} 
}
and the equations of motion for the Nakanishi-Lautrup field read
\muu{
N^j = -\frac{1}{\lambda} \partial_\mu V^{\mu j}, \ \ N^\gamma = -\frac{1}{\alpha} \partial_\mu V^{\mu \gamma}, \nonumber \\
 N^a = -\frac{1}{\xi} D_\mu^{ab}X^{\mu b} + igf^{abJ} \bar{\omega}^b C^J. \label{BEOM}
}
This completes the ``gauge fixing'' and leaves us with a BRST invariant Lagrangian
\muu{
\mathcal{L}=\mathcal{L}_{YM}+\mathcal{L}^{red}_{GF}+\mathcal{L}^{res}_{GF}.
}

\begin{figure}[h!] 
  \begin{minipage}[b]{0.4\linewidth}
    \includegraphics[width=1\linewidth]{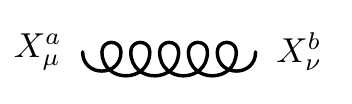} 
  \end{minipage} 
  \begin{minipage}[b]{0.4\linewidth}
    \includegraphics[width=1\linewidth]{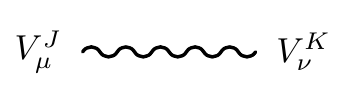} 
  \end{minipage} 
  \begin{minipage}[b]{0.4\linewidth}
    \includegraphics[width=1\linewidth]{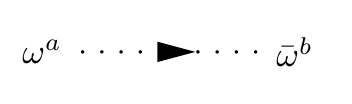} 
  \end{minipage}
  \begin{minipage}[b]{0.4\linewidth}
    \includegraphics[width=1\linewidth]{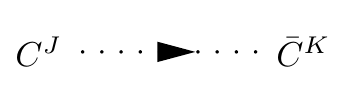}  
  \end{minipage}
  \caption{Propagators} \label{propic} 
\end{figure}

\begin{figure}[h!]
\begin{minipage}{0.2\textwidth}
\includegraphics[width=1\linewidth]{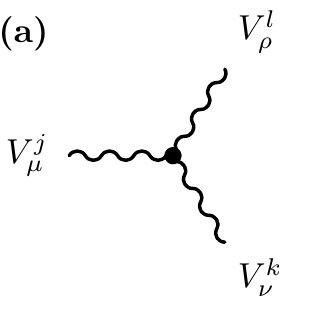}
\end{minipage}%
\begin{minipage}{0.2\textwidth}
\includegraphics[width=1\linewidth]{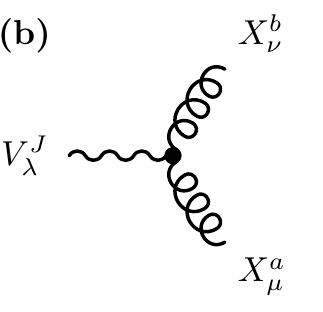}
\end{minipage}
\begin{minipage}{0.2\textwidth}
\includegraphics[width=1\linewidth]{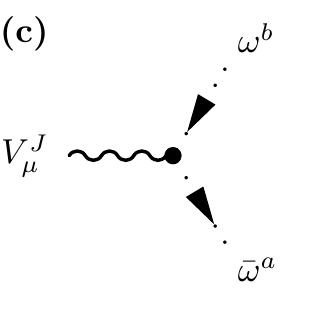}
\end{minipage}%
\begin{minipage}{0.2\textwidth}
\includegraphics[width=1\linewidth]{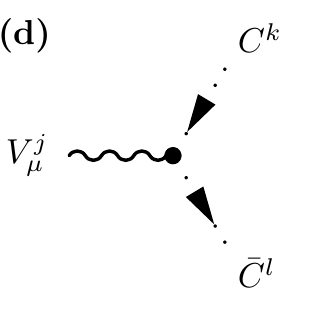}
\end{minipage}
\begin{minipage}{0.2\textwidth}
\includegraphics[width=1\linewidth]{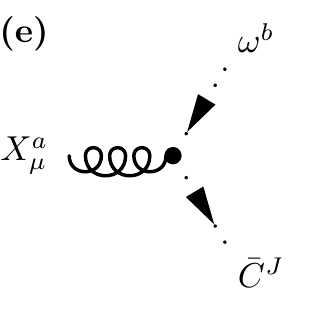}
\end{minipage}
\begin{minipage}{0.2\textwidth}
\hfill
\caption{Three-field vertices.}\label{threefvert}
\end{minipage}
\end{figure}

We proceed with the one-loop analysis of our theory,
\small
\muu{
\mathcal{L}&=\mathcal{L}_{YM}+i\delta_B \bar{\delta}_B \left(\frac{1}{2}X_\mu^a X^{\mu a} - i \frac{\xi}{2} \omega^a \bar{\omega}^a \right)\nonumber \\
&-i\delta_B \left[\bar{C}^j \left(\partial_\mu V^{\mu j}+ \frac{\lambda}{2} N^j \right)\right] -i\delta_B \left[\bar{C}^\gamma \left(\partial_\mu V^{\mu \gamma}+ \frac{\alpha}{2} N^\gamma \right)\right].
}
The induced propagators are shown in Fig.~\ref{propic},

and imply the Feynman rules
\muu{
\langle X_\mu^a X_\nu^b \rangle&= -\delta_{ab} \frac{i}{p^2}\Big(g^{\mu \nu}-(1-\xi) \frac{p_\mu p_\nu}{p^2}\Big),  \\
\langle V_\mu^j V_\nu^k \rangle&=  -\delta_{jk} \frac{i}{p^2}\Big(g^{\mu \nu}-(1-\lambda) \frac{p_\mu p_\nu}{p^2}\Big), \\
\langle V_\mu^\gamma V_\nu^\gamma \rangle&=  - \frac{i}{p^2}\Big(g^{\mu \nu}-(1-\alpha) \frac{p_\mu p_\nu}{p^2}\Big), \\
\langle \omega^a \bar{\omega}^b \rangle &= -\delta^{ab} \frac{1}{p^2},   \\
\langle C^J \bar{C}^K \rangle&= -\delta^{JK} \frac{1}{p^2}.
}
Furthermore, there exist five three-field vertices, see Fig.~\ref{threefvert},

with the Feynman rules
\muu{
&\textbf{(a)}\ \  i\langle V_\mu^j(p)V_\nu^k(q) V_\rho^l(r) \rangle =   \nonumber \\
&\delta_{p+q+r}  gf^{jkl} \left\{ g_{\mu \nu} (p-q)_\rho  + g_{\mu \rho} (r-p)_\nu +g_{\nu \rho} (q-r)_\mu \right\},\\
&\textbf{(b)}\ \ i\langle X_\mu^a(p) X_\nu^b(q) V_\lambda^J(r) \rangle =   \nonumber\\
& \delta_{p+q+r}  g f^{abJ}  \{ g_{\mu \nu} (q-p)_\lambda + g_{\lambda \nu} \left( r -q-\xi^{-1}p  \right)_\mu \nonumber \\
& \ \ \ \ \ \ \ \ \ \ - g_{\lambda \mu} \left( r -p -\xi^{-1}q   \right)_\nu   \} ,\\
&\textbf{(c)}\ \ i\langle V_\mu^J(r) \omega^b(q) \bar{\omega}^a(p)\rangle = - \delta_{q+r-p} igf^{aJb} (q+p)_\mu, \\
&\textbf{(d)}\ \ i\langle V_\mu^j(r) C^k(q) \bar{C}^l(p)\rangle = -\delta_{q+r-p}i g f^{jkl} p_\mu, \\
&\textbf{(e)}\ \ i\langle X_\mu^a(r) \omega^b(q) \bar{C}^J(p)=-\delta_{r+q-p}igf^{abJ} p_\mu,
}
where we have defined $\delta_q := (2\pi)^D \delta(q)$.

\begin{figure}[h!]
\begin{minipage}{0.2\textwidth}
\includegraphics[width=1\linewidth]{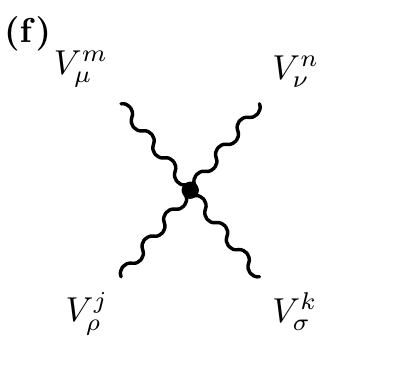}
\end{minipage}%
\begin{minipage}{0.2\textwidth}
\includegraphics[width=1\linewidth]{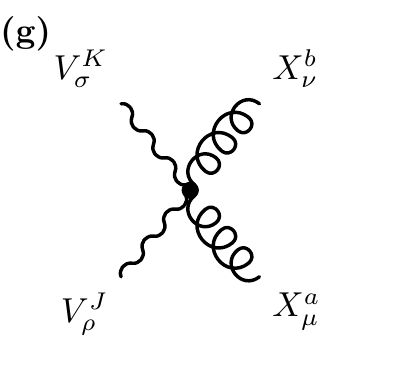}
\end{minipage}
\begin{minipage}{0.2\textwidth}
\includegraphics[width=1\linewidth]{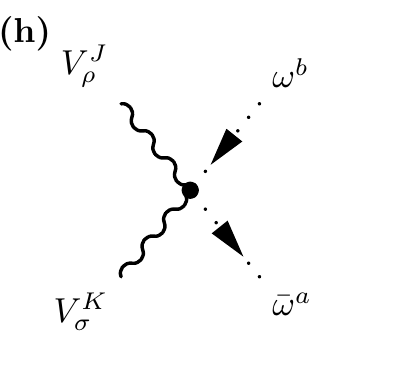}
\end{minipage}%
\begin{minipage}{0.2\textwidth}
\includegraphics[width=1\linewidth]{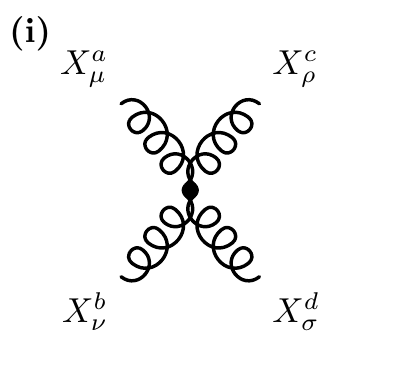}
\end{minipage}
\begin{minipage}{0.2\textwidth}
\includegraphics[width=1\linewidth]{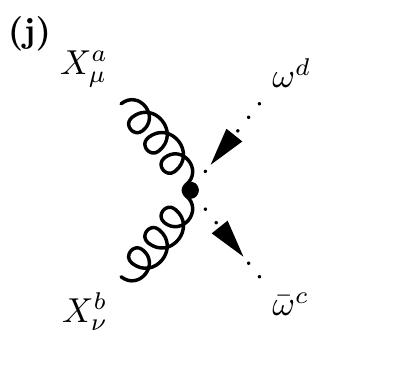}
\end{minipage}
\begin{minipage}{0.2\textwidth}
\includegraphics[width=1\linewidth]{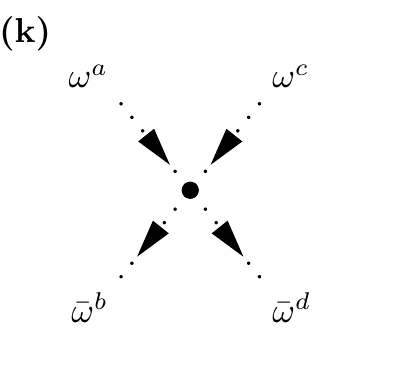}
\end{minipage}
\caption{Four-field vertices.}\label{4fvert}
\end{figure}

Moreover, we find six four-field vertices, see Fig.~\ref{4fvert},
and with defining $I_{\mu \nu, \rho \sigma}=g_{\mu \rho}g_{\nu \sigma}-g_{\mu \sigma}g_{\nu \rho}$ we obtain the Feynman rules
\muu{
&\textbf{(f)}\ \  i\langle V_\mu^m(p) V_\nu^n(q) V_\rho^j(r) V_\sigma^k(s) \rangle = \nonumber \\
&-ig^2 \delta_{p+q+r+s} \left\{  f^{mnl}f^{jkl} I_{\mu \nu, \rho \sigma}  \right. \nonumber \\
&\left. +f^{mjl}f^{nkl} I_{\mu \rho, \nu \sigma} +f^{mkl} f^{njl} I_{\mu \sigma, \nu \rho}\right\},\\
&\textbf{(g)}\ \ i\langle X_\mu^a(p) X_\nu^b(q) V_\rho^J(r) V_\sigma^K(s) \rangle = \nonumber \\
&-ig^2 \delta_{p+q+r+s} \left\{ 2 f^{abM}f^{JKM} I_{\mu \nu, \rho \sigma} \right. \nonumber \\
& +f^{cKa}f^{bJc} \left( \left( 1 -\xi^{-1}\right) g_{\mu \sigma} g_{\rho \nu}-g_{\mu \nu} g_{\rho \sigma} \right) \nonumber \\
&\left. +f^{cJa} f^{bKc} \left(\left(1-\xi^{-1}\right) g_{\mu \rho}g_{\nu \sigma}-g_{\mu \nu} g_{\rho \sigma} \right)\right\},\label{4gluonxi}\\
&\textbf{(h)}\ \ i\langle V_\rho^{J}(r)V_\sigma^{K}(s)\omega^b(q) \bar{\omega}^a(p) \rangle= \nonumber \\
&-g^2 \delta_{q-p+r+s}g_{\rho \sigma} (f^{aJc}f^{cKb}+f^{aKc}f^{cJb}),\label{ghostgluon1}\\
&\textbf{(i)}\ \ i\langle  X_\mu^a(p) X_\nu^b(q) X_\rho^c(r) X_\sigma^d(s) \rangle = \nonumber \\
&-ig^2 \delta_{p+q+r+s} \left\{  f^{acJ}f^{bdJ} I_{\mu \rho,  \nu \sigma}  \right. \nonumber \\
&\left. +f^{adJ}f^{bcJ} I_{\mu \sigma,\nu \rho} +f^{abJ} f^{cdJ} I_{\mu \nu ,\rho \sigma}\right\},\\
&\textbf{(j)}\ \ i\langle X_\mu^a(r) X_\nu^b(s) \omega^d(q) \bar{\omega}^c(p) \rangle= \nonumber \\
&\delta_{r+s+q-p}g^2 g_{\mu \nu} \left\{ f^{aJc} f^{bdJ} + f^{bJc} f^{adJ} \right\},\label{ghostgluon2}\\
&\textbf{(k)}\ \ i\langle \omega^a(q) \bar{\omega}^b(p) \omega^c(r) \bar{\omega}^d(s) \rangle = \nonumber \\
&-i\xi g^2 \delta_{q+r-p-s} f^{acJ}f^{Jbd}\label{fourghost}. 
}

Let us point out the differences to some other common gauges. As mentioned before, our ``gauge condition'' $D_\mu^{ab}X^{\mu b}=0$ is non-linear unlike for example the standard Lorenz gauge. This requires the four-ghost vertex \eqref{fourghost} in order to render our theory renormalizable. Furthermore, the non-linearity gives rise to the $\xi$-dependent corrections in the four-gluon interaction \eqref{4gluonxi} as well as in the two-gluon-two-ghost interactions \eqref{ghostgluon1} and \eqref{ghostgluon2}. These features are also observed in the MAG. However, in the MAG the coset field takes value in $SU(N)/U(1)^{N-1}$, where the quotient is Abelian and thus $f^{JKL}=0$. In our decomposition however, the coset field takes value in $SU(N)/U(N-1)$, where the quotient $U(N-1)=SU(N-1)\times U(1)$ is non-Abelian, $f^{JKL}\neq 0$. In fact, only the $U(1)$ generator $T^{\gamma}$ commutes with all other generators of the quotient, $f^{\gamma KL}=0$ while $f^{jkl}\neq 0$. On the other hand, in the MAG one has $f^{abc}\neq 0$ while in our decomposition $f^{abc}=0$. Once more we emphasize that in the case of $SU(2)$ our decomposition and the MAG are equivalent.

Finally, we state some color-algebra relations that are required in the upcoming one-loop calculations. Starting from the $SU(N)$ and $SU(N-1)$ identities 
\muu{
f^{ACD}f^{BCD}&=N \delta^{AB},\nonumber\\
f^{jmn}f^{kmn}&=(N-1)\delta^{jk},
}
and using the fact that only $f^{abj}$, $f^{ab\gamma}$ and $f^{jkl}$ are non-vanishing structure constants one derives using the Jacobi identity
\begin{align}
f^{\gamma ab}f^{\gamma ab}&=N, \nonumber\\
f^{iab}f^{jab}&=\delta^{ij},\nonumber\\
f^{imn}f^{jmn}&=(N-1)\delta^{ij},\nonumber\\
f^{ab\gamma}f^{cb\gamma}&=\frac{N}{2(N-1)}\delta^{ac},\nonumber\\
f^{abj}f^{cbj}&=\frac{N(N-2)}{2(N-1)}\delta^{ac},\nonumber\\
f^{\gamma ab}f^{jab}&=0.
\end{align}

\subsection{One-loop analysis}
\label{sec:oneloop}

We start our one-loop analysis with the introduction of the renormalization factors

\muu{
\begin{array}{cccc}
X_\mu^a = Z_X^{\frac{1}{2}} X_{\mu R}^a, & V_\mu^j = Z_V^{\frac{1}{2}} V_{\mu R}^j, & V_\mu^\gamma = \tilde{Z}_V^{\frac{1}{2}} V_{\mu R}^\gamma, &g= Z_g g_R,\\
C^j = \zdia^\frac{1}{2} C^j_R, & \bar{C}^j = \zadia^\frac{1}{2} \bar{C}^j_R, & C^\gamma = \tilde{Z}_{C}^\frac{1}{2} C^\gamma_R, & \bar{C}^\gamma = \tilde{Z}_{\bar{C}}^\frac{1}{2} \bar{C}^\gamma_R, \\
\omega^a = \zogh^\frac{1}{2} \omega^a_R, & \bar{\omega}^a = \zogh^\frac{1}{2} \bar{\omega}^a_R, & \xi = Z_\xi \xi_R, & \alpha = Z_\alpha \alpha_R, \\
\lambda = Z_\lambda \lambda_R. 
\end{array}
}
Note that we took the same renormalization factor for the coset ghost and anti-ghost, while for the residual ghost they must be chosen independently \cite{shinohara2003most}. Furthermore, one has to distinguish between the $SU(N-1)$ and $U(1)$ gauge field and ghosts. Hereafter, the subscript $R$ is dropped again. The corresponding counterterm Lagrangian is given in appendix \ref{appA:counter} together with the relation between the counterterms and the renormalization factors. The renormalization is done within dimensional regularization. 

We begin with the coset gluon self-energy. The one-loop correction reads\footnote{Here and in the following, labelling an internal residual field line as $V^j+V^\gamma$ means one has to calculate the sum of two diagrams, one with the propagator of the $SU(N-1)$ field $V^j$ and one with the propagator of the $U(1)$ field $V^\gamma$.}

\begin{equation}
\begin{gathered}
\begin{fmfgraph*}(50,50)
\fmfleft{i1}
\fmflabel{$\phantom{X_\mu^a}$}{i1}
\fmfright{o1}
\fmflabel{$\phantom{X_\nu^b}$}{o1}
\fmf{gluon}{i1,v1}
\fmf{gluon}{v1,o1}
\fmfblob{.30w}{v1}
\end{fmfgraph*}
\end{gathered}\  =   \ 
\begin{gathered}
\begin{fmfgraph*}(50,50)
\fmfleft{i1}
\fmfright{o1}
\fmf{gluon}{i1,v1}
\fmf{photon,label=$V^j+V^\gamma$}{v1,v1}
\fmf{gluon}{v1,o1}
\fmfdotn{v}{1}
\end{fmfgraph*}
\end{gathered}
+ \ \
\begin{gathered}
\begin{fmfgraph*}(50,50)
\fmfleft{i1}
\fmfright{o1}
\fmf{gluon}{i1,v1}
\fmf{gluon,right,label=$\phantom{x}$}{v1,v1}
\fmf{gluon}{v1,o1}
\fmfdotn{v}{1}
\end{fmfgraph*}
\end{gathered}\nonumber
\end{equation}
\begin{equation}+ \ \ \ \ \ 
\begin{gathered}
\begin{fmfgraph*}(60,60)
\fmfleft{i1}
\fmfright{o1}
\fmf{gluon}{i1,v1}
\fmf{ghost,label=$\omega$}{v1,v1}
\fmf{gluon}{v1,o1}
\fmfdotn{v}{1}
\end{fmfgraph*}
\end{gathered}+ \ \ \ \ \ \
\begin{gathered}
\begin{fmfgraph*}(60,60)
\fmfleft{i1}
\fmfright{o1}
\fmf{gluon}{i1,v1}
\fmf{photon,left,tension=0.4,label=$V^j +V^\gamma$}{v1,v2}
\fmf{gluon,left,tension=0.4,label=$\phantom{x}$}{v2,v1}
\fmf{gluon}{v2,o1}
\fmfdotn{v}{2}
\end{fmfgraph*}
\end{gathered}\ \ .
\end{equation}
\noindent
In dimensional regularization only the last diagram will contribute. The divergent part is calculated as 
\begin{equation}
\begin{gathered}
\begin{fmfgraph*}(50,50)
\fmfleft{i1}
\fmflabel{$X_\mu^a$}{i1}
\fmfright{o1}
\fmflabel{$X_\nu^b$}{o1}
\fmf{gluon}{i1,v1}
\fmf{photon,left,tension=0.4,label=$V^j+V^\gamma$}{v1,v2}
\fmf{gluon,left,tension=0.4,label=$\phantom{x}$}{v2,v1}
\fmf{gluon}{v2,o1}
\fmfdotn{v}{2}
\end{fmfgraph*}
\end{gathered}\nonumber
\end{equation}
\begin{align} = i\delta^{ab}\frac{g^2 \mu^{-2\epsilon}}{(4\pi)^2\epsilon} \frac{N}{2} \left[\left( \frac{17-3\xi }{6}-\frac{\alpha + (N-2)\lambda}{N-1}\right) \left[ g^{\mu \nu} p^2-p^\mu p^\nu \right]\right. \nonumber \\
\left. + \frac{1}{\xi} \left(\frac{6+\xi(\xi+3)}{2 \xi} -\frac{\alpha + (N-2)\lambda}{N-1} \right) p^\mu p^\nu \right], \label{hilf1}
\end{align}
where $\epsilon = \frac{4-D}{2}$. The renormalization factors are expanded according to $Z=1+Z^{(1)}+O(\hbar^2)$. Then, equation \eqref{hilf1} implies
\muu{
\Delta_1^{(1)}&=Z_X^{(1)}=\frac{g^2\mu^{-2\epsilon}}{(4\pi)^2\epsilon}\frac{N}{2}\left(\frac{17}{6}-\frac{\xi}{2} -\frac{\alpha + (N-2)\lambda}{N-1} \right),\label{ZX}\\
\Delta_2^{(1)}&=Z_X^{(1)}-Z_\xi^{(1)}\nonumber\\
&\ \ \ \ \ = \frac{g^2\mu^{-2\epsilon}}{(4\pi)^2\epsilon}\frac{N}{2}\left(\frac{6+\xi(\xi+3)}{2\xi}-\frac{\alpha + (N-2)\lambda}{N-1} \right)\label{ZXminusZxi},
}
and consequently,
\muu{
Z_\xi^{(1)}=\frac{g^2\mu^{-2\epsilon}}{(4\pi)^2\epsilon}\frac{N}{2}\left(\frac{4}{3}-\xi -\frac{3}{\xi}\right).\label{Zxi}
}

We proceed with the self-energy of the residual field, starting with the $SU(N-1)$ part,
\begin{equation}
\begin{gathered}
\begin{fmfgraph*}(50,50)
\fmfleft{i1}
\fmflabel{$V_\mu^j$}{i1}
\fmfright{o1}
\fmflabel{$V_\nu^k$}{o1}
\fmf{photon}{i1,v1}
\fmf{photon}{v1,o1}
\fmfblob{.30w}{v1}
\end{fmfgraph*}
\end{gathered}\ \ \ \ \ \ \ = \ \
\begin{gathered}
\begin{fmfgraph*}(50,50)
\fmfleft{i1}
\fmfright{o1}
\fmf{photon}{i1,v1}
\fmf{photon,label=$V^l$}{v1,v1}
\fmf{photon}{v1,o1}
\fmfdotn{v}{1} 
\end{fmfgraph*}
\end{gathered}+ \ \ 
\begin{gathered}
\begin{fmfgraph*}(50,50)
\fmfleft{i1}
\fmfright{o1}
\fmf{photon}{i1,v1}
\fmf{ghost,label=$\omega$}{v1,v1}
\fmf{photon}{v1,o1}
\fmfdotn{v}{1} 
\end{fmfgraph*}
\end{gathered}\nonumber
\end{equation}
\begin{equation}
+ \ \
\begin{gathered}
\begin{fmfgraph*}(50,50)
\fmfleft{i1}
\fmfright{o1}
\fmf{photon}{i1,v1}
\fmf{gluon,label=$\phantom{x}$}{v1,v1}
\fmf{photon}{v1,o1}
\fmfdotn{v}{1} 
\end{fmfgraph*}
\end{gathered}
+ \ \ \ \ \ \ 
\begin{gathered}
\begin{fmfgraph*}(50,50)
\fmfleft{i1}
\fmflabel{$\phantom{x}$}{i1}
\fmfright{o1}
\fmflabel{$\phantom{x}$}{o1}
\fmf{photon,label=$\phantom{x}$}{i1,v1}
\fmf{photon,left,label=$V^l$,tension=0.4}{v1,v2}
\fmf{photon,left,label=$\phantom{x}$,tension=0.4}{v2,v1}
\fmf{photon,label=$\phantom{x}$}{v2,o1}
\fmfdotn{v}{2}
\end{fmfgraph*}
\end{gathered}+ \ \ \ \ \ 
\begin{gathered}
\begin{fmfgraph*}(50,50)
\fmfleft{i1}
\fmflabel{$\phantom{x}$}{i1}
\fmfright{o1}
\fmflabel{$\phantom{x}$}{o1}
\fmf{photon,label=$\phantom{x}$}{i1,v1}
\fmf{gluon,left,label=$\phantom{x}$,tension=0.4}{v1,v2}
\fmf{gluon,left,label=$\phantom{x}$,tension=0.4}{v2,v1}
\fmf{photon,label=$\phantom{x}$}{v2,o1}
\fmfdotn{v}{2}
\end{fmfgraph*}
\end{gathered}\nonumber
\end{equation}
\begin{equation}+ \ \ \ \ \ 
\begin{gathered}
\begin{fmfgraph*}(50,50)
\fmfleft{i1}
\fmflabel{$\phantom{x}$}{i1}
\fmfright{o1}
\fmflabel{$\phantom{x}$}{o1}
\fmf{photon,label=$\phantom{x}$}{i1,v1}
\fmf{ghost,left,label=$C^l$,tension=0.4}{v1,v2}
\fmf{ghost,left,label=$\phantom{x}$,tension=0.4}{v2,v1}
\fmf{photon,label=$\phantom{x}$}{v2,o1}
\fmfdotn{v}{2}
\end{fmfgraph*}
\end{gathered}+ \ \ \ \ \ 
\begin{gathered}
\begin{fmfgraph*}(50,50)
\fmfleft{i1}
\fmflabel{$\phantom{x}$}{i1}
\fmfright{o1}
\fmflabel{$\phantom{x}$}{o1}
\fmf{photon,label=$\phantom{x}$}{i1,v1}
\fmf{ghost,left,label=$\omega$,tension=0.4}{v1,v2}
\fmf{ghost,left,label=$\phantom{x}$,tension=0.4}{v2,v1}
\fmf{photon,label=$\phantom{x}$}{v2,o1}
\fmfdotn{v}{2}
\end{fmfgraph*}
\end{gathered}\ \ .
\end{equation}
\noindent
Here, only the last four diagrams contribute in dimensional regularization. Their divergent parts read as follows,
\begin{equation}
\begin{gathered}
\begin{fmfgraph*}(60,60)
\fmfleft{i1}
\fmflabel{$V_\mu^j$}{i1}
\fmfright{o1}
\fmflabel{$V_\nu^k$}{o1}
\fmf{photon,label=$\phantom{p}$}{i1,v1}
\fmf{photon,left,label=$V^l$,tension=0.4}{v1,v2}
\fmf{photon,left,label=$\phantom{q_1}$,tension=0.4}{v2,v1}
\fmf{photon,label=$\phantom{q}$}{v2,o1}
\fmfdotn{v}{2}
\end{fmfgraph*}
\end{gathered}\nonumber
\end{equation}
\muu{= i\delta^{jk}\frac{(N-1)}{2} \frac{g^2 \mu^{-2\epsilon}}{(4\pi)^2\epsilon} \left[\left(\frac{25}{6}-\lambda \right) p^2 g^{\mu \nu }-\left(\frac{14 }{3}-\lambda\right)p^{\mu } p^{\nu }\right],
}
\begin{equation}
\begin{fmfgraph*}(60,60)
\fmfleft{i1}
\fmflabel{$V_\mu^j$}{i1}
\fmfright{o1}
\fmflabel{$V_\nu^k$}{o1}
\fmf{photon,label=$\phantom{x}$}{i1,v1}
\fmf{gluon,left,label=$\phantom{x}$,tension=0.4}{v1,v2}
\fmf{gluon,left,label=$\phantom{x}$,tension=0.4}{v2,v1}
\fmf{photon,label=$\phantom{x}$}{v2,o1}
\fmfdotn{v}{2}
\end{fmfgraph*}\nonumber
\end{equation}
\muu{=  i \delta^{jk}\frac{g^2 \mu^{-2\epsilon}}{(4\pi)^2\epsilon} \frac{10}{3} \left[p^2 g^{\mu \nu }-p^{\mu } p^{\nu }\right],
}

\begin{equation}
\begin{gathered}
\begin{fmfgraph*}(60,60)
\fmfleft{i1}
\fmflabel{$V_\mu^j$}{i1}
\fmfright{o1}
\fmflabel{$V_\nu^k$}{o1}
\fmf{photon,label=$\phantom{x}$}{i1,v1}
\fmf{ghost,left,label=$\omega$,tension=0.4}{v1,v2}
\fmf{ghost,left,label=$\phantom{x}$,tension=0.4}{v2,v1}
\fmf{photon,label=$\phantom{x}$}{v2,o1}
\fmfdotn{v}{2}
\end{fmfgraph*}
\end{gathered}\nonumber
\end{equation}
\muu{= i\delta^{jk}\frac{1}{3} \frac{g^2\mu^{-2\epsilon}}{(4 \pi)^2\epsilon} \left[ g^{\mu \nu} p^2 -  p^\mu p^\nu \right],
} 
\begin{equation}
\begin{fmfgraph*}(60,60)
\fmfleft{i1}
\fmflabel{$V_\mu^j$}{i1}
\fmfright{o1}
\fmflabel{$V_\nu^k$}{o1}
\fmf{photon,label=$\phantom{x}$}{i1,v1}
\fmf{ghost,left,label=$C^l$,tension=0.4}{v1,v2}
\fmf{ghost,left,label=$\phantom{x}$,tension=0.4}{v2,v1}
\fmf{photon,label=$\phantom{x}$}{v2,o1}
\fmfdotn{v}{2}
\end{fmfgraph*}\nonumber
\end{equation}
\muu{= i\delta^{jk}\frac{(N-1)}{2} \frac{g^2\mu^{-2\epsilon}}{(4 \pi)^2\epsilon} \left[\frac{1}{6}p^2 g^{\mu \nu}  +\frac{1}{3} p^\mu p^\nu \right].
}
The sum is transverse and we therefore find
\muu{
\Delta_3^{(1)}&=Z_V^{(1)}=\frac{g^2\mu^{-2\epsilon}}{(4\pi)^2\epsilon}\left[\frac{13N+9}{6}-\frac{\lambda}{2}(N-1) \right],\label{ZV}\\
\Delta_4^{(1)}&=Z_V^{(1)}-Z_\lambda^{(1)}=0,
}
or
\muu{
Z_\lambda^{(1)}=\frac{g^2\mu^{-2\epsilon}}{(4\pi)^2\epsilon}\left[\frac{13N+9}{6}-\frac{\lambda}{2}(N-1) \right].
}

The case of the $U(1)$ part is more simple, since $f^{\gamma KL}=0$. The one-loop correction is given by
\begin{equation}
\begin{gathered}
\begin{fmfgraph*}(50,50)
\fmfleft{i1}
\fmflabel{$V_\mu^\gamma$}{i1}
\fmfright{o1}
\fmflabel{$V_\nu^\gamma$}{o1}
\fmf{photon}{i1,v1}
\fmf{photon}{v1,o1}
\fmfblob{.30w}{v1}
\end{fmfgraph*}
\end{gathered}\ \ \ \ \ \ \ = \ \
\begin{gathered}
\begin{fmfgraph*}(50,50)
\fmfleft{i1}
\fmfright{o1}
\fmf{photon}{i1,v1}
\fmf{ghost,label=$\omega$}{v1,v1}
\fmf{photon}{v1,o1}
\fmfdotn{v}{1} 
\end{fmfgraph*}
\end{gathered}
+ \
\begin{gathered}
\begin{fmfgraph*}(50,50)
\fmfleft{i1}
\fmfright{o1}
\fmf{photon}{i1,v1}
\fmf{gluon,label=$\phantom{x}$}{v1,v1}
\fmf{photon}{v1,o1}
\fmfdotn{v}{1} 
\end{fmfgraph*}
\end{gathered}\nonumber
\end{equation}
\begin{equation}
+ \ \ \ \ \ 
\begin{gathered}
\begin{fmfgraph*}(80,60)
\fmfleft{i1}
\fmflabel{$\phantom{x}$}{i1}
\fmfright{o1}
\fmflabel{$\phantom{x}$}{o1}
\fmf{photon,label=$\phantom{x}$}{i1,v1}
\fmf{gluon,left,label=$\phantom{x}$,tension=0.4}{v1,v2}
\fmf{gluon,left,label=$\phantom{x}$,tension=0.4}{v2,v1}
\fmf{photon,label=$\phantom{x}$}{v2,o1}
\fmfdotn{v}{2}
\end{fmfgraph*}
\end{gathered} \ \ \ + \ \ \
\begin{gathered}
\begin{fmfgraph*}(60,60)
\fmfleft{i1}
\fmflabel{$\phantom{x}$}{i1}
\fmfright{o1}
\fmflabel{$\phantom{x}$}{o1}
\fmf{photon,label=$\phantom{x}$}{i1,v1}
\fmf{ghost,left,label=$\omega$,tension=0.4}{v1,v2}
\fmf{ghost,left,label=$\phantom{x}$,tension=0.4}{v2,v1}
\fmf{photon,label=$\phantom{x}$}{v2,o1}
\fmfdotn{v}{2}
\end{fmfgraph*}
\end{gathered}\ \ .
\end{equation}
Again, only the last two diagrams have to be calculated and contain the divergent parts
\begin{equation}
\begin{gathered}
\begin{fmfgraph*}(60,60)
\fmfleft{i1}
\fmflabel{$V_\mu^\gamma$}{i1}
\fmfright{o1}
\fmflabel{$V_\nu^\gamma$}{o1}
\fmf{photon,label=$\phantom{p}$}{i1,v1}
\fmf{gluon,left,label=$\phantom{V^l}$,tension=0.4}{v1,v2}
\fmf{gluon,left,label=$\phantom{q_1}$,tension=0.4}{v2,v1}
\fmf{photon,label=$\phantom{q}$}{v2,o1}
\fmfdotn{v}{2}
\end{fmfgraph*}
\end{gathered}\nonumber
\end{equation}
\muu{ = i\frac{g^2 \mu^{-2\epsilon}}{(4\pi)^2\epsilon}N\frac{10}{3}\left[ g^{\mu \nu}p^2 -p^\mu p^\nu\right],
}
\begin{equation}
\begin{fmfgraph*}(60,60)
\fmfleft{i1}
\fmflabel{$V_\mu^\gamma$}{i1}
\fmfright{o1}
\fmflabel{$V_\nu^\gamma$}{o1}
\fmf{photon,label=$\phantom{x}$}{i1,v1}
\fmf{ghost,left,label=$\omega$,tension=0.4}{v1,v2}
\fmf{ghost,left,label=$\phantom{x}$,tension=0.4}{v2,v1}
\fmf{photon,label=$\phantom{x}$}{v2,o1}
\fmfdotn{v}{2}
\end{fmfgraph*}\nonumber
\end{equation}
\muu{ =  i \frac{g^2 \mu^{-2\epsilon}}{(4\pi)^2\epsilon} \frac{N}{3} \left[p^2 g^{\mu \nu }-p^{\mu } p^{\nu }\right],
}
yielding a purely transverse correction and thus 
\muu{
\Delta_5^{(1)}&=\tilde{Z}_V^{(1)}=\frac{g^2\mu^{-2\epsilon}}{(4\pi)^2\epsilon}N\frac{11}{3},\label{ZVtilde}\\
\Delta_6^{(1)}&=\tilde{Z}_V^{(1)}-Z_\alpha^{(1)}=0,
}
or 
\muu{
Z_\alpha^{(1)}=\frac{g^2\mu^{-2\epsilon}}{(4\pi)^2\epsilon}N\frac{11}{3}.
}

Next, we turn to the ghost self-energy, starting with the coset ghosts. The one-loop correction is given by
\begin{equation}
\begin{gathered}
\begin{fmfgraph*}(50,50)
\fmfleft{i1}
\fmflabel{$\omega^a$}{i1}
\fmfright{o1}
\fmflabel{$\bar{\omega}^b$}{o1}
\fmf{ghost}{i1,v1}
\fmf{ghost}{v1,o1}
\fmfblob{.30w}{v1}
\end{fmfgraph*}
\end{gathered}\ \ \ \ \ \ \ = \ \ \ \ \
\begin{gathered}
\begin{fmfgraph*}(50,50)
\fmfleft{i1}
\fmfright{o1}
\fmf{ghost}{i1,v1}
\fmf{ghost,label=$\omega$}{v1,v1}
\fmf{ghost}{v1,o1}
\fmfdotn{v}{1} 
\end{fmfgraph*}
\end{gathered}+\ \ \ \ \ 
\begin{gathered}
\begin{fmfgraph*}(50,50)
\fmfleft{i1}
\fmfright{o1}
\fmf{ghost}{i1,v1}
\fmf{photon,label=$V^j+V^\gamma$}{v1,v1}
\fmf{ghost}{v1,o1}
\fmfdotn{v}{1} 
\end{fmfgraph*}
\end{gathered}\nonumber
\end{equation}
\begin{equation}+ \ \ \ \ \ \ 
\begin{gathered}
\begin{fmfgraph*}(50,50)
\fmfleft{i1}
\fmfright{o1}
\fmf{ghost}{i1,v1}
\fmf{gluon,label=$\phantom{x}$}{v1,v1}
\fmf{ghost}{v1,o1}
\fmfdotn{v}{1} 
\end{fmfgraph*}
\end{gathered}+ \ \ \ \ \ 
\begin{gathered}
\begin{fmfgraph*}(50,50)
\fmfleft{i1}
\fmflabel{$\phantom{x}$}{i1}
\fmfright{o1}
\fmflabel{$\phantom{x}$}{o1}
\fmf{ghost,label=$\phantom{x}$}{i1,v1}
\fmf{photon,left,label=$V^j+V^\gamma$,tension=0.4}{v1,v2}
\fmf{ghost,right,label=$\omega$,tension=0.4}{v1,v2}
\fmf{ghost,label=$\phantom{x}$}{v2,o1}
\fmfdotn{v}{2}
\end{fmfgraph*}
\end{gathered}\ \ .
\end{equation}
Only the last diagram contributes in dimensional regularization and the divergent part reads
\begin{equation}
\begin{gathered}
\begin{fmfgraph*}(50,50)
\fmfleft{i1}
\fmflabel{$\omega^a$}{i1}
\fmfright{o1}
\fmflabel{$\bar{\omega}^b$}{o1}
\fmf{ghost,label=$\phantom{x}$}{i1,v1}
\fmf{photon,left,label=$V^j+V^\gamma$,tension=0.4}{v1,v2}
\fmf{ghost,right,label=$\omega$,tension=0.4}{v1,v2}
\fmf{ghost,label=$\phantom{x}$}{v2,o1}
\fmfdotn{v}{2}
\end{fmfgraph*}
\end{gathered}
\end{equation}
\muu{
= -\frac{g^2 \mu^{-2\epsilon}}{(4\pi)^2\epsilon}\frac{N}{2}\left( 3-\frac{\alpha + (N-2)\lambda}{N-1}\right)\delta^{ab} p^2,
}
which yields 
\muu{
\Delta_7^{(1)}=Z_\omega^{(1)}= \frac{g^2 \mu^{-2\epsilon}}{(4\pi)^2\epsilon}\frac{N}{2}\left( 3-\frac{\alpha + (N-2)\lambda}{N-1}\right).\label{zomega}
}

For the residual part we find that the $U(1)$ ghost $C^\gamma$ does not enter any interaction vertex. Therefore, the self-energy correction is zero and we immediately obtain
\muu{
\Delta_9^{(1)}=\frac{1}{2}\left(\tilde{Z}_C^{(1)}+\tilde{Z}_{\bar{C}}^{(1)}\right)=0.\label{gammarela}
}

We complete the self-energy analysis by considering the $SU(N-1)$ ghosts. Their one-loop correction is given by only one diagram 
\begin{equation}
\begin{gathered}
\begin{fmfgraph*}(60,60)
\fmfleft{i1}
\fmflabel{$C^j$}{i1}
\fmfright{o1}
\fmflabel{$\bar{C}^k$}{o1}
\fmf{ghost}{i1,v1}
\fmf{ghost}{v1,o1}
\fmfblob{.30w}{v1}
\end{fmfgraph*}
\end{gathered}\ \ \ \ \ \ \ = \ \ \ \ \
\begin{gathered}
\begin{fmfgraph*}(60,60)
\fmfleft{i1}
\fmflabel{$C^j$}{i1}
\fmfright{o1}
\fmflabel{$\bar{C}^k$}{o1}
\fmf{ghost,label=$\phantom{x}$}{i1,v1}
\fmf{photon,left,tension=0.4,label=$V^l$}{v1,v2}
\fmf{ghost,right,tension=0.4,label=$C^l$}{v1,v2}
\fmf{ghost,label=$\phantom{x}$}{v2,o1}
\fmfdotn{v}{2}
\end{fmfgraph*}
\end{gathered}\nonumber
\end{equation}
\muu{ = - \frac{g^2 \mu^{-2\epsilon}}{(4\pi)^2\epsilon} \frac{N-1}{4}(3-\lambda)\delta^{jk}p^2.
}
Thus, we find the relation
\muu{
\Delta_8^{(1)}=\frac{1}{2}\left(Z_C^{(1)}+Z_{\bar{C}}^{(1)}\right)= \frac{g^2 \mu^{-2\epsilon}}{(4\pi)^2\epsilon}\frac{N-1}{4}(3-\lambda).\label{SUN-1ghostrela}
}

The next step is to obtain the renormalization factor of the Yang-Mills coupling by renormalizing the $V^j C^k \bar{C}^l$ vertex. Its one-loop correction reads
$\;$ \\
$\;$ \\
\begin{equation}
\begin{gathered}
\begin{fmfgraph*}(50,50)
\fmfleft{i1,i2}
\fmflabel{$C^k$}{i2}
\fmflabel{$\bar{C}^l$}{i1}
\fmfright{o1}
\fmflabel{$V_\mu^j$}{o1}
\fmf{ghost}{v1,i1}
\fmf{ghost}{i2,v1}
\fmf{photon}{v1,o1}
\fmfblob{.30w}{v1}
\end{fmfgraph*}
\end{gathered} \ \ \ \ \ \ \ \ =   
\begin{gathered}
\begin{fmfgraph*}(130,50)
    \fmfstraight
    \fmfleft{i1,i2}
    \fmfright{o1,h,o2}
    \fmf{ghost}{t1,i1}
    \fmf{ghost}{i2,t2}
    \fmf{phantom,tension=0.5}{t1,o1}
    \fmf{phantom,tension=0.5}{t2,o2}
    \fmffreeze
    \fmf{ghost,tension=1,label=$C^m$,label.side=left}{t2,t3}
    \fmf{ghost,tension=1,label=$C^m$,label.side=left}{t3,t1}
    \fmf{photon,tension=2.0}{t3,h}
    \fmf{photon,label=$V^m$,label.side=left}{t1,t2}
    \fmflabel{$\phantom{x}$}{i1}
    \fmflabel{$\phantom{x}$}{i2}
    \fmflabel{$\phantom{x}$}{h}
  \end{fmfgraph*}
\end{gathered}\nonumber
\end{equation}
$\;$ \\
\begin{equation} + \ \ \begin{gathered}
\begin{fmfgraph*}(130,50)
    \fmfstraight
    \fmfleft{i1,i2}
    \fmfright{o1,h,o2}
    \fmf{ghost}{t1,i1}
    \fmf{ghost}{i2,t2}
    \fmf{phantom,tension=0.5}{t1,o1}
    \fmf{phantom,tension=0.5}{t2,o2}
    \fmffreeze
    \fmf{photon,tension=1,label=$V^m$,label.side=left}{t2,t3}
    \fmf{photon,tension=1,label=$V^m$,label.side=left}{t3,t1}
    \fmf{photon,tension=2.0}{t3,h}
    \fmf{ghost,label=$C^m$}{t2,t1}
    \fmflabel{$\phantom{x}$}{i1}
    \fmflabel{$\phantom{x}$}{i2}
    \fmflabel{$\phantom{x}$}{h}
  \end{fmfgraph*}
\end{gathered}\ \ \ \ .
\end{equation}
The divergent parts of the diagrams read as follows,
\begin{equation}
\begin{gathered}
\begin{fmfgraph*}(130,50)
    \fmfstraight
    \fmfleft{i1,i2}
    \fmfright{o1,h,o2}
    \fmf{ghost}{t1,i1}
    \fmf{ghost}{i2,t2}
    \fmf{phantom,tension=0.5}{t1,o1}
    \fmf{phantom,tension=0.5}{t2,o2}
    \fmffreeze
    \fmf{ghost,tension=1,label=$C^m$,label.side=left}{t2,t3}
    \fmf{ghost,tension=1,label=$C^m$,label.side=left}{t3,t1}
    \fmf{photon,tension=2.0,label=$p$}{t3,h}
    \fmf{photon,label=$V^m$,label.side=left}{t1,t2}
    \fmflabel{$C^k$}{i2}
    \fmflabel{$\bar{C}^l$}{i1}
    \fmflabel{$V_\mu^j$}{h}
  \end{fmfgraph*}
  \end{gathered}\nonumber
\end{equation}
\muu{
=-\frac{g^2\mu^{-2\epsilon}}{(4\pi)^2\epsilon}\frac{N-1}{2} \frac{3\lambda}{4}igf^{jkl}p^\mu,
}
and 
\begin{equation}
\begin{gathered}
\begin{fmfgraph*}(130,50)
    \fmfstraight
    \fmfleft{i1,i2}
    \fmfright{o1,h,o2}
    \fmf{ghost}{t1,i1}
    \fmf{ghost}{i2,t2}
    \fmf{phantom,tension=0.5}{t1,o1}
    \fmf{phantom,tension=0.5}{t2,o2}
    \fmffreeze
    \fmf{photon,tension=1,label=$V^m$,label.side=left}{t2,t3}
    \fmf{photon,tension=1,label=$V^m$,label.side=left}{t3,t1}
    \fmf{photon,tension=2.0,label=$p$}{t3,h}
    \fmf{ghost,label=$C^m$}{t2,t1}
    \fmflabel{$\bar{C}^l$}{i1}
    \fmflabel{$C^k$}{i2}
    \fmflabel{$V_\mu^j$}{h}
  \end{fmfgraph*}
\end{gathered}
\end{equation}
\muu{
=-\frac{g^2\mu^{-2\epsilon}}{(4\pi)^2\epsilon}\frac{N-1}{2} \frac{\lambda}{4}igf^{jkl}p^\mu.
}
Therefore, we obtain the relation
\muu{
\Delta_{25}^{(1)}&=\frac{1}{2}\left(Z_V^{(1)}+ Z_C^{(1)}+Z_{\bar{C}}^{(1)} \right)+Z_g^{(1)}\nonumber \\
&=-\frac{g^2\mu^{-2\epsilon}}{(4\pi)^2\epsilon}\frac{N-1}{2}\lambda.
}
Using the equations \eqref{ZV} and \eqref{SUN-1ghostrela} we find
\muu{
Z_g^{(1)}=-\frac{g^2\mu^{-2\epsilon}}{(4\pi)^2\epsilon}\frac{11}{6}N.
}
As expected, this is the standard result for the Yang-Mills coupling in pure Yang-Mills theory. 

Finally, we consider the $X^a \omega^b \bar{C}^J$ vertex. The one-loop correction reads
$\;$ \\
$\;$ \\
\begin{equation}
\begin{gathered}
\begin{fmfgraph*}(50,50)
\fmfleft{i1,i2}
\fmflabel{$\omega^b$}{i2}
\fmflabel{$\bar{C}^J$}{i1}
\fmfright{o1}
\fmflabel{$X_\mu^a$}{o1}
\fmf{ghost}{v1,i1}
\fmf{ghost}{i2,v1}
\fmf{gluon}{v1,o1}
\fmfblob{.30w}{v1}
\end{fmfgraph*}
\end{gathered} \ \ \ \ \ \ \ \ =   
\begin{gathered}
\begin{fmfgraph*}(130,50)
    \fmfstraight
    \fmfleft{i1,i2}
    \fmfright{o1,h,o2}
    \fmf{ghost}{t1,i1}
    \fmf{ghost}{i2,t2}
    \fmf{phantom,tension=0.5}{t1,o1}
    \fmf{phantom,tension=0.5}{t2,o2}
    \fmffreeze
    \fmf{photon,tension=1,label=$V^k+V^\gamma$,label.side=left}{t2,t3}
    \fmf{gluon,tension=1,label=$\phantom{x}$,label.side=left}{t3,t1}
    \fmf{gluon,tension=2.0}{t3,h}
    \fmf{ghost,label=$\omega$}{t2,t1}
    \fmflabel{$\phantom{x}$}{i1}
    \fmflabel{$\phantom{x}$}{i2}
    \fmflabel{$\phantom{x}$}{h}
  \end{fmfgraph*}
\end{gathered}\nonumber
\end{equation}
$\;$ \\
$\;$ \\
\begin{equation}
+ \begin{gathered}
\begin{fmfgraph*}(100,50)
    \fmfstraight
    \fmfleft{i1,i2}
    \fmfright{o1,h,o2}
    \fmf{ghost}{t1,i1}
    \fmf{ghost}{i2,t2}
    \fmf{phantom,tension=0.5}{t1,o1}
    \fmf{phantom,tension=0.5}{t2,o2}
    \fmffreeze
    \fmf{gluon,tension=1,label=$\phantom{x}$,label.side=left}{t2,t3}
    \fmf{photon,tension=1,label=$V^k$,label.side=left}{t3,t1}
    \fmf{gluon,tension=2.0}{t3,h}
    \fmf{ghost,label=$C^n$}{t2,t1}
    \fmflabel{$\bar{C}^j$}{i1}
    \fmflabel{$\phantom{x}$}{i2}
    \fmflabel{$\phantom{x}$}{h}
  \end{fmfgraph*}
\end{gathered}\ \ \ \ +   \begin{gathered}
\begin{fmfgraph*}(100,50)
    \fmfstraight
    \fmfleft{i1,i2}
    \fmfright{o1,h,o2}
    \fmf{ghost}{t1,i1}
    \fmf{ghost}{i2,t2}
    \fmf{phantom,tension=0.5}{t1,o1}
    \fmf{phantom,tension=0.5}{t2,o2}
    \fmffreeze
    \fmf{ghost,tension=1,label=$\omega$,label.side=left}{t2,t3}
    \fmf{ghost,tension=1,label=$C^l$,label.side=left}{t3,t1}
    \fmf{gluon,tension=2.0}{t3,h}
    \fmf{photon,label=$V^k$}{t2,t1}
    \fmflabel{$\bar{C}^j$}{i1}
    \fmflabel{$\phantom{x}$}{i2}
    \fmflabel{$\phantom{x}$}{h}
  \end{fmfgraph*}
\end{gathered}
\nonumber
\end{equation}
$\;$ \\
$\;$ \\
\begin{equation}\ \ \ \ + \ \ \ \ \ \begin{gathered}
\begin{fmfgraph*}(80,60)
\fmfleft{i1,i2}
\fmflabel{$\phantom{x}$}{i1}
\fmflabel{$\phantom{x}$}{i2}
\fmfright{o1}
\fmflabel{$\phantom{x}$}{o1}
\fmf{ghost,tension=0.4,label=$\phantom{x}$}{v2,o1}
\fmf{gluon,tension=0.4,label=$\phantom{x}$}{i2,v1}
\fmf{ghost,tension=0.5,label=$\phantom{x}$,label.side=left}{i1,v1}
\fmf{ghost,right,tension=0.3,label=$\omega$,label.side=right}{v1,v2}
\fmf{gluon,left,tension=0.3,label=$\phantom{x}$}{v1,v2}
\fmfdotn{v}{2}
\end{fmfgraph*} 
\end{gathered}\ \ \ \ .
\end{equation}
$\;$ \\
$\;$ \\
Note that the second and third diagram only contribute to the vertex with an external $SU(N-1)$ antighost, while the first and last diagram contribute in both cases. The first diagram's divergent part is found to be 
\begin{equation}
\begin{gathered}
\begin{fmfgraph*}(130,50)
    \fmfstraight
    \fmfleft{i1,i2}
    \fmfright{o1,h,o2}
    \fmf{ghost,label=$p$}{t1,i1}
    \fmf{ghost}{i2,t2}
    \fmf{phantom,tension=0.5}{t1,o1}
    \fmf{phantom,tension=0.5}{t2,o2}
    \fmffreeze
    \fmf{photon,tension=1,label=$V^k+V^\gamma$,label.side=left}{t2,t3}
    \fmf{gluon,tension=1,label=$\phantom{x}$,label.side=left}{t3,t1}
    \fmf{gluon,tension=2.0}{t3,h}
    \fmf{ghost,label=$\omega$}{t2,t1}
    \fmflabel{$\bar{C}^J$}{i1}
    \fmflabel{$\omega^b$}{i2}
    \fmflabel{$X_\mu^a$}{h}
  \end{fmfgraph*}
\end{gathered}\nonumber
\end{equation}
\muu{
=\frac{g^2 \mu^{-2\epsilon}}{(4 \pi)^2 \epsilon}igf^{ceJ}\left(\alpha f^{e\gamma b}f^{ca\gamma} +\lambda f^{ekb}f^{cak} \right)p^\mu,
}
while the second diagram has the divergent part\\
$\;$ \\
\begin{equation}
\begin{gathered}
\begin{fmfgraph*}(80,60)
\fmfleft{i1,i2}
\fmflabel{$X_\mu^a$}{i2}
\fmflabel{$\omega^b$}{i1}
\fmfright{o1}
\fmflabel{$\bar{C}^J$}{o1}
\fmf{ghost,tension=0.4,label=$p$}{v2,o1}
\fmf{gluon,tension=0.4,label=$\phantom{x}$}{i2,v1}
\fmf{ghost,tension=0.5,label=$\phantom{x}$,label.side=left}{i1,v1}
\fmf{ghost,right,tension=0.3,label=$\omega$,label.side=right}{v1,v2}
\fmf{gluon,left,tension=0.3,label=$\phantom{x}$}{v1,v2}
\fmfdotn{v}{2}
\end{fmfgraph*} 
\end{gathered}\nonumber
\end{equation}
\muu{= ig\left[ \left(f^{eKc} f^{abK}+f^{aKc} f^{ebK} \right) f^{Jec} \right] \frac{\xi+3}{4} \frac{g^2 \mu^{-2\epsilon}}{(4 \pi)^2 \epsilon} p^\mu. 
}
The two diagrams that only contribute in the $J=j$ case have the divergent parts
\begin{equation}
\begin{gathered}
\begin{fmfgraph*}(100,50)
    \fmfstraight
    \fmfleft{i1,i2}
    \fmfright{o1,h,o2}
    \fmf{ghost}{t1,i1}
    \fmf{ghost}{i2,t2}
    \fmf{phantom,tension=0.5}{t1,o1}
    \fmf{phantom,tension=0.5}{t2,o2}
    \fmffreeze
    \fmf{gluon,tension=1,label=$\phantom{x}$,label.side=left}{t2,t3}
    \fmf{photon,tension=1,label=$V^k$,label.side=left}{t3,t1}
    \fmf{gluon,tension=2.0}{t3,h}
    \fmf{ghost,label=$C^n$}{t2,t1}
    \fmflabel{$\bar{C}^j$}{i1}
    \fmflabel{$\omega^b$}{i2}
    \fmflabel{$X_\mu^a$}{h}
  \end{fmfgraph*}
\end{gathered}\nonumber
\end{equation}
\muu{
=ig\left[f^{ebn}f^{knj}f^{aek} \right]\frac{\lambda+3(\xi+1)}{4}\frac{g^2\mu^{-2\epsilon}}{(4\pi)^2\epsilon}p_\mu,
}
and
\begin{equation}
\begin{gathered}
\begin{fmfgraph*}(100,50)
    \fmfstraight
    \fmfleft{i1,i2}
    \fmfright{o1,h,o2}
    \fmf{ghost}{t1,i1}
    \fmf{ghost}{i2,t2}
    \fmf{phantom,tension=0.5}{t1,o1}
    \fmf{phantom,tension=0.5}{t2,o2}
    \fmffreeze
    \fmf{ghost,tension=1,label=$\omega$,label.side=left}{t2,t3}
    \fmf{ghost,tension=1,label=$C^l$,label.side=left}{t3,t1}
    \fmf{gluon,tension=2.0}{t3,h}
    \fmf{photon,label=$V^k$}{t2,t1}
    \fmflabel{$\bar{C}^j$}{i1}
    \fmflabel{$\omega^b$}{i2}
    \fmflabel{$X_\mu^a$}{h}
  \end{fmfgraph*}
\end{gathered}\nonumber
\end{equation}
\muu{
=ig \left[ f^{ekb}f^{klj}f^{ael} \right]\frac{\lambda}{4}\frac{g^2\mu^{-2\epsilon}}{(4\pi)^2\epsilon}p_\mu.
}
Adding up all contributions we find after some color algebra in the $SU(N-1)$ case $J=j$
\muu{
&\Delta_{26}^{(1)}=\left(\frac{1}{2} Z_X^{(1)} + \frac{1}{2} Z_{\bar{C}}^{(1)}+\frac{1}{2} Z_\omega^{(1)} + Z_g^{(1)}\right) \nonumber \\
&= -\frac{g^2}{(4\pi)^2\epsilon}\left(\frac{9+3\xi}{8}+\frac{\alpha N-\lambda}{2(N-1)}+\frac{N-1}{8}[2\lambda+3(\xi+1)]  \right).\label{xwcv1}
}
Together with \eqref{SUN-1ghostrela} this implies 
\muu{
Z_{\bar{C}}^{(1)}&=-\frac{g^2\mu^{-2\epsilon}}{(4\pi)^2\epsilon}\frac{1}{2}[N\xi+3-\lambda(N-1)],\\
Z_C^{(1)}&=\frac{g^2\mu^{-2\epsilon}}{(4\pi)^2\epsilon}\left[\frac{N}{2}(3+\xi)-\lambda(N-1)\right].
} 
In the $U(1)$ case $J=\gamma$ we obtain 
\muu{
\Delta_{27}^{(1)}=\left(\frac{1}{2} Z_X^{(1)} + \frac{1}{2} \tilde{Z}_{\bar{C}}^{(1)}+\frac{1}{2} Z_\omega^{(1)} + Z_g^{(1)}\right) \nonumber \\
= -\frac{g^2 \mu^{-2\epsilon}}{(4\pi)^2\epsilon}N \left(\frac{9+3\xi}{8}+\frac{\alpha +(N-2)\lambda}{2(N-1)} \right),\label{xwcv2}
} 
which implies using \eqref{gammarela}
\muu{
\tilde{Z}_{\bar{C}}^{(1)}=-\tilde{Z}_C^{(1)}&=-\frac{g^2\mu^{-2\epsilon}}{(4\pi)^2 \epsilon} \frac{N}{2}(3+\xi) \label{Zgamma}.
}
This completes the one-loop analysis. 

We want to finish this section by summarizing all the corresponding RG functions. We define them by
\muu{
\gamma_A &=\frac{\mu}{A_R} \frac{\partial A_R }{\partial \mu}=  -\frac{1}{2} \mu \frac{\partial }{\partial \mu}\log Z_A = -\frac{1}{2} \mu \frac{\partial }{\partial \mu}Z_A^{(1)} +O(\hbar^2),\\
\gamma_B &=\frac{\mu}{B_R} \frac{\partial B_R }{\partial \mu}=-  \mu \frac{\partial }{\partial \mu}\log Z_B=  -  \mu \frac{\partial }{\partial \mu}Z_B^{(1)}+O(\hbar^2),\\
\beta_g &= \mu \frac{\partial g_R}{\partial \mu}=-g_R \ \mu \frac{\partial}{\partial \mu} \log Z_g = -g_R \ \mu \frac{\partial}{\partial \mu} Z_g^{(1)}+O(\hbar^2),
} 
for the fields $A$, the parameters $B$ and the Yang-Mills coupling, respectively.  Then we obtain:
\muu{
\gamma_X &= \frac{g^2}{(4\pi)^2}\frac{N}{2}\left( \frac{17}{6}-\frac{\xi}{2}-\frac{\alpha+(N-2)\lambda}{N-1} \right), \nonumber\\
\gamma_V &= \frac{g^2}{(4\pi)^2}\left(\frac{13N+9}{6}-\frac{\lambda}{2}(N-1)\right), \nonumber \\
\tilde{\gamma}_V&= \frac{g^2}{(4\pi)^2}\frac{11}{3}N, \nonumber\\
\gamma_\omega &= \gamma_{\bar{\omega}}=  \frac{g^2}{(4\pi)^2}\frac{N}{2}\left( 3-\frac{\alpha +(N-2)\lambda}{N-1}\right), \nonumber\\
\gamma_C &=\frac{g^2}{(4\pi)^2}\left[\frac{N}{2}(3+\xi)-\lambda(N-1)\right], \nonumber\\ 
\gamma_{\bar{C}}&= -\frac{g^2}{(4\pi)^2}\frac{1}{2}[N\xi+3-\lambda(N-1)], \nonumber\\
\tilde{\gamma}_C&=-\tilde{\gamma}_{\bar{C}}= \frac{g^2}{(4\pi)^2}\frac{N}{2}(3+\xi), \nonumber\\
 \gamma_\xi&= \frac{g^2}{(4\pi)^2}\left(\frac{4}{3} -\xi-\frac{3}{\xi}\right)N, \nonumber\\
\gamma_\lambda &=  \frac{g^2}{(4\pi)^2}\left(\frac{13N+9}{3}-\lambda(N-1)\right)\nonumber,\\
\gamma_\alpha &=  \frac{g^2}{(4\pi)^2}\frac{22}{3}N \nonumber ,\\
\beta_g &= - \frac{g^3}{(4\pi)^2}\frac{11}{3}N.
\label{RGfunctions}
}
We find that the running of $\alpha$ and $\lambda$ according to $\mu \frac{\partial \alpha}{\partial \mu}=\alpha \gamma_\alpha$ and $\mu \frac{\partial \lambda}{\partial \mu}=\lambda \gamma_\lambda$ implies the existence of both the ''symmetric`` as well as the ''asymmetric`` fixed point, 
\muu{
(\alpha,\lambda)=(0,0);\ \ \ \ (\alpha,\lambda)=\left(0,\frac{13N+9}{3(N-1)}\right). 
}
Even though the latter one implies an ''asymmetric`` gauge fixing of the $U(1)$ and $SU(N-1)$ part of the residual field, no problem occurs as the invariance of the residual gauge fixing Lagrangian under global $U(N-1)$ color transformations is completely independent of the parameters $\alpha$ and $\lambda$.

Moreover, as mentioned before, for $N=2$ our decomposition coincides with the MAG. In that case, our results are in full agreement with the existing literature, see for example \cite{shinohara2003most, kondo2001renormalizable,shinohara2003renormalizable, ellwanger2003massive,kondo2003implications}. In particular, note that the $\lambda$ dependent terms in $\gamma_X$ and $\gamma_\omega$ coming from the $SU(N-1)$ part of $U(N-1)$ vanish in this case, which reflects the fact that for $N=2$ we have the decomposition $SU(2)/U(1) \times U(1)$, i.e. the $SU(N-1)$ part of the residual field is absent.

\section{BRST invariance and multiplicative renormalizability of the composite operator}\label{sec:the composite}

\subsection{BRST invariance of the composite operator}
The first step in the proper introduction of the composite operator $\mathcal{O}=\frac{1}{2}X_\mu^a X^{\mu a}-i\xi \omega^a \bar{\omega}^a$ is to add a source term to the action,

\muu{
S= \int_x \mathcal{L} + J \mathcal{O},
}
and to show that the BRST invariance of the action is preserved. The source shall satisfy $\delta_B J =0$. Using \eqref{BRST} the composite operator transforms as
\muu{
\delta_B \mathcal{O} = X_\mu^a \delta_B X^{\mu a} -i\xi (-gf^{abJ}\omega^b C^J)\bar{\omega}^a+i\xi \omega^a \left(iN^a\right). 
}
Replacing $N^a$ by its equation of motion \eqref{BEOM} we find 
\muu{
\delta_B \mathcal{O} &=X_\mu^a \left(D^{\mu ab} \omega^b+g f^{abJ}X^{\mu b} C^J \right)+i\xi gf^{abJ}\omega^b C^J \bar{\omega}^a \nonumber\\
&-\xi \omega^a \left(-\frac{1}{\xi}D_\mu^{ab}X^{\mu b}+igf^{abJ}\bar{\omega}^b C^J \right)\nonumber\\
&=X^{\mu a} D_\mu^{ ab} \omega^b +\omega^a D_\mu^{ab} X^{\mu b}\nonumber \\
&=\partial^\mu \left( X_\mu^a \omega^a \right),
}
and therefore the (on-shell) BRST invariance is maintained after introducing the source term for the composite operator. Yet two problems remain to be solved. The first is the proof of multiplicative renormalizability of the composite operator, at least to one-loop level. This will be given below. The second and more involved problem are the divergences proportional to $J^2$ which are generated by the source term. This will be postponed to the next section. 

\subsection{Multiplicative renormalizability of the composite operator}
\label{sec:compop}
\begin{figure}[h!]
\begin{minipage}{0.4\textwidth}
 \includegraphics[width=1\linewidth]{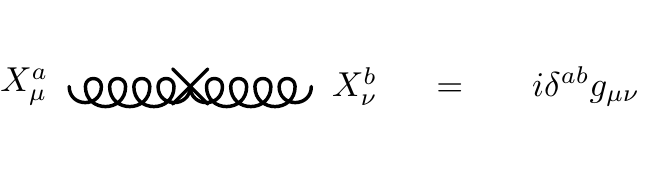} 
\end{minipage}
\begin{minipage}{0.4\textwidth}
 \includegraphics[width=1\linewidth]{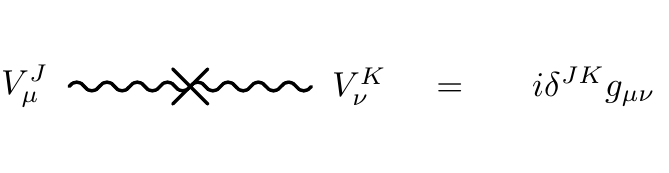}
\end{minipage}
\begin{minipage}{0.4\textwidth}
 \includegraphics[width=1\linewidth]{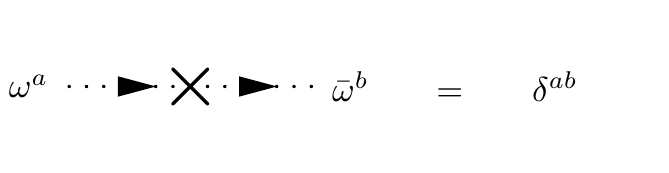}
\end{minipage}
\begin{minipage}{0.4\textwidth}
 \includegraphics[width=1\linewidth]{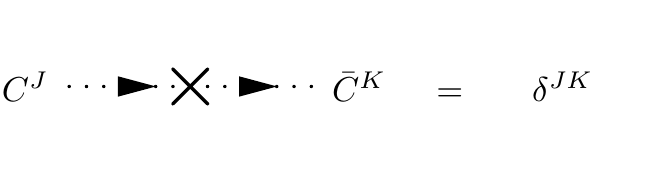} 
\end{minipage}
\caption{Feynman rules for operator insertions.}\label{insertpic}
\end{figure}

The composite operator $\mathcal{O}_R = \left[\frac{1}{2}X_\mu^a X^{\mu a} \right]_R - \xi \left[i\omega^a\bar{\omega}^a \right]_R$ can in principle mix with any condensate that has the same mass dimension and quantum number. We therefore have to set up the renormalization matrix
\muu{
&\mmo{\left[\frac{1}{2}X_\mu^a X^\mu_a \right]_R \\ \left[\frac{1}{2}V_\mu^j V^\mu_j \right]_R \\ \left[i \omega^a \bar{\omega}^a\right]_R \\ \left[i C^j \bar{C}^j \right]_R \\ \left[\frac{1}{2}V_\mu^\gamma V^\mu_\gamma \right]_R \\ \left[i C^\gamma \bar{C}^\gamma \right]_R} =\nonumber\\
& \mmo{Z_1 & Z_2 & Z_3 & Z_4 & Z_5 & Z_6 \\ Z_7 & Z_8 & Z_9 & Z_{10} & Z_{11} & Z_{12}\\ Z_{13} & Z_{14} & Z_{15} & Z_{16} & Z_{17} & Z_{18} \\ Z_{19} & Z_{20} & Z_{21} & Z_{22} & Z_{23} & Z_{24} \\ Z_{25} & Z_{26} & Z_{27} & Z_{28} & Z_{29} & Z_{30} \\ Z_{31} & Z_{32} & Z_{33} &Z_{34} & Z_{35} & Z_{36} } \mmo{\left[\frac{1}{2}X_\mu^a X^\mu_a \right] \\ \left[\frac{1}{2}V_\mu^j V^\mu_j \right] \\ \left[i \omega^a \bar{\omega}^a\right] \\ \left[i C^j \bar{C}^j \right] \\ \left[\frac{1}{2}V_\mu^\gamma V^\mu_\gamma \right] \\  \left[i C^\gamma \bar{C}^\gamma \right]}.\label{renmatrix}
}
The matrix elements are calculated by inserting $\mathcal{O}_R$ into the various two-point functions and requiring the cancellation of the resulting divergences \cite{kondo2002renormalizing}, using the Feynman rules for the operator insertions as shown in Fig.~\ref{insertpic}.

At this point we only state the result, since the calculations are quite lengthy. The details are presented in appendix \ref{appB}. The renormalization matrix is shown to have the form
\muu{
\mathcal{Z}= \mathbbm{1} + \mathcal{Z}^{(1)},
}
where the one-loop part $\mathcal{Z}^{(1)}$ contains 10 non-vanishing elements given by

\muu{
Z_1^{(1)}&=-\frac{g^2 \mu^{-2\epsilon}}{(4\pi)^2\epsilon}\frac{N}{2}\left(\frac{3}{2}(\xi+1)+\frac{\alpha+(N-2)\lambda}{N-1} \right), \nonumber\\
Z_3^{(1)}&=\frac{g^2\mu^{-2\epsilon}}{(4\pi)^2\epsilon}N\frac{3+\xi^2}{2},\nonumber\\
Z_7^{(1)} &=-\frac{g^2\mu^{-2\epsilon}}{(4\pi)^2\epsilon} \frac{N(N-2)}{2(N-1)}\frac{3}{4}(\xi+3),\nonumber\\
Z_8^{(1)}&=-\frac{g^2\mu^{-2\epsilon}}{(4\pi)^2\epsilon}\frac{3}{4}(N-1)(1+\lambda),\nonumber\\
Z_9^{(1)}&=-3\frac{g^2\mu^{-2\epsilon}}{(4\pi)^2\epsilon}\frac{N(N-2)}{2(N-1)},\nonumber\\
Z_{13}^{(1)}&=-\frac{g^2\mu^{-2\epsilon}}{(4\pi)^2\epsilon}N,\nonumber\\
Z_{15}^{(1)}&=\frac{g^2 \mu^{-2\epsilon}}{(4\pi)^2\epsilon}\frac{N}{2}\left(\xi -\frac{\alpha+(N-2)\lambda}{N-1} \right),\nonumber\\
Z_{20}^{(1)}&=-\frac{g^2\mu^{-2\epsilon}}{(4\pi)^2\epsilon}(N-1)\frac{1}{2},\nonumber \\
Z_{25}^{(1)}&=-\frac{g^2\mu^{-2\epsilon}}{(4\pi)^2\epsilon}\frac{N}{2(N-1)}\frac{3}{4}(\xi+3),\nonumber \\
Z_{27}^{(1)}&=-3\frac{g^2 \mu^{-2\epsilon}}{(4\pi)^2\epsilon} \frac{N}{2(N-1)}.\label{matrixentries}
}
Using the fact that to one-loop level the inverse of the renormalization matrix reads
\muu{
\mathcal{Z}^{-1}= \mathbbm{1} - \mathcal{Z}^{(1)},
}
we can invert equation \eqref{renmatrix}, obtaining
\muu{
&\mmo{\left[\frac{1}{2}X_\mu^a X^\mu_a \right] \\ \left[\frac{1}{2}V_\mu^j V^\mu_j \right] \\ \left[i \omega^a \bar{\omega}^a\right] \\ \left[i C^j \bar{C}^j \right] \\ \left[\frac{1}{2}V_\mu^\gamma V^\mu_\gamma \right] \\  \left[i C^\gamma \bar{C}^\gamma \right]} = \nonumber \\
&\mmo{1-Z_1^{(1)} & 0 & -Z_3^{(1)} & 0 &0 & 0 \\ -Z_7^{(1)} & 1-Z_8^{(1)} & -Z_{9}^{(1)} & 0 & 0& 0\\ -Z_{13}^{(1)} & 0 & 1-Z_{15}^{(1)} & 0 & 0 & 0 \\ 0 & -Z_{20}^{(1)} & 0 & 1 & 0 & 0 \\ -Z_{25}^{(1)} & 0 & -Z_{27}^{(1)} & 0 & 1 & 0 \\ 0 & 0 & 0 &0 & 0 & 1}   \mmo{\left[\frac{1}{2}X_\mu^a X^\mu_a \right]_R \\ \left[\frac{1}{2}V_\mu^j V^\mu_j \right]_R \\ \left[i \omega^a \bar{\omega}^a\right]_R \\ \left[i C^j \bar{C}^j \right]_R\\ \left[\frac{1}{2}V_\mu^\gamma V^\mu_\gamma \right]_R \\  \left[i C^\gamma \bar{C}^\gamma \right]_R}.\label{renmacalc}
}

The composite operator is thus renormalized as
\muu{
\mathcal{O}&=\left(1+Z_X^{(1)}\right)\frac{1}{2}X^\mu_a{}_R X_\mu^a{}_R -\left(1+Z_\xi^{(1)}\right)\left(1+Z_\omega^{(1)} \right)\xi i \omega^a_R \bar{\omega}^a_R\nonumber \\
&= \left(1+Z_X^{(1)}\right) \left\{\left(1-Z_1^{(1)}\right)\left[\frac{1}{2}X_\mu^a X^\mu_a \right]_R - Z_3^{(1)}\left[ i\omega^a \bar{\omega}^a\right]_R \right\}\nonumber \\
& - \left(1+Z_\xi^{(1)}\right)\left(1+Z_\omega^{(1)} \right)\xi \left\{-Z_{13}^{(1)}\left[\frac{1}{2}X_\mu^a X^\mu_a \right]_R \right. \nonumber \\
&\left. \phantom{\left[\frac{1}{2}X_\mu^a X^\mu_a \right]_R}  +\left(1-Z_{15}^{(1)}\right) \left[i\omega^a \bar{\omega}^a \right]_R \right\}\nonumber \\
&\overset{!}{=}\left(1+\frac{1}{2} Z_{\mathcal{O}}^{(1)}\right) \left( \left[\frac{1}{2}X_\mu^a X^\mu_a\right]_R - \xi \left[ i\omega^a \bar{\omega}^a\right]_R \right). 
}
This yields the condition
\muu{
-Z_1^{(1)}+Z_X^{(1)}+\xi Z_{13}^{(1)}=Z_\xi^{(1)}-Z_{15}^{(1)}+Z_\omega^{(1)}+\frac{1}{\xi}Z_3^{(1)}.\label{condi1}
}
Indeed, we find
\muu{
-Z_1^{(1)}+Z_X^{(1)}+\xi Z_{13}^{(1)}&= \frac{g^2\mu^{-2\epsilon}}{(4\pi)^2\epsilon}\left[ \frac{N}{6}(13-3\xi) \right], \\
Z_\xi^{(1)}-Z_{15}^{(1)}+Z_\omega^{(1)}+\frac{1}{\xi}Z_3^{(1)}&= \frac{g^2\mu^{-2\epsilon}}{(4\pi)^2\epsilon}\left[ \frac{N}{6}(13-3\xi) \right]. 
}
Thus, the composite operator is one-loop multiplicatively renormalizable, $\mathcal{O}= Z_\mathcal{O}^{1/2} \mathcal{O}_R$ with the renormalization factor
\muu{
Z_{\mathcal{O}}^{(1)}&=2\left( -Z_1^{(1)}+Z_X^{(1)}+\xi Z_{13}^{(1)} \right)\nonumber \\
&=2\left( Z_\xi^{(1)}-Z_{15}^{(1)}+Z_\omega^{(1)}+\frac{1}{\xi}Z_3^{(1)} \right)\nonumber \\
&=\frac{g^2 \mu^{-2\epsilon}}{(4\pi)^2\epsilon}\frac{N}{3}(13-3\xi).
} 
Again, this result is in agreement with the existing $N=2$ MAG results, e.g. \cite{dudal2004analytic}. 
According to equation \eqref{renmacalc} the existence of the coset gluon condensate seems to induce a residual field condensate $V_\mu^j V^{\mu j}$ ($V_\mu^\gamma V^{\mu \gamma}$) due to a non-vanishing of the matrix entries $Z_7^{(1)}$ and $Z_9^{(1)}$ ($Z_{25}^{(1)}$ and $Z_{27}^{(1)}$). However, no BRST invariant combination of mass dimension two operators including the residual field condensate can be constructed. This renders such a condensate non-physical and thus we continue to discuss the composite operator $\mathcal{O}$ only.

Finally, for later use we furthermore introduce the composite operator anomalous dimension,
\muu{
\gamma_\mathcal{O}=\frac{\mu}{\mathcal{O}_R}\frac{\partial \mathcal{O}_R}{\partial \mu}=-\frac{1}{2}\mu \frac{\partial}{\partial \mu}\log Z_\mathcal{O},
}
which reads to one-loop level
\muu{
\gamma_{\mathcal{O}}=\frac{g^2}{(4\pi)^2}\frac{N}{3}(13-3\xi).\label{compositeano}
}

\section{LCO formalism}
\label{sec:LCO}
As mentioned before, the introduction of the composite operator source term leads to new divergences quadratic in the source. To treat these divergences the so-called LCO formalism has been developed in \cite{verschelde1995perturbative, knecht2001new} and also has been applied to similar gluon-ghost composite operators for example in the usual Lorenz gauge and in the MAG \cite{verschelde2001non, dudal2004dynamical, dudal2004analytic}. In order to make this paper self-contained, we briefly introduce the LCO formalism, thereby mainly following the lines of \cite{dudal2004analytic}.

In order to cure the aforementioned divergences we extend the Lagrangian by adding
\muu{
\frac{1}{2}\kappa J^2 + \frac{1}{2}\delta \kappa J^2,
} 
where $\kappa$ is an a priori arbitrary parameter and the second term is understood to be a pure counterterm. Since we already proved the multiplicative renormalizability of the composite operator we define $J_0 = Z_\mathcal{O}^{-1/2}J$ such that $J_0 \mathcal{O}_0 = J \mathcal{O}$. The running of the generating functional then becomes
\muu{
\left[\mu \frac{\partial}{\partial \mu} + \beta_{g^2} \frac{\partial}{\partial g^2} +\xi \gamma_\xi \frac{\partial}{\partial \xi}-\gamma_\mathcal{O} J \frac{\partial}{\partial J} + \eta \frac{\partial}{\partial \kappa}\right]W[J]=0,\label{callan}
}
where $\eta = \mu \frac{\partial}{\partial \mu} \kappa$. Its running behaviour will allow us to determine $\kappa$ if we assume that it only runs implicitly through its dependence on $g$ and $\xi$, as shown below. By noting that $\kappa$ and $\delta \kappa$ have mass dimension $[\kappa]=[\delta \kappa]=D-4=-2\epsilon$ we find that starting from 
\muu{
0&= \mu \frac{\partial}{\partial \mu}\left[ \frac{1}{2}(\kappa +\delta \kappa) J^2 \mu^{-2\epsilon} \right],
}
the RG function of $\kappa$ can be written as
\muu{
\mu \frac{\partial \kappa}{\partial \mu}=(2\epsilon +2 \gamma_\mathcal{O})\kappa + \delta \label{prerunning},
}
with the inhomogeneity
\muu{
\delta = (2\epsilon + 2 \gamma_\mathcal{O})\delta \kappa - \mu \frac{\partial }{\partial \mu}\delta \kappa. \label{prerunningdelta}
}
Next, we use the assumption that the auxiliary parameter $\kappa=\kappa(g^2,\xi,\mu)$ depends on $\mu$ only implicitly via $g^2(\mu)$ and $\xi(\mu)$. The equation \eqref{prerunning} then becomes
\muu{
\left[2\epsilon + 2 \gamma_\mathcal{O}-\beta_{g^2}\frac{\partial}{\partial g^2}- \xi \gamma_\xi \frac{\partial}{\partial \xi} \right](\kappa +\delta \kappa)=0. \label{runningkappa}
}
Expanding in $g^2$ this implies that the solution can be written as
\muu{
\kappa(g^2, \xi)= \frac{\kappa_0}{g^2}+\hbar \kappa_1 + \hbar^2 \kappa_2 g^2 +\dots,
}
where we temporarily introduced $\hbar$. At this stage it becomes obvious that we unfortunately need to perform $(n+1)$-loop calculations in order to determine $\kappa$ to $n$-loop. For example, assuming all quantities have been determined to two-loop level we have the expansions
\muu{
\beta_{g^2} &= -2\epsilon g^2 + \beta_1\  g^4 +\beta_2\  g^6, \nonumber\\
\delta\kappa &= \frac{\delta \kappa_0}{\epsilon} + \left(\frac{\delta \kappa_{1,1}}{\epsilon}+\frac{\delta \kappa_{1,2}}{\epsilon^2}\right) g^2,   \nonumber\\
\gamma_\mathcal{O}&=\gamma_{\mathcal{O},0}\ g^2 + \gamma_{\mathcal{O},1}\ g^4,  \nonumber\\
 \gamma_\xi &= \gamma_{\xi, 0}\ g^2 +\gamma_{\xi, 1}\ g^4,
}
and equation \eqref{runningkappa} implies 
\muu{
&\frac{1}{g^2}:\ \ \ \ 2 \epsilon \kappa_0  - 2\epsilon \kappa_0 =0,\\
&g^0: \ \ \ \  2 \gamma_{\mathcal{O},0}\  \kappa_0 + \kappa_0 \beta_1-\xi \gamma_{\xi ,0}\ \partial_\xi \kappa_0 +2  \delta \kappa_0 =0,\\
&g^2: \ \ \ \ \xi  \gamma_{\xi,0}\ \partial_\xi \kappa_1 -2\gamma_{\mathcal{O},0}\ \kappa_1 = \delta_1, \label{g2}
}
where
\muu{
\delta_1&=- \xi \gamma_{\xi,1}\ \partial_\xi \kappa_0 +2\gamma_{\mathcal{O},1}\ \kappa_0 +\beta_2 \kappa_0 + 4\delta \kappa_{1,1}\nonumber \\
&+\frac{1}{\epsilon}[4\delta \kappa_{1,2}+2\gamma_{\mathcal{O},0}\ \delta \kappa_0 -\xi \gamma_{\xi,0}\ \partial_\xi \delta \kappa_0].\label{inhom}
}
The first equation is satisfied identically while the second equation implies the ODE for $\kappa_0$, 
\muu{
\left[\xi \gamma_{\xi,0}\  \partial_\xi -2\gamma_{\mathcal{O},0}-\beta_1 \right]\kappa_0 = 2\delta\kappa_0. \label{kappa0ODE}
}
Therefore, knowledge of the one-loop quantities $\gamma_{\xi,0}$, $\gamma_{\mathcal{O},0}$, $\beta_1$ and $\delta \kappa_0$ is necessary to obtain the tree-level part $\kappa_0$. The solution of this ODE is plugged into equation \eqref{g2} to obtain $\kappa_1$. However, one comment needs to be made about the inhomogeneity $\delta_1$. When taking the limit $\epsilon \to 0$ the last term in equation \eqref{inhom} can only be finite if the bracket vanishes identically. 

This is guaranteed from the fact that if the theory is renormalizable, the finiteness of equations \eqref{callan}-\eqref{prerunning} implies the finiteness of $\delta$ and therefore there is no need to consider the terms proportional to $1/\epsilon$ in $\delta$, as they must vanish by construction \cite{dudalcorr}. In fact, based on the results in \cite{dudal2004dynamical} for the case of the ``full'' gluon composite operator $A_\mu^A A^{\mu A}$ and in Lorenz gauge with arbitrary gauge parameter this condition can be explicitly checked and is found to be satisfied. It should be remarked on the other hand that if one is interested in the mere existence of the condensate, knowledge of $\kappa_0$ is sufficient, thus avoiding this subtlety in determining $\kappa_1$.

Before we turn to the calculation of the last ingredient for the ODE \eqref{kappa0ODE}, that is the one-loop part of $\delta \kappa$, let us note that there actually exist two ways of calculating this quantity and also $\gamma_{\mathcal{O}}$, depending on the interpretation of the composite operator source $J$. One possibility is to regard $J$ as a constant parameter and therefore treat $\gamma_\mathcal{O}$ as a mass renormalization. Hence, all calculations are performed using a massive gluon propagator, which is quite cumbersome especially in higher-loop calculations. This has been adopted in the original version of the LCO formalism. Alternatively, in \cite{browne2003two} it has been suggested to treat $J$ as a non-dynamical field that interacts with the gluon. In this case, the calculations can be performed using massless propagators and the renormalization is done by inserting the composite operator into two-point functions in order to obtain $\gamma_\mathcal{O}$, while $\delta \kappa$ is obtained by inserting the composite operator into the vacuum bubbles, requiring the quantity $\langle \mathcal{O}(x)\mathcal{O}(y)\rangle$ to be finite. It actually was the second viewpoint that we used to prove the one-loop multiplicative renormalizability of our composite operator in section \ref{sec:compop}. Both approaches seem to be equivalent as for example the results derived in \cite{browne2003two} agree with those in \cite{verschelde2001non}. 

To obtain $\delta \kappa_0$ it is convenient to return to the viewpoint of $J$ being a mass, then the one-loop correction to the generating functional is given by
\muu{
&-\frac{i}{2}\Tr \log \left[\delta^{ab}\left(-p^2 g^{\mu \nu}+(1-\xi^{-1}) p^\mu p^\nu +g^{\mu \nu} J\right) \right] \nonumber \\
&\ \ \ \ \ \ \ +i \Tr \log \left[ \delta^{ab}\left( -p^2 +\xi J\right) \right]\nonumber\\
&=i\frac{2(N-1)}{2}\left[(D-1) \Tr \log (-p^2 +J) + \Tr \log (-p^2 +\xi J) \right]\nonumber \\
&\ \ \ \ \ \ \ +i \Tr \log \left[ \delta^{ab}\left( -p^2 +\xi J\right) \right],
}
where the second line is obtained using the orthonormality of the transverse and longitudinal gluon propagator. Adopting dimensional regularization and taking the derivative with respect to $J$ twice we find the $\epsilon$-divergent part proportional to $J^2$,
\muu{
\frac{2(N-1)}{2}\frac{(3-\xi^2)}{(4\pi)^2 \epsilon},
}
and therefore
\muu{
\frac{\delta \kappa_0}{\epsilon}=-\frac{2(N-1)}{2}\frac{(3-\xi^2)}{(4\pi)^2 \epsilon}.
}
We are now ready to solve the differential equation \eqref{kappa0ODE} for $\kappa_0$. A particular solution is given by
\muu{
\kappa_0^{(p)}= \frac{2(N-1)}{N} \xi.
}
The homogeneous part is solved as 
\muu{
\int \frac{d\kappa_0}{\kappa_0}&=\int \frac{2 \gamma_{\mathcal{O},0}+\beta_1}{ \xi \gamma_{\xi,0}}\nonumber \\
&=\int \frac{4-6\xi}{4\xi -3\xi^2 -9} d \xi = \int \frac{\frac{d}{d\xi}[4\xi -3\xi^2 -9]}{4\xi -3\xi^2 -9}  d\xi,
}
and therefore 
\muu{
\kappa_0^{(h)}= C (4\xi - 3 \xi^2 -9),
}
which implies the general solution
\muu{
\kappa_0 =  \frac{2(N-1)}{N} \xi + C (4\xi - 3 \xi^2 -9).\label{kappa0sol}
}
As discussed in \cite{dudal2004analytic} the minimum of the effective potential should be independent of the gauge fixing parameter, allowing us to choose the integration constant $C$ arbitrarily. In practice, the result for the vacuum energy may explicitly depend on $\xi$ due to the mixing between different orders of perturbation theory. This could only be avoided if one knew the potential up to infinite order. Nevertheless, in the next section we will motivate a reasonable choice for $C$.

\section{Effective potential and existence of the condensate}
\label{sec:effpot}

Before we calculate the one-loop effective potential there is still one problem left. Because of the introduction of the terms proportional to $J^2$ the generating functional lost its usual interpretation as an energy density. However, following \cite{dudal2004analytic} this can easily be circumvented by performing a Hubbard-Stratonovich transformation, introducing the auxiliary field $\sigma$ as 
\muu{
1= \int d\sigma \ \ \text{exp}\left[-i\frac{1}{2g^2 \kappa}( \sigma + A \mathcal{O}+B J)^2\right].\label{hubbard}
} 
Here, a normalization constant was absorbed into the path integral measure. The parameters $A$ and $B$ are chosen such that the $J^2$ term and the $J \mathcal{O}$ term of the original Lagrangian are cancelled, for example $A=-g$ and $B=-g \kappa$. The modified Lagrangian then reads
\muu{
\mathcal{L}_{mod}=\mathcal{L}_{YM}+\mathcal{L}_{GF}^{red}+\mathcal{L}_{GF}^{res}+\mathcal{L}_\sigma +\frac{\sigma}{g}J,
}
where 
\muu{
\mathcal{L}_{\sigma}=-\frac{\sigma^2}{2g^2\kappa}+\frac{1}{\kappa}\frac{\sigma}{g}\mathcal{O}-\frac{1}{2\kappa}\mathcal{O}^2.
}
From equation \eqref{hubbard} we also find that the vacuum expectation values of $\mathcal{O}$ and the auxiliary field $\sigma$ at $J=0$ are related as
\muu{
\langle \sigma \rangle = g \langle \mathcal{O}\rangle.
}
Provided the auxiliary field has a non-zero vacuum expectation value and using
\muu{
\frac{1}{g^2 \kappa}=\frac{1}{\kappa_0}-\frac{\kappa_1}{\kappa_0^2}g^2 +O(g^4, \hbar^2)
}
we note that $\mathcal{L}_\sigma$ contains the mass term for the coset gluon and ghosts, with the tree-level masses
\muu{
m^2_X = \frac{g\langle \sigma \rangle}{\kappa_0}; \ \ \ \ m^2_{\bar{\omega}\omega}=\frac{\xi g \langle\sigma\rangle}{\kappa_0}.\label{treemasses}
}
Thus, to answer whether the condensate exists or not, we need to calculate the effective potential for the auxiliary field. Decomposing the potential into $V=V_0+V_1$ with the tree part $V_0$ and the one-loop part $V_1$ we immediately find the tree-level part
\muu{
V_0(\sigma)=\frac{\sigma^2}{2 \kappa_0}.\label{treepot}
}
For the one-loop correction we have
\muu{
V_1(\sigma)&= -\frac{\kappa_1}{2\kappa_0^2}g^2\sigma^2 +i \Tr \log \left[ \delta^{ab}\left( -p^2 +\frac{\xi g \sigma}{ \kappa_0}\right) \right]  \nonumber \\
&-\frac{i}{2}\Tr \log \left[\delta^{ab}\left(-p^2 g^{\mu \nu}+(1-\xi^{-1}) p^\mu p^\nu + g^{\mu \nu} \frac{g\sigma}{\kappa_0}\right) \right].
}
Within dimensional regularization, the calculation of the logarithms can be done analogously to section \ref{sec:LCO}. Adopting the $\overline{MS}$-scheme we find
\muu{
V_1(\sigma)&= -\frac{\kappa_1}{2\kappa_0^2}g^2\sigma^2 - \frac{3}{64 \pi^2} 2(N-1) \frac{g^2 \sigma^2}{ \kappa_0^2}  \left( \frac{5}{6} - \log \left[ \frac{g \sigma}{\kappa_0 \bar{\mu}^2} \right] \right)\nonumber \\
&+\frac{1}{64 \pi^2}2(N-1)\frac{\xi^2 g^2 \sigma^2}{ \kappa_0^2}  \left( \frac{3}{2}-\log\left[ \frac{\xi g\sigma}{\kappa_0 \bar{\mu}^2} \right]\right),\label{onelooppot}
}
where $\bar{\mu}^2=4\pi \mu^2 e^{-\gamma}$. Next we are looking for the stationary points, 
\muu{
\frac{dV}{d\sigma}&=\frac{\sigma}{\kappa_0} \left(1-\frac{g^2 \kappa_1}{\kappa_0}\right)\nonumber\\
&-\frac{3}{32\pi^2} 2(N-1) \frac{g^2\sigma}{\kappa_0^2}\left(\frac{1}{3}-\log \left[\frac{g\sigma}{\kappa_0 \bar{\mu}^2} \right] \right)\nonumber \\
& +\frac{1}{32\pi^2} 2(N-1) \frac{\xi^2 g^2\sigma}{\kappa_0^2}\left(1- \log \left[\frac{\xi g\sigma}{\kappa_0 \bar{\mu}^2} \right] \right).
}
Besides the solution $\sigma=0$ we find another stationary point $\sigma_*$, providing the squared mass $m_X^2$ given by
\muu{
m_X^2=\frac{g\sigma_*}{\kappa_0 }=\bar{\mu}^2\ \text{Exp}\left[\frac{H_1}{g^2}+H_2 \right],
}
with
\muu{
H_1(\xi,\kappa_0)&=-\frac{1}{(3-\xi^2)}\frac{32\pi^2}{2(N-1)}\kappa_0,\\
H_2(\xi,\kappa_1)&=\frac{1}{(3-\xi^2)}\left( \frac{32\pi^2}{2(N-1)}\kappa_1 +1+\frac{1}{2}\xi^2 \log \xi^2 -\xi^2  \right).
}
Based on these results, we want to discuss the open issue of fixing the integration constant in the solution for $\kappa_0(\xi)$, equation \eqref{kappa0sol}. First of all, we need to recover the correct UV limit, $\sigma_* \to 0$ as $g^2 \to 0$,which implies that $H_1$ must be negative and thus $\xi^2 < 3$. This is consistent with the fact that the ``physical'' region for $\xi$ is the close vicinity of $\xi=0$.\footnote{See the discussion at the end of this section.} Then, from the tree potential \eqref{treepot} we learn that $\kappa_0$ should be positive in order to have a bounded-from-below tree part. The choice
\muu{
C_0=-\frac{1}{11}\frac{N-1}{N}
}
guarantees that $\kappa_0$ is positive for all $\xi$ within the close vicinity of $\xi=0$. Moreover, for this choice we find that for $\xi=0$ the function $H_2$ becomes an irrelevant constant, while for $H_1$ we obtain,
\muu{
H_1\left(\xi=0,\kappa_0=-9C_0\right)=-(4\pi)^2\frac{3}{11N}.
} 
Introducing the experimentally accessible and RG invariant QCD scale $\Lambda_{QCD}$ as usual,
\muu{
\Lambda_{QCD}=\bar{\mu}\ \text{Exp} \left[-\int^g \frac{dg'}{\beta_g(g')} \right],
}
we find that at $\xi=0$ and to one-loop order the coset gluon mass becomes proportional to $\Lambda_{QCD}^2$,
\muu{
m_X^2=e^{H_2(\xi=0,\kappa_1)}\Lambda_{QCD}^2.
}
Therefore, assuming that $\xi$ only changes marginally with $\bar{\mu}$ around $\xi=0$ we obtain an RG invariant coset gluon mass. More explicitly, to one-loop order
\muu{
\bar{\mu} \frac{d}{d\bar{\mu}}m_X^2= \text{const.} \times \bar{\mu} \frac{d}{d\bar{\mu}}\left( \bar{\mu}^2\  \text{Exp}\left[-\frac{3}{11N}\frac{(4\pi)^2}{g^2}\right]\right)=0.
}
Consequently, the vacuum energy is calculated as 
\muu{
V(\sigma_*)&=\frac{\sigma_*^2}{2\kappa_0}\left(1-\frac{g^2 \kappa_1}{\kappa_0} \right)\nonumber\\
&-\frac{3\cdot 2(N-1)}{64\pi^2}\frac{g^2 \sigma_*^2}{\kappa_0^2}\left(\frac{5}{6}-\log \left[\frac{g\sigma_*}{\kappa_0 \bar{\mu}^2}\right] \right)\nonumber \\
& +\frac{2(N-1)}{64\pi^2}\frac{g^2 \sigma_*^2 \xi^2}{\kappa_0^2}\left(\frac{3}{2}-\frac{1}{2}\log \xi^2-\log \left[\frac{g\sigma_*}{\kappa_0 \bar{\mu}^2}\right] \right).\label{inter1}
}
The first term in equation \eqref{inter1} is replaced using $\frac{dV}{d\sigma} \Big \vert_{\sigma_*}=0$, which yields
\muu{
\frac{\sigma_*^2}{2\kappa_0}\left(1-\frac{g^2 \kappa_1}{\kappa_0} \right)&=\frac{3\cdot 2(N-1)}{64\pi^2}\frac{g^2 \sigma_*^2}{\kappa_0^2}\left(\frac{1}{3}-\log \left[\frac{g\sigma_*}{\kappa_0 \bar{\mu}^2}\right] \right)\nonumber \\
&-\frac{2(N-1)}{64\pi^2}\frac{g^2 \sigma_*^2 \xi^2}{\kappa_0^2}\left(1-\log \left[\frac{\xi g\sigma_*}{\kappa_0 \bar{\mu}^2}\right] \right).
}
Plugging this into the equation \eqref{inter1} we obtain
\muu{
V(\sigma_*)&=-(3-\xi^2)\frac{2(N-1)}{128\pi^2}\frac{g^2\sigma_*^2}{\kappa_0^2}\nonumber\\
&=-(3-\xi^2)\frac{2(N-1)}{128\pi^2}m_X^4,
}
where in the last line we used that the gluon mass is given by $m_X^2=\frac{g\sigma_*}{\kappa_0}$ and the result is in full agreement with the $N=2$ MAG case \cite{dudal2004analytic}. Together with the condition $\xi^2<3$ we indeed find that the energy for this vacuum is negative and therefore the condensate is energetically favoured. At first sight, the dependence of the vacuum energy on the parameter $\xi$ is problematic, as one should obtain the gauge independent result. However, our ``gauge'' is different from the usual treatment in the sense that it removes superfluous degrees of freedom from the extended Yang-Mills theory, in order to recover the theory equipollent to the $SU(N)$ Yang-Mills theory. This suggests we set $\xi=0$ and thus our result hints at the existence of the non-zero coset field condensate, at least to one-loop level.

\section{Conclusion}

\label{sec:conclusions}

In this paper we investigated the decomposition of the $G=SU(N)$ Yang-Mills theory with respect to the stability group $H=U(N-1)$. We proved the one-loop renormalizability of this theory and explicitly obtained all the involved RG functions.

An important feature of this theory is the fact that one can introduce a gauge invariant mass term for the coset-field $\mathcal{X}_\mu \in \text{Lie}(G/H)$. This is interesting from the viewpoint that the existence of a non-zero condensate $\langle \mathcal{X}_\mu \mathcal{X}^\mu \rangle\neq 0$ directly leads to many implications such as quark confinement at low temperature. While it is true that the mass term for the coset gluon is gauge invariant within the original version of the reformulated Yang-Mills theory, it loses its gauge invariance because the color-field is considered to be fixed within this paper. However, we showed that one can at least introduce the on-shell BRST invariant composite operator $\mathcal{O}=\Tr_{G/H}\left(\mathcal{X}_\mu \mathcal{X}^\mu -2i\xi \mathcal{C}\bar{\mathcal{C}} \right)$ to investigate the possibility of a coset gluon condensate. As an intermediate step, by taking into account the mixing with condensates of the same quantum number, we obtained the one-loop renormalizability of this composite operator.

In the second part of the paper we used these results to discuss the existence of the condensate by means of the local composite operator formalism. Consequently, after performing a Hubbard-Stratonovich transformation, we obtained the one-loop effective potential $V(\sigma)$ for the auxiliary field $\sigma$, where the vacuum expectation values of $\sigma$ and $\mathcal{O}$ are related as $\langle \sigma \rangle = g\langle\mathcal{O}\rangle$. Indeed, we found a non-zero stationary point $\sigma_*$ away from the origin. However, the corresponding vacuum energy $V(\sigma_*)$ explicitly depends on the parameter $\xi$. This would be a problem in the usual gauge fixing framework, but our reduction condition has a different meaning as it reduces the enlarged color-field extended gauge symmetry back to the theory equipollent to the $SU(N)$ Yang-Mills theory. In other words, even though the reduction condition is imposed, the full $SU(N)$ gauge symmetry is preserved. But again, due to the fixing of the color-field, the situation changes. The reduction condition appears as a gauge fixing-like term for the coset gluon. Nevertheless, we take the standpoint that according to the previous argument, we should adopt the ``physical'' choice $\xi=0$ in order to incorporate the reduction condition in an $\delta$-function like manner. In this case, the value $V(\sigma_*)$ is negative and a non-zero coset gluon condensate is energetically favoured. Certainly, these considerations need to be improved, for example by discussing the existence of the condensate within a non-perturbative approach such as the functional renormalization group.

$\;$ \\
\section*{Acknowledgments}

M.W. was supported by the Ministry of Education, Culture, Sports, Science and Technology, Japan (MEXT scholarship). R. M. was  supported by Grant-in-Aid for JSPS Research Fellow Grant Number 17J04780. S.N. would like to thank Nakamura Sekizen-kai for a scholarship. K.-I. K. was  supported by Grant-in-Aid for Scientific Research, JSPS KAKENHI Grant Number (C) No.15K05042. The authors thank J.A. Gracey for pointing out missing contributions to the one-loop correction of the $\langle X\omega\bar{C}^j \rangle$ - vertex.

\appendix

\section{Counterterm Lagrangian}
\label{appA:counter}
In this appendix we set up the counterterm Lagrangian corresponding to 
\begin{widetext}
\muu{
\mathcal{L}=\mathcal{L}_{YM}+i\delta_B \bar{\delta}_B\left(\frac{1}{2}X_\mu^a X^{\mu a}-i\frac{\xi}{2}\omega^a \bar{\omega}^a \right)-i\delta_B \left[\bar{C}^j \left(\partial_\mu V^{\mu j}+ \frac{\lambda}{2} N^j \right)\right]-i\delta_B \left[\bar{C}^\gamma \left(\partial_\mu V^{\mu \gamma}+ \frac{\alpha}{2} N^\gamma \right)\right].
}
Then the counterterm Lagrangian is written as
\muu{
&\mathcal{L}_{c.t.} =\nonumber\\
&\Delta_1 \frac{1}{2} X_\mu^a \left( g^{\mu \nu} \partial^2 - \partial^\mu \partial^\nu \right) X_\nu^a + \Delta_2 \frac{1}{2\xi}  X_\mu^a \partial^\mu \partial^\nu X_\nu^a  +\Delta_3 \frac{1}{2} V_\mu^j \left( g^{\mu \nu} \partial^2 - \partial^\mu \partial^\nu \right) V_\nu^j + \Delta_4 \frac{1}{2\lambda}  V_\mu^j \partial^\mu \partial^\nu V_\nu^j+\Delta_5 \frac{1}{2} V_\mu^\gamma \left( g^{\mu \nu} \partial^2 - \partial^\mu \partial^\nu \right) V_\nu^\gamma \nonumber \\
& + \Delta_6 \frac{1}{2\alpha}  V_\mu^\gamma \partial^\mu \partial^\nu V_\nu^\gamma+ \Delta_7 i  \bar{\omega}^a \partial^2 \omega^a +\Delta_8 i \bar{C}^j \partial^2 C^j +\Delta_9 i \bar{C}^\gamma \partial^2 C^\gamma     \nonumber\\
&-\Delta_{10} \frac{g}{2}  f^{jkl} (\partial_\mu V_\nu^j -\partial_\nu V_\mu^j) V^{\mu k} V^{\nu l} -\Delta_{11} \frac{g}{\xi} f^{ajb}  \partial_\mu X^{\mu a} V_\nu^j X^{\nu b} -\Delta_{12} \frac{g}{\xi} f^{a\gamma b}  \partial_\mu X^{\mu a} V_\nu^\gamma X^{\nu b}   \nonumber \\
&+ \Delta_{13} \frac{g}{2} f^{ajb} \left\{ X^{\mu a} X^{\nu b} (\partial_\mu V_\nu^j - \partial_\nu V_\mu^j) + V^{\mu j} X^{\nu b}(\partial_\nu X_\mu^a -\partial_\mu X_\nu^a )+V^{\nu j} X^{\mu b}(\partial_\mu X_\nu^a - \partial_\nu X_\mu^a) \right\}\nonumber \\
&+\Delta_{14} \frac{g}{2} f^{a\gamma b} \left\{ X^{\mu a} X^{\nu b} (\partial_\mu V_\nu^\gamma - \partial_\nu V_\mu^\gamma) + V^{\mu \gamma} X^{\nu b}(\partial_\nu X_\mu^a -\partial_\mu X_\nu^a )+V^{\nu \gamma} X^{\mu b}(\partial_\mu X_\nu^a - \partial_\nu X_\mu^a) \right\}         \nonumber\\
& -\Delta_{15}\frac{g^2}{4} f^{abJ} f^{cdJ} X_\mu^a X_\nu^b X^{\mu c} X^{\nu d} - \Delta_{16} \frac{g^2}{4}  f^{jkl}f^{jmn} V_\mu^k V_\nu^l V^{\mu m} V^{\nu n} \nonumber \\
&+\Delta_{17} g^2  \left\{ \frac{1}{2} f^{akc} f^{cjb} \left(  X_\mu^a X^{\mu b} V_\nu^k V^{\nu j} - X_\mu^a X_\nu^b V^{\mu k} V^{\nu j} \right) - f^{abj} f^{jkl} V_\mu^k V_\nu^l X^{\mu a} X^{\nu b} \right\}\nonumber\\
&+\Delta_{18} g^2  \left\{ \frac{1}{2} f^{a\gamma c} f^{c\gamma b} \left(  X_\mu^a X^{\mu b} V_\nu^\gamma V^{\nu \gamma} - X_\mu^a X_\nu^b V^{\mu \gamma} V^{\nu \gamma} \right)  \right\}\nonumber\\
&+\Delta_{19} g^2 f^{akc}f^{c\gamma b} \left\{ X_\mu^a X^{\mu b} V_\nu^k V^{\nu \gamma} -\frac{1}{2} X_\mu^a X^{\nu b}V^{\mu k} V_\nu^\gamma -\frac{1}{2}X_\mu^b X^{\nu a} V^{\mu \gamma} V_\nu^k \right\}\nonumber\\
&-\Delta_{20}\frac{g^2}{2\xi} f^{ajb} f^{akc}  V^{\mu j} X_\mu^b V^k_\nu X^{\nu c} -\Delta_{21}\frac{g^2}{2\xi} \frac{N}{2(N-1)}  V^{\mu \gamma} X_\mu^a V^\gamma_\nu X^{\nu a} -\Delta_{22}\frac{g^2}{2\xi} f^{a\gamma b} f^{akc} \left\{   V^{\mu \gamma} X_\mu^b V^k_\nu X^{\nu c} + V^{\mu k} X_\mu^c X_\nu^b V^{\nu \gamma}  \right\}\nonumber\\
&+\Delta_{23}i  g f^{ajb} \left\{\bar{\omega}^a \partial_\mu (V^{\mu j} \omega^b)+ \bar{\omega}^a V^{\mu j} \partial_\mu \omega^b\right\}+\Delta_{24}i  g f^{a\gamma b} \left\{\bar{\omega}^a \partial_\mu (V^{\mu \gamma} \omega^b)+ \bar{\omega}^a V^{\mu \gamma} \partial_\mu \omega^b\right\}\nonumber\\
&+\Delta_{25}ig f^{jkl} \bar{C}^j \partial_\mu (V^{\mu k} C^l)+\Delta_{26}ig f^{jab} \bar{C}^j \partial_\mu (X^{\mu a} \omega^b)+\Delta_{27}ig f^{\gamma ab} \bar{C}^\gamma \partial_\mu (X^{\mu a} \omega^b)\nonumber \\
&+\Delta_{28} i  g^2 f^{akc} f^{cjb} \bar{\omega}^a \omega^b V^{\mu k} V_\mu^j-i \Delta_{29}  g^2 \frac{N}{2(N-1)} \bar{\omega}^a \omega^a V^{\mu \gamma} V_\mu^\gamma\nonumber\\
&+\Delta_{30}i  g^2 f^{akc} f^{c\gamma b} V_\mu^k V^{\mu \gamma}\left\{\bar{\omega}^a \omega^b + \bar{\omega}^b \omega^a \right\}+\Delta_{31}i g^2 f^{abJ} f^{cdJ} \bar{\omega}^b \omega^d X_\mu^a X^\mu_c +\Delta_{32} \frac{\xi g^2}{4} f^{abJ} f^{cdJ} \bar{\omega}^a \bar{\omega}^b \omega^c \omega^d,
}
where the coefficients $\Delta_i$ are expressed in terms of the renormalization factors as

\begin{eqnarray*}
\Delta_1=Z_X-1, &
\Delta_2=Z_X Z_\xi^{-1}-1, &
\Delta_3=Z_V-1,\\
\Delta_4=Z_V Z_\lambda^{-1}-1, &
\Delta_5=\tilde{Z}_V-1,&
\Delta_6=\tilde{Z}_V Z_\alpha^{-1}-1, \\
\Delta_7=Z_\omega-1, &
\Delta_8=Z_C^{1/2}Z_{\bar{C}}^{1/2}-1,&
\Delta_9=\tilde{Z}_C^{1/2}\tilde{Z}_{\bar{C}}^{1/2}-1,\\
\Delta_{10}=Z_V^{3/2}Z_g-1, &
\Delta_{11}=Z_X Z_V^{1/2}Z_g Z_\xi^{-1}-1,&
\Delta_{12}=Z_X \tilde{Z}_V^{1/2}Z_g Z_\xi^{-1}-1,\\
\Delta_{13}=Z_X Z_V^{1/2}Z_g-1,&
\Delta_{14}=Z_X \tilde{Z}_V^{1/2}Z_g-1,&
\Delta_{15}=Z_g^2 Z_X^2 -1, \\
\Delta_{16}=Z_V^2 Z_g^2-1 ,&
\Delta_{17}=Z_X Z_V Z_g^2 -1,&
\Delta_{18}=Z_X \tilde{Z}_V Z_g^2 -1,\\
\Delta_{19}=Z_X Z_V^{1/2}\tilde{Z}_V^{1/2} Z_g^2 -1,&
\Delta_{20}=Z_X Z_V Z_g^2 Z_\xi^{-1}-1,&
\Delta_{21}=Z_X \tilde{Z}_V Z_g^2 Z_\xi^{-1}-1,\\
\Delta_{22}=Z_X Z_V^{1/2}\tilde{Z}_V^{1/2} Z_g^2 Z_\xi^{-1}-1,&
\Delta_{23}=Z_\omega Z_V^{1/2} Z_g-1,&
\Delta_{24}=Z_\omega \tilde{Z}_V^{1/2} Z_g-1,\\
\Delta_{25}=Z_C^{1/2}Z_{\bar{C}}^{1/2}Z_V^{1/2}Z_g-1,&
\phantom{xx} \Delta_{26}=Z_\omega^{1/2} Z_{\bar{C}}^{1/2} Z_X^{1/2}Z_g-1,\phantom{xx}&
\Delta_{27}=Z_\omega^{1/2} \tilde{Z}_{\bar{C}}^{1/2} Z_X^{1/2}Z_g-1,\\
\Delta_{28}=Z_\omega Z_V Z_g^2-1,&
\Delta_{29}=Z_\omega \tilde{Z}_V Z_g^2-1,&
\Delta_{30}=Z_\omega Z_V^{1/2} \tilde{Z}_V^{1/2} Z_g^2-1,\\
\Delta_{31}=Z_\omega Z_X Z_g^2-1,&
\Delta_{32}=Z_\omega^2 Z_g^2 Z_\xi -1.
\end{eqnarray*}
\muu{
\phantom{x} \label{counterterms}
}
In the main text we determine $\Delta_1$, $\Delta_2$, $\Delta_3$, $\Delta_4$, $\Delta_5$, $\Delta_6$, $\Delta_7$, $\Delta_8$ and  $\Delta_9$ by considering one-loop self-energy corrections. Next, we consider corrections to the $V^jC^k\bar{C}^l$-vertex and the $X\omega \bar{C}^J$-vertex, yielding $\Delta_{25}$, $\Delta_{26}$ and $\Delta_{27}$, respectively. This is sufficient to determine all the renormalization factors to one-loop level.
\end{widetext}

\section{Determining the renormalization Matrix of the composite operator}
\label{appB}

In this appendix we briefly explain how to determine the renormalization matrix of the composite operator renormalization, deriving the diagrammatic equations for the renormalization matrix elements. This is done by inserting the composite operator into the propagators of the fields.

\subsubsection{Insertion into $\langle X X \rangle$}

\begin{equation}
\begin{gathered}
\begin{fmfgraph*}(80,40)
\fmfleft{i1}
\fmflabel{$\phantom{x}$}{i1}
\fmfright{o1}
\fmflabel{$\phantom{x}$}{o1}
\fmf{gluon,label=$\phantom{x}$}{i1,v}
\fmf{gluon,label=$\phantom{x}$}{v,o1}
\fmfv{decor.shape=cross}{v}
\end{fmfgraph*} 
\end{gathered}\equiv 
\begin{gathered}
\langle XX \left[\frac{1}{2}XX\right]_R \rangle = \nonumber
\end{gathered} 
\end{equation}
\begin{equation}
Z_1 \Bigg \{
\begin{gathered}
\begin{fmfgraph*}(40,40)
\fmfleft{i1}
\fmflabel{$\phantom{x}$}{i1}
\fmfright{o1}
\fmflabel{$\phantom{x}$}{o1}
\fmf{gluon,label=$\phantom{x}$}{i1,v}
\fmf{gluon,label=$\phantom{x}$}{v,o1}
\fmfv{decor.shape=cross}{v}
\end{fmfgraph*} 
\end{gathered}+ 
\begin{gathered}
\begin{fmfgraph*}(80,40)
   \fmfleft{i1,i2}
\fmflabel{$\phantom{x}$}{i1}
\fmfright{o1,o2}
\fmflabel{$\phantom{x}$}{o1}
\fmf{phantom,tension=1.5}{i2,v2}
\fmf{phantom,tension=1.5}{o1,v2}
\fmf{gluon,label=$\phantom{x}$,tension=1.5,label.side=right}{i1,v1}
\fmf{gluon,label=$\phantom{x}$,left,tension=0.05}{v1,v2}
\fmf{gluon,label=$\phantom{x}$,left,tension=0.05}{v2,v1}
\fmf{gluon,label=$\phantom{x}$,tension=1.5}{v1,o1}
\fmfdotn{v}{1}
\fmfv{decor.shape=cross}{v2}
  \end{fmfgraph*}
\end{gathered} + 
\begin{gathered}
\begin{fmfgraph*}(80,40)
\fmfleft{i1}
\fmflabel{$\phantom{x}$}{i1}
\fmfright{o1}
\fmflabel{$\phantom{x}$}{o1}
\fmf{gluon,label=$\phantom{x}$}{v1,i1}
\fmf{photon,left,tension=0.4,label=$V^j+V^\gamma$}{v1,v2}
\fmf{gluon,label=$\phantom{x}$}{v3,v1}
\fmf{gluon,label=$\phantom{x}$}{v2,v3}
\fmfv{decor.shape=cross}{v3}
\fmf{gluon,label=$\phantom{x}$}{o1,v2}
\fmfdotn{v}{2}
  \end{fmfgraph*}
\end{gathered}\Bigg \} \nonumber 
\end{equation}
\begin{equation}
+Z_2 \Bigg \{\begin{gathered}
\begin{fmfgraph*}(70,40)
\fmfleft{i1}
\fmflabel{$\phantom{x}$}{i1}
\fmfright{o1}
\fmflabel{$\phantom{x}$}{o1}
\fmf{gluon,label=$\phantom{x}$}{v1,i1}
\fmf{photon,label.side=left,label=$\phantom{x}$}{v1,v3}
\fmf{photon,label.side=left,label=$\phantom{x}$}{v3,v2}
\fmf{gluon,left,tension=0.4,label=$\phantom{x}$}{v2,v1}
\fmfv{decor.shape=cross}{v3}
\fmfv{label=$V^j$,label.angle=90,label.dist=0.3cm}{v3}
\fmf{gluon,label=$\phantom{x}$}{o1,v2}
\fmfdotn{v}{2}
\end{fmfgraph*} 
\end{gathered}\ \ +
\begin{gathered}
\begin{fmfgraph*}(50,60)
\fmfleft{i1,i2}
\fmflabel{$\phantom{x}$}{i1}
\fmfright{o1,o2}
\fmflabel{$\phantom{x}$}{o1}
\fmflabel{$\phantom{x}$}{v2}
\fmf{phantom,tension=1.5}{i2,v2}
\fmf{phantom,tension=1.5}{o1,v2}
\fmf{gluon,label=$\phantom{x}$,tension=1.5,label.side=right}{i1,v1}
\fmf{photon,label=$\phantom{x}$,left,tension=0.05}{v1,v2}
\fmf{photon,label=$\phantom{x}$,left,tension=0.05}{v2,v1}
\fmf{gluon,label=$\phantom{x}$,tension=1.5}{v1,o1}
\fmfdotn{v}{1}
\fmfv{decor.shape=cross}{v2}
\fmfv{label=$V^j$,label.angle=90,label.dist=0.3cm}{v2}
\end{fmfgraph*} 
\end{gathered}\Bigg \} 
+Z_3 \Bigg \{\begin{gathered}
\begin{fmfgraph*}(50,60)
\fmfleft{i1,i2}
\fmflabel{$\phantom{x}$}{i1}
\fmfright{o1,o2}
\fmflabel{$\phantom{x}$}{o1}
\fmf{phantom,tension=1.5}{i2,v2}
\fmf{phantom,tension=1.5}{o1,v2}
\fmf{gluon,label=$\phantom{x}$,tension=1.5,label.side=right}{i1,v1}
\fmf{ghost,label=$\phantom{x}$,left,tension=0.05}{v1,v2}
\fmf{ghost,label=$\phantom{x}$,left,tension=0.05}{v2,v1}
\fmf{gluon,label=$\phantom{x}$,tension=1.5}{v1,o1}
\fmfdotn{v}{1}
\fmfv{decor.shape=cross}{v2}
\fmfv{label=$\omega$,label.angle=90,label.dist=0.3cm}{v2}
\end{fmfgraph*} 
\end{gathered}
\Bigg \} \nonumber
\end{equation}
\begin{equation}
+Z_4  \{ 0  \} +Z_5 \Bigg \{\begin{gathered}
\begin{fmfgraph*}(80,40)
\fmfleft{i1}
\fmflabel{$\phantom{x}$}{i1}
\fmfright{o1}
\fmflabel{$\phantom{x}$}{o1}
\fmf{gluon,label=$\phantom{x}$}{v1,i1}
\fmf{photon,label.side=left,label=$\phantom{x}$}{v1,v3}
\fmf{photon,label.side=left,label=$\phantom{x}$}{v3,v2}
\fmf{gluon,left,tension=0.4,label=$\phantom{x}$}{v2,v1}
\fmfv{decor.shape=cross}{v3}
\fmfv{label=$V^\gamma$,label.angle=90,label.dist=0.3cm}{v3}
\fmf{gluon,label=$\phantom{x}$}{o1,v2}
\fmfdotn{v}{2}
\end{fmfgraph*} 
\end{gathered}+
\begin{gathered}
\begin{fmfgraph*}(50,50)
\fmfleft{i1,i2}
\fmflabel{$\phantom{x}$}{i1}
\fmfright{o1,o2}
\fmflabel{$\phantom{x}$}{o1}
\fmflabel{$\phantom{x}$}{v2}
\fmf{phantom,tension=1.5}{i2,v2}
\fmf{phantom,tension=1.5}{o1,v2}
\fmf{gluon,label=$\phantom{x}$,tension=1.5,label.side=right}{i1,v1}
\fmf{photon,label=$\phantom{x}$,left,tension=0.05}{v1,v2}
\fmf{photon,label=$\phantom{x}$,left,tension=0.05}{v2,v1}
\fmf{gluon,label=$\phantom{x}$,tension=1.5}{v1,o1}
\fmfdotn{v}{1}
\fmfv{decor.shape=cross}{v2}
\fmfv{label=$V^\gamma$,label.angle=90,label.dist=0.3cm}{v2}
\end{fmfgraph*} 
\end{gathered}\Bigg \} + Z_6  \{ 0  \}.   \label{eq1} 
\end{equation}

From this it follows that $Z_1$ has the tree part.

\begin{equation}
0 \equiv \begin{gathered}
\langle XX \left[\frac{1}{2}V^j V^j \right]_R \rangle = \nonumber  
\end{gathered}
\end{equation}
\begin{equation} 
Z_7 \Bigg \{
\begin{gathered}
\begin{fmfgraph*}(40,40)
\fmfleft{i1}
\fmflabel{$\phantom{x}$}{i1}
\fmfright{o1}
\fmflabel{$\phantom{x}$}{o1}
\fmf{gluon,label=$\phantom{x}$}{i1,v}
\fmf{gluon,label=$\phantom{x}$}{v,o1}
\fmfv{decor.shape=cross}{v}
\end{fmfgraph*} 
\end{gathered} + 
\begin{gathered}
\begin{fmfgraph*}(50,50)
   \fmfleft{i1,i2}
\fmflabel{$\phantom{x}$}{i1}
\fmfright{o1,o2}
\fmflabel{$\phantom{x}$}{o1}
\fmf{phantom,tension=1.5}{i2,v2}
\fmf{phantom,tension=1.5}{o1,v2}
\fmf{gluon,label=$\phantom{x}$,tension=1.5,label.side=right}{i1,v1}
\fmf{gluon,label=$\phantom{x}$,left,tension=0.05}{v1,v2}
\fmf{gluon,label=$\phantom{x}$,left,tension=0.05}{v2,v1}
\fmf{gluon,label=$\phantom{x}$,tension=1.5}{v1,o1}
\fmfdotn{v}{1}
\fmfv{decor.shape=cross}{v2}
  \end{fmfgraph*}
\end{gathered} + 
\begin{gathered}
\begin{fmfgraph*}(80,40)
\fmfleft{i1}
\fmflabel{$\phantom{x}$}{i1}
\fmfright{o1}
\fmflabel{$\phantom{x}$}{o1}
\fmf{gluon,label=$\phantom{x}$}{v1,i1}
\fmf{photon,left,tension=0.4,label=$V^j+V^\gamma$}{v1,v2}
\fmf{gluon,label=$\phantom{x}$}{v3,v1}
\fmf{gluon,label=$\phantom{x}$}{v2,v3}
\fmfv{decor.shape=cross}{v3}
\fmf{gluon,label=$\phantom{x}$}{o1,v2}
\fmfdotn{v}{2}
  \end{fmfgraph*}
\end{gathered}\Bigg \} \nonumber 
\end{equation}
\begin{equation}
+Z_8 \Bigg \{\begin{gathered}
\begin{fmfgraph*}(80,40)
\fmfleft{i1}
\fmflabel{$\phantom{x}$}{i1}
\fmfright{o1}
\fmflabel{$\phantom{x}$}{o1}
\fmf{gluon,label=$\phantom{x}$}{v1,i1}
\fmf{photon,label.side=left,label=$\phantom{x}$}{v1,v3}
\fmf{photon,label.side=left,label=$\phantom{x}$}{v3,v2}
\fmf{gluon,left,tension=0.4,label=$\phantom{x}$}{v2,v1}
\fmfv{decor.shape=cross}{v3}
\fmfv{label=$V^j$,label.angle=90,label.dist=0.3cm}{v3}
\fmf{gluon,label=$\phantom{x}$}{o1,v2}
\fmfdotn{v}{2}
\end{fmfgraph*} 
\end{gathered}+
\begin{gathered}
\begin{fmfgraph*}(50,50)
\fmfleft{i1,i2}
\fmflabel{$\phantom{x}$}{i1}
\fmfright{o1,o2}
\fmflabel{$\phantom{x}$}{o1}
\fmflabel{$\phantom{x}$}{v2}
\fmf{phantom,tension=1.5}{i2,v2}
\fmf{phantom,tension=1.5}{o1,v2}
\fmf{gluon,label=$\phantom{x}$,tension=1.5,label.side=right}{i1,v1}
\fmf{photon,label=$\phantom{x}$,left,tension=0.05}{v1,v2}
\fmf{photon,label=$\phantom{x}$,left,tension=0.05}{v2,v1}
\fmf{gluon,label=$\phantom{x}$,tension=1.5}{v1,o1}
\fmfdotn{v}{1}
\fmfv{decor.shape=cross}{v2}
\fmfv{label=$V^j$,label.angle=90,label.dist=0.3cm}{v2}
\end{fmfgraph*} 
\end{gathered}\Bigg \} 
+Z_9 \Bigg \{\begin{gathered}
\begin{fmfgraph*}(50,50)
\fmfleft{i1,i2}
\fmflabel{$\phantom{x}$}{i1}
\fmfright{o1,o2}
\fmflabel{$\phantom{x}$}{o1}
\fmf{phantom,tension=1.5}{i2,v2}
\fmf{phantom,tension=1.5}{o1,v2}
\fmf{gluon,label=$\phantom{x}$,tension=1.5,label.side=right}{i1,v1}
\fmf{ghost,label=$\phantom{x}$,left,tension=0.05}{v1,v2}
\fmf{ghost,label=$\phantom{x}$,left,tension=0.05}{v2,v1}
\fmf{gluon,label=$\phantom{x}$,tension=1.5}{v1,o1}
\fmfdotn{v}{1}
\fmfv{decor.shape=cross}{v2}
\fmfv{label=$\omega$,label.angle=90,label.dist=0.3cm}{v2}
\end{fmfgraph*} 
\end{gathered}
\Bigg \} \nonumber
\end{equation}
\begin{equation}
+Z_{10}  \{ 0 \} +Z_{11} \Bigg \{\begin{gathered}
\begin{fmfgraph*}(80,40)
\fmfleft{i1}
\fmflabel{$\phantom{x}$}{i1}
\fmfright{o1}
\fmflabel{$\phantom{x}$}{o1}
\fmf{gluon,label=$\phantom{x}$}{v1,i1}
\fmf{photon,label.side=left,label=$\phantom{x}$}{v1,v3}
\fmf{photon,label.side=left,label=$\phantom{x}$}{v3,v2}
\fmf{gluon,left,tension=0.4,label=$\phantom{x}$}{v2,v1}
\fmfv{decor.shape=cross}{v3}
\fmfv{label=$V^\gamma$,label.angle=90,label.dist=0.3cm}{v3}
\fmf{gluon,label=$\phantom{x}$}{o1,v2}
\fmfdotn{v}{2}
\end{fmfgraph*} 
\end{gathered}+
\begin{gathered}
\begin{fmfgraph*}(50,50)
\fmfleft{i1,i2}
\fmflabel{$\phantom{x}$}{i1}
\fmfright{o1,o2}
\fmflabel{$\phantom{x}$}{o1}
\fmflabel{$\phantom{x}$}{v2}
\fmf{phantom,tension=1.5}{i2,v2}
\fmf{phantom,tension=1.5}{o1,v2}
\fmf{gluon,label=$\phantom{x}$,tension=1.5,label.side=right}{i1,v1}
\fmf{photon,label=$\phantom{x}$,left,tension=0.05}{v1,v2}
\fmf{photon,label=$\phantom{x}$,left,tension=0.05}{v2,v1}
\fmf{gluon,label=$\phantom{x}$,tension=1.5}{v1,o1}
\fmfdotn{v}{1}
\fmfv{decor.shape=cross}{v2}
\fmfv{label=$V^\gamma$,label.angle=90,label.dist=0.3cm}{v2}
\end{fmfgraph*} 
\end{gathered}\Bigg \} + Z_{12}  \{ 0  \}.   \label{eq2} 
\end{equation}
From this it follows that $Z_7$ does not have the tree part.

\begin{equation}
0 \equiv \begin{gathered}
\langle XX \left[i\omega \bar{\omega} \right]_R \rangle =  \nonumber
\end{gathered} 
\end{equation}
\begin{equation}
Z_{13} \Bigg \{
\begin{gathered}
\begin{fmfgraph*}(40,40)
\fmfleft{i1}
\fmflabel{$\phantom{x}$}{i1}
\fmfright{o1}
\fmflabel{$\phantom{x}$}{o1}
\fmf{gluon,label=$\phantom{x}$}{i1,v}
\fmf{gluon,label=$\phantom{x}$}{v,o1}
\fmfv{decor.shape=cross}{v}
\end{fmfgraph*} 
\end{gathered} + 
\begin{gathered}
\begin{fmfgraph*}(40,40)
   \fmfleft{i1,i2}
\fmflabel{$\phantom{x}$}{i1}
\fmfright{o1,o2}
\fmflabel{$\phantom{x}$}{o1}
\fmf{phantom,tension=1.5}{i2,v2}
\fmf{phantom,tension=1.5}{o1,v2}
\fmf{gluon,label=$\phantom{x}$,tension=1.5,label.side=right}{i1,v1}
\fmf{gluon,label=$\phantom{x}$,left,tension=0.05}{v1,v2}
\fmf{gluon,label=$\phantom{x}$,left,tension=0.05}{v2,v1}
\fmf{gluon,label=$\phantom{x}$,tension=1.5}{v1,o1}
\fmfdotn{v}{1}
\fmfv{decor.shape=cross}{v2}
  \end{fmfgraph*}
\end{gathered} + 
\begin{gathered}
\begin{fmfgraph*}(80,40)
\fmfleft{i1}
\fmflabel{$\phantom{x}$}{i1}
\fmfright{o1}
\fmflabel{$\phantom{x}$}{o1}
\fmf{gluon,label=$\phantom{x}$}{v1,i1}
\fmf{photon,left,tension=0.4,label=$V^j+V^\gamma$}{v1,v2}
\fmf{gluon,label=$\phantom{x}$}{v3,v1}
\fmf{gluon,label=$\phantom{x}$}{v2,v3}
\fmfv{decor.shape=cross}{v3}
\fmf{gluon,label=$\phantom{x}$}{o1,v2}
\fmfdotn{v}{2}
  \end{fmfgraph*}
\end{gathered}\Bigg \} \nonumber 
\end{equation}
\begin{equation}
+Z_{14} \Bigg \{\begin{gathered}
\begin{fmfgraph*}(80,40)
\fmfleft{i1}
\fmflabel{$\phantom{x}$}{i1}
\fmfright{o1}
\fmflabel{$\phantom{x}$}{o1}
\fmf{gluon,label=$\phantom{x}$}{v1,i1}
\fmf{photon,label.side=left,label=$\phantom{x}$}{v1,v3}
\fmf{photon,label.side=left,label=$\phantom{x}$}{v3,v2}
\fmf{gluon,left,tension=0.4,label=$\phantom{x}$}{v2,v1}
\fmfv{decor.shape=cross}{v3}
\fmfv{label=$V^j$,label.angle=90,label.dist=0.3cm}{v3}
\fmf{gluon,label=$\phantom{x}$}{o1,v2}
\fmfdotn{v}{2}
\end{fmfgraph*} 
\end{gathered}+
\begin{gathered}
\begin{fmfgraph*}(50,50)
\fmfleft{i1,i2}
\fmflabel{$\phantom{x}$}{i1}
\fmfright{o1,o2}
\fmflabel{$\phantom{x}$}{o1}
\fmflabel{$\phantom{x}$}{v2}
\fmf{phantom,tension=1.5}{i2,v2}
\fmf{phantom,tension=1.5}{o1,v2}
\fmf{gluon,label=$\phantom{x}$,tension=1.5,label.side=right}{i1,v1}
\fmf{photon,label=$\phantom{x}$,left,tension=0.05}{v1,v2}
\fmf{photon,label=$\phantom{x}$,left,tension=0.05}{v2,v1}
\fmf{gluon,label=$\phantom{x}$,tension=1.5}{v1,o1}
\fmfdotn{v}{1}
\fmfv{decor.shape=cross}{v2}
\fmfv{label=$V^j$,label.angle=90,label.dist=0.3cm}{v2}
\end{fmfgraph*} 
\end{gathered}\Bigg \} 
+Z_{15} \Bigg \{\begin{gathered}
\begin{fmfgraph*}(50,50)
\fmfleft{i1,i2}
\fmflabel{$\phantom{x}$}{i1}
\fmfright{o1,o2}
\fmflabel{$\phantom{x}$}{o1}
\fmf{phantom,tension=1.5}{i2,v2}
\fmf{phantom,tension=1.5}{o1,v2}
\fmf{gluon,label=$\phantom{x}$,tension=1.5,label.side=right}{i1,v1}
\fmf{ghost,label=$\phantom{x}$,left,tension=0.05}{v1,v2}
\fmf{ghost,label=$\phantom{x}$,left,tension=0.05}{v2,v1}
\fmf{gluon,label=$\phantom{x}$,tension=1.5}{v1,o1}
\fmfdotn{v}{1}
\fmfv{decor.shape=cross}{v2}
\fmfv{label=$\omega$,label.angle=90,label.dist=0.3cm}{v2}
\end{fmfgraph*} 
\end{gathered}
\Bigg \} \nonumber
\end{equation}
\begin{equation}
+Z_{16}   \{ 0  \} +Z_{17} \Bigg \{\begin{gathered}
\begin{fmfgraph*}(80,40)
\fmfleft{i1}
\fmflabel{$\phantom{x}$}{i1}
\fmfright{o1}
\fmflabel{$\phantom{x}$}{o1}
\fmf{gluon,label=$\phantom{x}$}{v1,i1}
\fmf{photon,label.side=left,label=$\phantom{x}$}{v1,v3}
\fmf{photon,label.side=left,label=$\phantom{x}$}{v3,v2}
\fmf{gluon,left,tension=0.4,label=$\phantom{x}$}{v2,v1}
\fmfv{decor.shape=cross}{v3}
\fmfv{label=$V^\gamma$,label.angle=90,label.dist=0.3cm}{v3}
\fmf{gluon,label=$\phantom{x}$}{o1,v2}
\fmfdotn{v}{2}
\end{fmfgraph*} 
\end{gathered}+
\begin{gathered}
\begin{fmfgraph*}(50,50)
\fmfleft{i1,i2}
\fmflabel{$\phantom{x}$}{i1}
\fmfright{o1,o2}
\fmflabel{$\phantom{x}$}{o1}
\fmflabel{$\phantom{x}$}{v2}
\fmf{phantom,tension=1.5}{i2,v2}
\fmf{phantom,tension=1.5}{o1,v2}
\fmf{gluon,label=$\phantom{x}$,tension=1.5,label.side=right}{i1,v1}
\fmf{photon,label=$\phantom{x}$,left,tension=0.05}{v1,v2}
\fmf{photon,label=$\phantom{x}$,left,tension=0.05}{v2,v1}
\fmf{gluon,label=$\phantom{x}$,tension=1.5}{v1,o1}
\fmfdotn{v}{1}
\fmfv{decor.shape=cross}{v2}
\fmfv{label=$V^\gamma$,label.angle=90,label.dist=0.3cm}{v2}
\end{fmfgraph*} 
\end{gathered}\Bigg \} + Z_{18}  \{ 0  \}.  \nonumber 
\end{equation}
\begin{equation}
\phantom{x} \label{eq3} 
\end{equation}
From this it follows that $Z_{13}$ does not have the tree part.

\begin{equation}
0 \equiv \begin{gathered}
\langle XX \left[iC^j \bar{C}^j \right]_R \rangle = 
\end{gathered}\nonumber \\
\end{equation}
\begin{equation} 
Z_{19} \Bigg \{
\begin{gathered}
\begin{fmfgraph*}(40,40)
\fmfleft{i1}
\fmflabel{$\phantom{x}$}{i1}
\fmfright{o1}
\fmflabel{$\phantom{x}$}{o1}
\fmf{gluon,label=$\phantom{x}$}{i1,v}
\fmf{gluon,label=$\phantom{x}$}{v,o1}
\fmfv{decor.shape=cross}{v}
\end{fmfgraph*} 
\end{gathered} + 
\begin{gathered}
\begin{fmfgraph*}(50,50)
   \fmfleft{i1,i2}
\fmflabel{$\phantom{x}$}{i1}
\fmfright{o1,o2}
\fmflabel{$\phantom{x}$}{o1}
\fmf{phantom,tension=1.5}{i2,v2}
\fmf{phantom,tension=1.5}{o1,v2}
\fmf{gluon,label=$\phantom{x}$,tension=1.5,label.side=right}{i1,v1}
\fmf{gluon,label=$\phantom{x}$,left,tension=0.05}{v1,v2}
\fmf{gluon,label=$\phantom{x}$,left,tension=0.05}{v2,v1}
\fmf{gluon,label=$\phantom{x}$,tension=1.5}{v1,o1}
\fmfdotn{v}{1}
\fmfv{decor.shape=cross}{v2}
  \end{fmfgraph*}
\end{gathered} + 
\begin{gathered}
\begin{fmfgraph*}(80,40)
\fmfleft{i1}
\fmflabel{$\phantom{x}$}{i1}
\fmfright{o1}
\fmflabel{$\phantom{x}$}{o1}
\fmf{gluon,label=$\phantom{x}$}{v1,i1}
\fmf{photon,left,tension=0.4,label=$V^j+V^\gamma$}{v1,v2}
\fmf{gluon,label=$\phantom{x}$}{v3,v1}
\fmf{gluon,label=$\phantom{x}$}{v2,v3}
\fmfv{decor.shape=cross}{v3}
\fmf{gluon,label=$\phantom{x}$}{o1,v2}
\fmfdotn{v}{2}
  \end{fmfgraph*}
\end{gathered}\Bigg \} \nonumber 
\end{equation}
\begin{equation}
+Z_{20} \Bigg \{\begin{gathered}
\begin{fmfgraph*}(80,40)
\fmfleft{i1}
\fmflabel{$\phantom{x}$}{i1}
\fmfright{o1}
\fmflabel{$\phantom{x}$}{o1}
\fmf{gluon,label=$\phantom{x}$}{v1,i1}
\fmf{photon,label.side=left,label=$\phantom{x}$}{v1,v3}
\fmf{photon,label.side=left,label=$\phantom{x}$}{v3,v2}
\fmf{gluon,left,tension=0.4,label=$\phantom{x}$}{v2,v1}
\fmfv{decor.shape=cross}{v3}
\fmfv{label=$V^j$,label.angle=90,label.dist=0.3cm}{v3}
\fmf{gluon,label=$\phantom{x}$}{o1,v2}
\fmfdotn{v}{2}
\end{fmfgraph*} 
\end{gathered}+
\begin{gathered}
\begin{fmfgraph*}(50,50)
\fmfleft{i1,i2}
\fmflabel{$\phantom{x}$}{i1}
\fmfright{o1,o2}
\fmflabel{$\phantom{x}$}{o1}
\fmflabel{$\phantom{x}$}{v2}
\fmf{phantom,tension=1.5}{i2,v2}
\fmf{phantom,tension=1.5}{o1,v2}
\fmf{gluon,label=$\phantom{x}$,tension=1.5,label.side=right}{i1,v1}
\fmf{photon,label=$\phantom{x}$,left,tension=0.05}{v1,v2}
\fmf{photon,label=$\phantom{x}$,left,tension=0.05}{v2,v1}
\fmf{gluon,label=$\phantom{x}$,tension=1.5}{v1,o1}
\fmfdotn{v}{1}
\fmfv{decor.shape=cross}{v2}
\fmfv{label=$Vj$,label.angle=90,label.dist=0.3cm}{v2}
\end{fmfgraph*} 
\end{gathered}\Bigg \} 
+Z_{21} \Bigg \{\begin{gathered}
\begin{fmfgraph*}(50,50)
\fmfleft{i1,i2}
\fmflabel{$\phantom{x}$}{i1}
\fmfright{o1,o2}
\fmflabel{$\phantom{x}$}{o1}
\fmf{phantom,tension=1.5}{i2,v2}
\fmf{phantom,tension=1.5}{o1,v2}
\fmf{gluon,label=$\phantom{x}$,tension=1.5,label.side=right}{i1,v1}
\fmf{ghost,label=$\phantom{x}$,left,tension=0.05}{v1,v2}
\fmf{ghost,label=$\phantom{x}$,left,tension=0.05}{v2,v1}
\fmf{gluon,label=$\phantom{x}$,tension=1.5}{v1,o1}
\fmfdotn{v}{1}
\fmfv{decor.shape=cross}{v2}
\fmfv{label=$\omega$,label.angle=90,label.dist=0.3cm}{v2}
\end{fmfgraph*} 
\end{gathered}
\Bigg \} \nonumber
\end{equation}
\begin{equation}
+Z_{22}  \{ 0  \} +Z_{23} \Bigg \{\begin{gathered}
\begin{fmfgraph*}(80,40)
\fmfleft{i1}
\fmflabel{$\phantom{x}$}{i1}
\fmfright{o1}
\fmflabel{$\phantom{x}$}{o1}
\fmf{gluon,label=$\phantom{x}$}{v1,i1}
\fmf{photon,label.side=left,label=$\phantom{x}$}{v1,v3}
\fmf{photon,label.side=left,label=$\phantom{x}$}{v3,v2}
\fmf{gluon,left,tension=0.4,label=$\phantom{x}$}{v2,v1}
\fmfv{decor.shape=cross}{v3}
\fmfv{label=$V^\gamma$,label.angle=90,label.dist=0.3cm}{v3}
\fmf{gluon,label=$\phantom{x}$}{o1,v2}
\fmfdotn{v}{2}
\end{fmfgraph*} 
\end{gathered}+
\begin{gathered}
\begin{fmfgraph*}(50,50)
\fmfleft{i1,i2}
\fmflabel{$\phantom{x}$}{i1}
\fmfright{o1,o2}
\fmflabel{$\phantom{x}$}{o1}
\fmflabel{$\phantom{x}$}{v2}
\fmf{phantom,tension=1.5}{i2,v2}
\fmf{phantom,tension=1.5}{o1,v2}
\fmf{gluon,label=$\phantom{x}$,tension=1.5,label.side=right}{i1,v1}
\fmf{photon,label=$\phantom{x}$,left,tension=0.05}{v1,v2}
\fmf{photon,label=$\phantom{x}$,left,tension=0.05}{v2,v1}
\fmf{gluon,label=$\phantom{x}$,tension=1.5}{v1,o1}
\fmfdotn{v}{1}
\fmfv{decor.shape=cross}{v2}
\fmfv{label=$V^\gamma$,label.angle=90,label.dist=0.3cm}{v2}
\end{fmfgraph*} 
\end{gathered}\Bigg \} + Z_{24} \{ 0 \}.   \label{eq4} 
\end{equation}
From this it follows that $Z_{19}$ does not have the tree part.

\begin{equation}
0 \equiv \begin{gathered}
\langle XX \left[\frac{1}{2}V^\gamma V^\gamma \right]_R \rangle = 
\end{gathered} \nonumber\\
\end{equation}
\begin{equation}
Z_{25} \Bigg \{
\begin{gathered}
\begin{fmfgraph*}(40,40)
\fmfleft{i1}
\fmflabel{$\phantom{x}$}{i1}
\fmfright{o1}
\fmflabel{$\phantom{x}$}{o1}
\fmf{gluon,label=$\phantom{x}$}{i1,v}
\fmf{gluon,label=$\phantom{x}$}{v,o1}
\fmfv{decor.shape=cross}{v}
\end{fmfgraph*} 
\end{gathered} + 
\begin{gathered}
\begin{fmfgraph*}(50,50)
   \fmfleft{i1,i2}
\fmflabel{$\phantom{x}$}{i1}
\fmfright{o1,o2}
\fmflabel{$\phantom{x}$}{o1}
\fmf{phantom,tension=1.5}{i2,v2}
\fmf{phantom,tension=1.5}{o1,v2}
\fmf{gluon,label=$\phantom{x}$,tension=1.5,label.side=right}{i1,v1}
\fmf{gluon,label=$\phantom{x}$,left,tension=0.05}{v1,v2}
\fmf{gluon,label=$\phantom{x}$,left,tension=0.05}{v2,v1}
\fmf{gluon,label=$\phantom{x}$,tension=1.5}{v1,o1}
\fmfdotn{v}{1}
\fmfv{decor.shape=cross}{v2}
  \end{fmfgraph*}
\end{gathered} + 
\begin{gathered}
\begin{fmfgraph*}(80,40)
\fmfleft{i1}
\fmflabel{$\phantom{x}$}{i1}
\fmfright{o1}
\fmflabel{$\phantom{x}$}{o1}
\fmf{gluon,label=$\phantom{x}$}{v1,i1}
\fmf{photon,left,tension=0.4,label=$V^j+V^\gamma$}{v1,v2}
\fmf{gluon,label=$\phantom{x}$}{v3,v1}
\fmf{gluon,label=$\phantom{x}$}{v2,v3}
\fmfv{decor.shape=cross}{v3}
\fmf{gluon,label=$\phantom{x}$}{o1,v2}
\fmfdotn{v}{2}
  \end{fmfgraph*}
\end{gathered}\Bigg \} \nonumber 
\end{equation}
\begin{equation}
+Z_{26} \Bigg \{\begin{gathered}
\begin{fmfgraph*}(80,40)
\fmfleft{i1}
\fmflabel{$\phantom{x}$}{i1}
\fmfright{o1}
\fmflabel{$\phantom{x}$}{o1}
\fmf{gluon,label=$\phantom{x}$}{v1,i1}
\fmf{photon,label.side=left,label=$\phantom{x}$}{v1,v3}
\fmf{photon,label.side=left,label=$\phantom{x}$}{v3,v2}
\fmf{gluon,left,tension=0.4,label=$\phantom{x}$}{v2,v1}
\fmfv{decor.shape=cross}{v3}
\fmfv{label=$V^j$,label.angle=90,label.dist=0.3cm}{v3}
\fmf{gluon,label=$\phantom{x}$}{o1,v2}
\fmfdotn{v}{2}
\end{fmfgraph*} 
\end{gathered}+
\begin{gathered}
\begin{fmfgraph*}(50,50)
\fmfleft{i1,i2}
\fmflabel{$\phantom{x}$}{i1}
\fmfright{o1,o2}
\fmflabel{$\phantom{x}$}{o1}
\fmflabel{$\phantom{x}$}{v2}
\fmf{phantom,tension=1.5}{i2,v2}
\fmf{phantom,tension=1.5}{o1,v2}
\fmf{gluon,label=$\phantom{x}$,tension=1.5,label.side=right}{i1,v1}
\fmf{photon,label=$\phantom{x}$,left,tension=0.05}{v1,v2}
\fmf{photon,label=$\phantom{x}$,left,tension=0.05}{v2,v1}
\fmf{gluon,label=$\phantom{x}$,tension=1.5}{v1,o1}
\fmfdotn{v}{1}
\fmfv{decor.shape=cross}{v2}
\fmfv{label=$V^j$,label.angle=90,label.dist=0.3cm}{v2}
\end{fmfgraph*} 
\end{gathered}\Bigg \} 
+Z_{27} \Bigg \{\begin{gathered}
\begin{fmfgraph*}(50,50)
\fmfleft{i1,i2}
\fmflabel{$\phantom{x}$}{i1}
\fmfright{o1,o2}
\fmflabel{$\phantom{x}$}{o1}
\fmf{phantom,tension=1.5}{i2,v2}
\fmf{phantom,tension=1.5}{o1,v2}
\fmf{gluon,label=$\phantom{x}$,tension=1.5,label.side=right}{i1,v1}
\fmf{ghost,label=$\phantom{x}$,left,tension=0.05}{v1,v2}
\fmf{ghost,label=$\phantom{x}$,left,tension=0.05}{v2,v1}
\fmf{gluon,label=$\phantom{x}$,tension=1.5}{v1,o1}
\fmfdotn{v}{1}
\fmfv{decor.shape=cross}{v2}
\fmfv{label=$\omega$,label.angle=90,label.dist=0.3cm}{v2}
\end{fmfgraph*} 
\end{gathered}
\Bigg \} \nonumber
\end{equation}
\begin{equation}
+Z_{28}   \{ 0  \} +Z_{29} \Bigg \{\begin{gathered}
\begin{fmfgraph*}(80,40)
\fmfleft{i1}
\fmflabel{$\phantom{x}$}{i1}
\fmfright{o1}
\fmflabel{$\phantom{x}$}{o1}
\fmf{gluon,label=$\phantom{x}$}{v1,i1}
\fmf{photon,label.side=left,label=$\phantom{x}$}{v1,v3}
\fmf{photon,label.side=left,label=$\phantom{x}$}{v3,v2}
\fmf{gluon,left,tension=0.4,label=$\phantom{x}$}{v2,v1}
\fmfv{decor.shape=cross}{v3}
\fmfv{label=$V^\gamma$,label.angle=90,label.dist=0.3cm}{v3}
\fmf{gluon,label=$\phantom{x}$}{o1,v2}
\fmfdotn{v}{2}
\end{fmfgraph*} 
\end{gathered}+
\begin{gathered}
\begin{fmfgraph*}(50,50)
\fmfleft{i1,i2}
\fmflabel{$\phantom{x}$}{i1}
\fmfright{o1,o2}
\fmflabel{$\phantom{x}$}{o1}
\fmflabel{$\phantom{x}$}{v2}
\fmf{phantom,tension=1.5}{i2,v2}
\fmf{phantom,tension=1.5}{o1,v2}
\fmf{gluon,label=$\phantom{x}$,tension=1.5,label.side=right}{i1,v1}
\fmf{photon,label=$\phantom{x}$,left,tension=0.05}{v1,v2}
\fmf{photon,label=$\phantom{x}$,left,tension=0.05}{v2,v1}
\fmf{gluon,label=$\phantom{x}$,tension=1.5}{v1,o1}
\fmfdotn{v}{1}
\fmfv{decor.shape=cross}{v2}
\fmfv{label=$V^\gamma$,label.angle=90,label.dist=0.3cm}{v2}
\end{fmfgraph*} 
\end{gathered}\Bigg \} + Z_{30}  \{ 0 \}.  \label{eq5} 
\end{equation}
From this it follows that $Z_{25}$ does not have the tree part.

\begin{equation}
0 \equiv \begin{gathered}
\langle XX \left[iC^\gamma \bar{C}^\gamma \right]_R \rangle = 
\end{gathered} \nonumber\\
\end{equation}
\begin{equation}
Z_{31} \Bigg \{
\begin{gathered}
\begin{fmfgraph*}(40,40)
\fmfleft{i1}
\fmflabel{$\phantom{x}$}{i1}
\fmfright{o1}
\fmflabel{$\phantom{x}$}{o1}
\fmf{gluon,label=$\phantom{x}$}{i1,v}
\fmf{gluon,label=$\phantom{x}$}{v,o1}
\fmfv{decor.shape=cross}{v}
\end{fmfgraph*} 
\end{gathered} + 
\begin{gathered}
\begin{fmfgraph*}(50,50)
   \fmfleft{i1,i2}
\fmflabel{$\phantom{x}$}{i1}
\fmfright{o1,o2}
\fmflabel{$\phantom{x}$}{o1}
\fmf{phantom,tension=1.5}{i2,v2}
\fmf{phantom,tension=1.5}{o1,v2}
\fmf{gluon,label=$\phantom{x}$,tension=1.5,label.side=right}{i1,v1}
\fmf{gluon,label=$\phantom{x}$,left,tension=0.05}{v1,v2}
\fmf{gluon,label=$\phantom{x}$,left,tension=0.05}{v2,v1}
\fmf{gluon,label=$\phantom{x}$,tension=1.5}{v1,o1}
\fmfdotn{v}{1}
\fmfv{decor.shape=cross}{v2}
  \end{fmfgraph*}
\end{gathered} + 
\begin{gathered}
\begin{fmfgraph*}(80,40)
\fmfleft{i1}
\fmflabel{$\phantom{x}$}{i1}
\fmfright{o1}
\fmflabel{$\phantom{x}$}{o1}
\fmf{gluon,label=$\phantom{x}$}{v1,i1}
\fmf{photon,left,tension=0.4,label=$V^j+V^\gamma$}{v1,v2}
\fmf{gluon,label=$\phantom{x}$}{v3,v1}
\fmf{gluon,label=$\phantom{x}$}{v2,v3}
\fmfv{decor.shape=cross}{v3}
\fmf{gluon,label=$\phantom{x}$}{o1,v2}
\fmfdotn{v}{2}
  \end{fmfgraph*}
\end{gathered}\Bigg \} \nonumber 
\end{equation}
$\;$ \\
\begin{equation}
+Z_{32} \Bigg \{\begin{gathered}
\begin{fmfgraph*}(80,40)
\fmfleft{i1}
\fmflabel{$\phantom{x}$}{i1}
\fmfright{o1}
\fmflabel{$\phantom{x}$}{o1}
\fmf{gluon,label=$\phantom{x}$}{v1,i1}
\fmf{photon,label.side=left,label=$\phantom{x}$}{v1,v3}
\fmf{photon,label.side=left,label=$\phantom{x}$}{v3,v2}
\fmf{gluon,left,tension=0.4,label=$\phantom{x}$}{v2,v1}
\fmfv{decor.shape=cross}{v3}
\fmfv{label=$V^j$,label.angle=90,label.dist=0.3cm}{v3}
\fmf{gluon,label=$\phantom{x}$}{o1,v2}
\fmfdotn{v}{2}
\end{fmfgraph*} 
\end{gathered}+
\begin{gathered}
\begin{fmfgraph*}(50,50)
\fmfleft{i1,i2}
\fmflabel{$\phantom{x}$}{i1}
\fmfright{o1,o2}
\fmflabel{$\phantom{x}$}{o1}
\fmflabel{$\phantom{x}$}{v2}
\fmf{phantom,tension=1.5}{i2,v2}
\fmf{phantom,tension=1.5}{o1,v2}
\fmf{gluon,label=$\phantom{x}$,tension=1.5,label.side=right}{i1,v1}
\fmf{photon,label=$\phantom{x}$,left,tension=0.05}{v1,v2}
\fmf{photon,label=$\phantom{x}$,left,tension=0.05}{v2,v1}
\fmf{gluon,label=$\phantom{x}$,tension=1.5}{v1,o1}
\fmfdotn{v}{1}
\fmfv{decor.shape=cross}{v2}
\fmfv{label=$V^j$,label.angle=90,label.dist=0.3cm}{v2}
\end{fmfgraph*} 
\end{gathered}\Bigg \} 
+Z_{33} \Bigg \{\begin{gathered}
\begin{fmfgraph*}(50,50)
\fmfleft{i1,i2}
\fmflabel{$\phantom{x}$}{i1}
\fmfright{o1,o2}
\fmflabel{$\phantom{x}$}{o1}
\fmf{phantom,tension=1.5}{i2,v2}
\fmf{phantom,tension=1.5}{o1,v2}
\fmf{gluon,label=$\phantom{x}$,tension=1.5,label.side=right}{i1,v1}
\fmf{ghost,label=$\phantom{x}$,left,tension=0.05}{v1,v2}
\fmf{ghost,label=$\phantom{x}$,left,tension=0.05}{v2,v1}
\fmf{gluon,label=$\phantom{x}$,tension=1.5}{v1,o1}
\fmfdotn{v}{1}
\fmfv{decor.shape=cross}{v2}
\fmfv{label=$\omega$,label.angle=90,label.dist=0.3cm}{v2}
\end{fmfgraph*} 
\end{gathered}
\Bigg \} \nonumber
\end{equation}
\begin{equation}
+Z_{34}   \{ 0  \} +Z_{35} \Bigg \{\begin{gathered}
\begin{fmfgraph*}(80,40)
\fmfleft{i1}
\fmflabel{$\phantom{x}$}{i1}
\fmfright{o1}
\fmflabel{$\phantom{x}$}{o1}
\fmf{gluon,label=$\phantom{x}$}{v1,i1}
\fmf{photon,label.side=left,label=$\phantom{x}$}{v1,v3}
\fmf{photon,label.side=left,label=$\phantom{x}$}{v3,v2}
\fmf{gluon,left,tension=0.4,label=$\phantom{x}$}{v2,v1}
\fmfv{decor.shape=cross}{v3}
\fmfv{label=$V^\gamma$,label.angle=90,label.dist=0.3cm}{v3}
\fmf{gluon,label=$\phantom{x}$}{o1,v2}
\fmfdotn{v}{2}
\end{fmfgraph*} 
\end{gathered}+
\begin{gathered}
\begin{fmfgraph*}(50,50)
\fmfleft{i1,i2}
\fmflabel{$\phantom{x}$}{i1}
\fmfright{o1,o2}
\fmflabel{$\phantom{x}$}{o1}
\fmflabel{$\phantom{x}$}{v2}
\fmf{phantom,tension=1.5}{i2,v2}
\fmf{phantom,tension=1.5}{o1,v2}
\fmf{gluon,label=$\phantom{x}$,tension=1.5,label.side=right}{i1,v1}
\fmf{photon,label=$\phantom{x}$,left,tension=0.05}{v1,v2}
\fmf{photon,label=$\phantom{x}$,left,tension=0.05}{v2,v1}
\fmf{gluon,label=$\phantom{x}$,tension=1.5}{v1,o1}
\fmfdotn{v}{1}
\fmfv{decor.shape=cross}{v2}
\fmfv{label=$V^\gamma$,label.angle=90,label.dist=0.3cm}{v2}
\end{fmfgraph*} 
\end{gathered}\Bigg \} + Z_{36}  \{ 0 \}.  \label{eq6} 
\end{equation}
From this it follows that $Z_{31}$ does not have the tree part.

\subsubsection{Insertion into $\langle V^j V^k \rangle$}

\begin{equation}
0 \equiv \begin{gathered}
\langle V^j V^k \left[\frac{1}{2} X X \right]_R \rangle = 
\end{gathered} \nonumber 
\end{equation}
\begin{equation}
Z_1 \Bigg \{
\begin{gathered}
\begin{fmfgraph*}(40,40)
\fmfleft{i1,i2}
\fmflabel{$V^j$}{i1}
\fmfright{o1,o2}
\fmflabel{$V^k$}{o1}
\fmf{phantom,tension=1.5}{i2,v2}
\fmf{phantom,tension=1.5}{o1,v2}
\fmf{photon,label=$\phantom{x}$,tension=1.5,label.side=right}{i1,v1}
\fmf{gluon,label=$\phantom{x}$,left,tension=0.05}{v1,v2}
\fmf{gluon,label=$\phantom{x}$,left,tension=0.05}{v2,v1}
\fmf{photon,label=$\phantom{x}$,tension=1.5}{v1,o1}
\fmfdotn{v}{1}
\fmfv{decor.shape=cross}{v2}
\end{fmfgraph*} 
\end{gathered} + \ \ \ \ \ \ 
\begin{gathered}
  \begin{fmfgraph*}(100,70)
\fmfleft{i1}
\fmflabel{$V^j$}{i1}
\fmfright{o1}
\fmflabel{$V^k$}{o1}
\fmf{photon,label=$\phantom{x}$}{i1,v1}
\fmf{gluon,label=$\phantom{x}$,label.side=left}{v1,v3}
\fmf{gluon,label=$\phantom{x}$,label.side=left}{v3,v2}
\fmf{gluon,left,label=$\phantom{x}$,tension=0.4}{v2,v1}
\fmf{photon,label=$\phantom{x}$}{v2,o1}
\fmfdotn{v}{2}
\fmfv{decor.shape=cross}{v3}
  \end{fmfgraph*}
\end{gathered}\ \ \ \ \ \ \ \Bigg \} \nonumber 
\end{equation}
\begin{equation}
+Z_2 \Bigg \{\ \ \ \ \ \ \ \  \begin{gathered}
\begin{fmfgraph*}(30,30)
\fmfleft{i1}
\fmflabel{$V^j$}{i1}
\fmfright{o1}
\fmflabel{$V^k$}{o1}
\fmf{photon,label=$\phantom{x}$}{i1,v}
\fmf{photon,label=$\phantom{x}$}{v,o1}
\fmfv{decor.shape=cross}{v}
\end{fmfgraph*}
\end{gathered}\ \ \ \ \ \ \ \ \  + \begin{gathered}
\begin{fmfgraph*}(40,40)
\fmfleft{i1,i2}
\fmflabel{$V^j$}{i1}
\fmfright{o1,o2}
\fmflabel{$V^k$}{o1}
\fmf{phantom,tension=1.5}{i2,v2}
\fmf{phantom,tension=1.5}{o1,v2}
\fmf{photon,label=$\phantom{x}$,tension=1.5,label.side=right}{i1,v1}
\fmf{photon,label=$\phantom{x}$,left,tension=0.05}{v1,v2}
\fmf{photon,label=$\phantom{x}$,left,tension=0.05}{v2,v1}
\fmf{photon,label=$\phantom{x}$,tension=1.5}{v1,o1}
\fmfdotn{v}{1}
\fmfv{decor.shape=cross}{v2}
\fmfv{label=$V^l$,label.angle=90,label.dist=0.3cm}{v2}
\end{fmfgraph*} 
\end{gathered}+\ \ \ \ \ \ \ \ 
\begin{gathered}
\begin{fmfgraph*}(60,40)
\fmfleft{i1}
\fmflabel{$V^j$}{i1}
\fmfright{o1}
\fmflabel{$V^k$}{o1}
\fmf{photon,label=$\phantom{x}$}{i1,v1}
\fmf{photon,right,label=$\phantom{x}$,tension=0.4}{v1,v2}
\fmf{photon,label.side=left,label=$\phantom{x}$}{v1,v3}
\fmf{photon,label.side=left,label=$\phantom{x}$}{v3,v2}
\fmf{photon,label=$\phantom{x}$}{v2,o1}
\fmfdotn{v}{2}
\fmfv{decor.shape=cross}{v3}
\fmfv{label=$V^l$,label.angle=90,label.dist=0.3cm}{v3}
\end{fmfgraph*} 
\end{gathered}\ \ \ \ \ \ \ \ \Bigg \}  \nonumber
\end{equation}
\begin{equation}
+Z_3 \ \ \ \ \ \ \Bigg \{\ \ \ \ \ \ \begin{gathered}
\begin{fmfgraph*}(30,40)
\fmfleft{i1,i2}
\fmflabel{$V^j$}{i1}
\fmfright{o1,o2}
\fmflabel{$V^k$}{o1}
\fmf{phantom,tension=1.5}{i2,v2}
\fmf{phantom,tension=1.5}{o1,v2}
\fmf{photon,label=$\phantom{x}$,tension=1.5,label.side=right}{i1,v1}
\fmf{ghost,label=$\phantom{x}$,left,tension=0.05}{v1,v2}
\fmf{ghost,label=$\phantom{x}$,left,tension=0.05}{v2,v1}
\fmf{photon,label=$\phantom{x}$,tension=1.5}{v1,o1}
\fmfdotn{v}{1}
\fmfv{decor.shape=cross}{v2}
\fmfv{label=$\omega$,label.angle=90,label.dist=0.3cm}{v2}
\end{fmfgraph*} 
\end{gathered}\ \ \ \ \ \ +\ \ \ \ \ \ \ \ \begin{gathered}
\begin{fmfgraph*}(60,40)
\fmfleft{i1}
\fmflabel{$V^j$}{i1}
\fmfright{o1}
\fmflabel{$V^k$}{o1}
\fmf{photon,label=$\phantom{x}$}{i1,v1}
\fmf{ghost,label=$\phantom{x}$}{v1,v3}
\fmf{ghost,label=$\phantom{x}$}{v3,v2}
\fmf{ghost,left,label=$\phantom{x}$,tension=0.4}{v2,v1}
\fmf{photon,label=$\phantom{x}$}{v2,o1}
\fmfdotn{v}{2}
\fmfv{decor.shape=cross}{v3}
\fmfv{label=$\omega$,label.angle=90,label.dist=0.3cm}{v3}
\end{fmfgraph*} 
\end{gathered} \ \ \ \ \ \ \ \  \Bigg \} \nonumber
\end{equation}
\begin{equation}
+Z_4 \Bigg  \{ \ \ \ \ \ \ \ \begin{gathered}
\begin{fmfgraph*}(60,40)
\fmfleft{i1}
\fmflabel{$V^j$}{i1}
\fmfright{o1}
\fmflabel{$V^k$}{o1}
\fmf{photon,label=$\phantom{x}$}{i1,v1}
\fmf{ghost,label=$\phantom{x}$}{v1,v3}
\fmf{ghost,label=$\phantom{x}$}{v3,v2}
\fmf{ghost,left,label=$\phantom{x}$,tension=0.4}{v2,v1}
\fmf{photon,label=$\phantom{x}$}{v2,o1}
\fmfdotn{v}{2}
\fmfv{decor.shape=cross}{v3}
\fmfv{label=$C^l$,label.angle=90,label.dist=0.3cm}{v3}
\end{fmfgraph*} 
\end{gathered} \ \ \ \ \ \ \ \Bigg \} +Z_5  \{ 0  \} + Z_6  \{ 0 \}. \label{eq7} 
\end{equation}
From this it follows that $Z_2$ does not have the tree part.

\begin{equation}
\begin{gathered}
\begin{fmfgraph*}(40,40)
\fmfleft{i1}
\fmflabel{$V^j$}{i1}
\fmfright{o1}
\fmflabel{$V^k$}{o1}
\fmf{photon,label=$\phantom{x}$}{i1,v}
\fmf{photon,label=$\phantom{x}$}{v,o1}
\fmfv{decor.shape=cross}{v}
\end{fmfgraph*}
\end{gathered} \ \ \ \ \ \ \ \  \equiv \ \ \ \begin{gathered}
\langle V^j V^k \left[\frac{1}{2} V^l V^l \right]_R \rangle = 
\end{gathered} \nonumber 
\end{equation}
\begin{equation}
Z_7 \Bigg \{
\begin{gathered}
\begin{fmfgraph*}(40,40)
\fmfleft{i1,i2}
\fmflabel{$V^j$}{i1}
\fmfright{o1,o2}
\fmflabel{$V^k$}{o1}
\fmf{phantom,tension=1.5}{i2,v2}
\fmf{phantom,tension=1.5}{o1,v2}
\fmf{photon,label=$\phantom{x}$,tension=1.5,label.side=right}{i1,v1}
\fmf{gluon,label=$\phantom{x}$,left,tension=0.05}{v1,v2}
\fmf{gluon,label=$\phantom{x}$,left,tension=0.05}{v2,v1}
\fmf{photon,label=$\phantom{x}$,tension=1.5}{v1,o1}
\fmfdotn{v}{1}
\fmfv{decor.shape=cross}{v2}
\end{fmfgraph*} 
\end{gathered} + \ \ \ \ \ \ 
\begin{gathered}
  \begin{fmfgraph*}(100,70)
\fmfleft{i1}
\fmflabel{$V^j$}{i1}
\fmfright{o1}
\fmflabel{$V^k$}{o1}
\fmf{photon,label=$\phantom{x}$}{i1,v1}
\fmf{gluon,label=$\phantom{x}$,label.side=left}{v1,v3}
\fmf{gluon,label=$\phantom{x}$,label.side=left}{v3,v2}
\fmf{gluon,left,label=$\phantom{x}$,tension=0.4}{v2,v1}
\fmf{photon,label=$\phantom{x}$}{v2,o1}
\fmfdotn{v}{2}
\fmfv{decor.shape=cross}{v3}
  \end{fmfgraph*}
\end{gathered}\ \ \ \ \ \ \ \Bigg \} \nonumber 
\end{equation}
\begin{equation}
+Z_8 \Bigg \{\ \ \ \ \ \ \ \  \begin{gathered}
\begin{fmfgraph*}(30,30)
\fmfleft{i1}
\fmflabel{$V^j$}{i1}
\fmfright{o1}
\fmflabel{$V^k$}{o1}
\fmf{photon,label=$\phantom{x}$}{i1,v}
\fmf{photon,label=$\phantom{x}$}{v,o1}
\fmfv{decor.shape=cross}{v}
\end{fmfgraph*}
\end{gathered}\ \ \ \ \ \ \ \ \  + \begin{gathered}
\begin{fmfgraph*}(40,40)
\fmfleft{i1,i2}
\fmflabel{$V^j$}{i1}
\fmfright{o1,o2}
\fmflabel{$V^k$}{o1}
\fmf{phantom,tension=1.5}{i2,v2}
\fmf{phantom,tension=1.5}{o1,v2}
\fmf{photon,label=$\phantom{x}$,tension=1.5,label.side=right}{i1,v1}
\fmf{photon,label=$\phantom{x}$,left,tension=0.05}{v1,v2}
\fmf{photon,label=$\phantom{x}$,left,tension=0.05}{v2,v1}
\fmf{photon,label=$\phantom{x}$,tension=1.5}{v1,o1}
\fmfdotn{v}{1}
\fmfv{decor.shape=cross}{v2}
\fmfv{label=$V^l$,label.angle=90,label.dist=0.3cm}{v2}
\end{fmfgraph*} 
\end{gathered}+\ \ \ \ \ \ \ \ 
\begin{gathered}
\begin{fmfgraph*}(60,40)
\fmfleft{i1}
\fmflabel{$V^j$}{i1}
\fmfright{o1}
\fmflabel{$V^k$}{o1}
\fmf{photon,label=$\phantom{x}$}{i1,v1}
\fmf{photon,right,label=$\phantom{x}$,tension=0.4}{v1,v2}
\fmf{photon,label.side=left,label=$\phantom{x}$}{v1,v3}
\fmf{photon,label.side=left,label=$\phantom{x}$}{v3,v2}
\fmf{photon,label=$\phantom{x}$}{v2,o1}
\fmfdotn{v}{2}
\fmfv{decor.shape=cross}{v3}
\fmfv{label=$V^l$,label.angle=90,label.dist=0.3cm}{v3}
\end{fmfgraph*} 
\end{gathered}\ \ \ \ \ \ \ \ \Bigg \}  \nonumber
\end{equation}
\begin{equation}
+Z_9 \ \ \ \ \ \ \Bigg \{\ \ \ \ \ \ \begin{gathered}
\begin{fmfgraph*}(30,40)
\fmfleft{i1,i2}
\fmflabel{$V^j$}{i1}
\fmfright{o1,o2}
\fmflabel{$V^k$}{o1}
\fmf{phantom,tension=1.5}{i2,v2}
\fmf{phantom,tension=1.5}{o1,v2}
\fmf{photon,label=$\phantom{x}$,tension=1.5,label.side=right}{i1,v1}
\fmf{ghost,label=$\phantom{x}$,left,tension=0.05}{v1,v2}
\fmf{ghost,label=$\phantom{x}$,left,tension=0.05}{v2,v1}
\fmf{photon,label=$\phantom{x}$,tension=1.5}{v1,o1}
\fmfdotn{v}{1}
\fmfv{decor.shape=cross}{v2}
\fmfv{label=$\omega$,label.angle=90,label.dist=0.3cm}{v2}
\end{fmfgraph*} 
\end{gathered}\ \ \ \ \ \ +\ \ \ \ \ \ \ \ \begin{gathered}
\begin{fmfgraph*}(60,40)
\fmfleft{i1}
\fmflabel{$V^j$}{i1}
\fmfright{o1}
\fmflabel{$V^k$}{o1}
\fmf{photon,label=$\phantom{x}$}{i1,v1}
\fmf{ghost,label=$\phantom{x}$}{v1,v3}
\fmf{ghost,label=$\phantom{x}$}{v3,v2}
\fmf{ghost,left,label=$\phantom{x}$,tension=0.4}{v2,v1}
\fmf{photon,label=$\phantom{x}$}{v2,o1}
\fmfdotn{v}{2}
\fmfv{decor.shape=cross}{v3}
\fmfv{label=$\omega$,label.angle=90,label.dist=0.3cm}{v3}
\end{fmfgraph*} 
\end{gathered} \ \ \ \ \ \ \ \  \Bigg \} \nonumber
\end{equation}
\begin{equation}
+Z_{10} \Bigg  \{ \ \ \ \ \ \ \ \begin{gathered}
\begin{fmfgraph*}(60,40)
\fmfleft{i1}
\fmflabel{$V^j$}{i1}
\fmfright{o1}
\fmflabel{$V^k$}{o1}
\fmf{photon,label=$\phantom{x}$}{i1,v1}
\fmf{ghost,label=$\phantom{x}$}{v1,v3}
\fmf{ghost,label=$\phantom{x}$}{v3,v2}
\fmf{ghost,left,label=$\phantom{x}$,tension=0.4}{v2,v1}
\fmf{photon,label=$\phantom{x}$}{v2,o1}
\fmfdotn{v}{2}
\fmfv{decor.shape=cross}{v3}
\fmfv{label=$C^l$,label.angle=90,label.dist=0.3cm}{v3}
\end{fmfgraph*} 
\end{gathered} \ \ \ \ \ \ \ \Bigg \} +Z_{11} \{ 0  \} + Z_{12}  \{ 0 \}  . \label{eq8} 
\end{equation}
From this it follows that $Z_8$ does have the tree part.

\begin{equation}
0 \equiv \begin{gathered}
\langle V^j V^k \left[i\omega \bar{\omega} \right]_R \rangle = 
\end{gathered} \nonumber 
\end{equation}
\begin{equation}
Z_{13} \Bigg \{
\begin{gathered}
\begin{fmfgraph*}(40,40)
\fmfleft{i1,i2}
\fmflabel{$V^j$}{i1}
\fmfright{o1,o2}
\fmflabel{$V^k$}{o1}
\fmf{phantom,tension=1.5}{i2,v2}
\fmf{phantom,tension=1.5}{o1,v2}
\fmf{photon,label=$\phantom{x}$,tension=1.5,label.side=right}{i1,v1}
\fmf{gluon,label=$\phantom{x}$,left,tension=0.05}{v1,v2}
\fmf{gluon,label=$\phantom{x}$,left,tension=0.05}{v2,v1}
\fmf{photon,label=$\phantom{x}$,tension=1.5}{v1,o1}
\fmfdotn{v}{1}
\fmfv{decor.shape=cross}{v2}
\end{fmfgraph*} 
\end{gathered} + \ \ \ \ \ \ 
\begin{gathered}
  \begin{fmfgraph*}(100,70)
\fmfleft{i1}
\fmflabel{$V^j$}{i1}
\fmfright{o1}
\fmflabel{$V^k$}{o1}
\fmf{photon,label=$\phantom{x}$}{i1,v1}
\fmf{gluon,label=$\phantom{x}$,label.side=left}{v1,v3}
\fmf{gluon,label=$\phantom{x}$,label.side=left}{v3,v2}
\fmf{gluon,left,label=$\phantom{x}$,tension=0.4}{v2,v1}
\fmf{photon,label=$\phantom{x}$}{v2,o1}
\fmfdotn{v}{2}
\fmfv{decor.shape=cross}{v3}
  \end{fmfgraph*}
\end{gathered}\ \ \ \ \ \ \ \Bigg \} \nonumber 
\end{equation}
\begin{equation}
+Z_{14} \Bigg \{\ \ \ \ \ \ \ \  \begin{gathered}
\begin{fmfgraph*}(30,30)
\fmfleft{i1}
\fmflabel{$V^j$}{i1}
\fmfright{o1}
\fmflabel{$V^k$}{o1}
\fmf{photon,label=$\phantom{x}$}{i1,v}
\fmf{photon,label=$\phantom{x}$}{v,o1}
\fmfv{decor.shape=cross}{v}
\end{fmfgraph*}
\end{gathered}\ \ \ \ \ \ \ \ \  +  \begin{gathered}
\begin{fmfgraph*}(40,40)
\fmfleft{i1,i2}
\fmflabel{$V^j$}{i1}
\fmfright{o1,o2}
\fmflabel{$V^k$}{o1}
\fmf{phantom,tension=1.5}{i2,v2}
\fmf{phantom,tension=1.5}{o1,v2}
\fmf{photon,label=$\phantom{x}$,tension=1.5,label.side=right}{i1,v1}
\fmf{photon,label=$\phantom{x}$,left,tension=0.05}{v1,v2}
\fmf{photon,label=$\phantom{x}$,left,tension=0.05}{v2,v1}
\fmf{photon,label=$\phantom{x}$,tension=1.5}{v1,o1}
\fmfdotn{v}{1}
\fmfv{decor.shape=cross}{v2}
\fmfv{label=$V^l$,label.angle=90,label.dist=0.3cm}{v2}
\end{fmfgraph*} 
\end{gathered}+\ \ \ \ \ \ \ \ 
\begin{gathered}
\begin{fmfgraph*}(60,40)
\fmfleft{i1}
\fmflabel{$V^j$}{i1}
\fmfright{o1}
\fmflabel{$V^k$}{o1}
\fmf{photon,label=$\phantom{x}$}{i1,v1}
\fmf{photon,right,label=$\phantom{x}$,tension=0.4}{v1,v2}
\fmf{photon,label.side=left,label=$\phantom{x}$}{v1,v3}
\fmf{photon,label.side=left,label=$\phantom{x}$}{v3,v2}
\fmf{photon,label=$\phantom{x}$}{v2,o1}
\fmfdotn{v}{2}
\fmfv{decor.shape=cross}{v3}
\fmfv{label=$V^l$,label.angle=90,label.dist=0.3cm}{v3}
\end{fmfgraph*} 
\end{gathered}\ \ \ \ \ \ \ \ \Bigg \}  \nonumber
\end{equation}
\begin{equation}
+Z_{15} \ \ \ \ \ \ \Bigg \{\ \ \ \ \ \ \begin{gathered}
\begin{fmfgraph*}(30,40)
\fmfleft{i1,i2}
\fmflabel{$V^j$}{i1}
\fmfright{o1,o2}
\fmflabel{$V^k$}{o1}
\fmf{phantom,tension=1.5}{i2,v2}
\fmf{phantom,tension=1.5}{o1,v2}
\fmf{photon,label=$\phantom{x}$,tension=1.5,label.side=right}{i1,v1}
\fmf{ghost,label=$\phantom{x}$,left,tension=0.05}{v1,v2}
\fmf{ghost,label=$\phantom{x}$,left,tension=0.05}{v2,v1}
\fmf{photon,label=$\phantom{x}$,tension=1.5}{v1,o1}
\fmfdotn{v}{1}
\fmfv{decor.shape=cross}{v2}
\fmfv{label=$\omega$,label.angle=90,label.dist=0.3cm}{v2}
\end{fmfgraph*} 
\end{gathered}\ \ \ \ \ \ +\ \ \ \ \ \ \ \ \begin{gathered}
\begin{fmfgraph*}(60,40)
\fmfleft{i1}
\fmflabel{$V^j$}{i1}
\fmfright{o1}
\fmflabel{$V^k$}{o1}
\fmf{photon,label=$\phantom{x}$}{i1,v1}
\fmf{ghost,label=$\phantom{x}$}{v1,v3}
\fmf{ghost,label=$\phantom{x}$}{v3,v2}
\fmf{ghost,left,label=$\phantom{x}$,tension=0.4}{v2,v1}
\fmf{photon,label=$\phantom{x}$}{v2,o1}
\fmfdotn{v}{2}
\fmfv{decor.shape=cross}{v3}
\fmfv{label=$\omega$,label.angle=90,label.dist=0.3cm}{v3}
\end{fmfgraph*} 
\end{gathered} \ \ \ \ \ \ \ \  \Bigg \} \nonumber
\end{equation}
\begin{equation}
+Z_{16} \Bigg  \{ \ \ \ \ \ \ \ \begin{gathered}
\begin{fmfgraph*}(60,40)
\fmfleft{i1}
\fmflabel{$V^j$}{i1}
\fmfright{o1}
\fmflabel{$V^k$}{o1}
\fmf{photon,label=$\phantom{x}$}{i1,v1}
\fmf{ghost,label=$\phantom{x}$}{v1,v3}
\fmf{ghost,label=$\phantom{x}$}{v3,v2}
\fmf{ghost,left,label=$\phantom{x}$,tension=0.4}{v2,v1}
\fmf{photon,label=$\phantom{x}$}{v2,o1}
\fmfdotn{v}{2}
\fmfv{decor.shape=cross}{v3}
\fmfv{label=$C^l$,label.angle=90,label.dist=0.3cm}{v3}
\end{fmfgraph*} 
\end{gathered} \ \ \ \ \ \ \ \Bigg \} +Z_{17}  \{ 0  \} + Z_{18}  \{ 0 \}.   \label{eq9} 
\end{equation}
From this it follows that $Z_{14}$ does not have the tree part.

\begin{equation}
0 \equiv \begin{gathered}
\langle V^j V^k \left[iC^l \bar{C}^l \right]_R \rangle = 
\end{gathered} \nonumber 
\end{equation}
\begin{equation}
Z_{19} \Bigg \{
\begin{gathered}
\begin{fmfgraph*}(40,40)
\fmfleft{i1,i2}
\fmflabel{$V^j$}{i1}
\fmfright{o1,o2}
\fmflabel{$V^k$}{o1}
\fmf{phantom,tension=1.5}{i2,v2}
\fmf{phantom,tension=1.5}{o1,v2}
\fmf{photon,label=$\phantom{x}$,tension=1.5,label.side=right}{i1,v1}
\fmf{gluon,label=$\phantom{x}$,left,tension=0.05}{v1,v2}
\fmf{gluon,label=$\phantom{x}$,left,tension=0.05}{v2,v1}
\fmf{photon,label=$\phantom{x}$,tension=1.5}{v1,o1}
\fmfdotn{v}{1}
\fmfv{decor.shape=cross}{v2}
\end{fmfgraph*} 
\end{gathered} + \ \ \ \ \ \ 
\begin{gathered}
  \begin{fmfgraph*}(100,70)
\fmfleft{i1}
\fmflabel{$V^j$}{i1}
\fmfright{o1}
\fmflabel{$V^k$}{o1}
\fmf{photon,label=$\phantom{x}$}{i1,v1}
\fmf{gluon,label=$\phantom{x}$,label.side=left}{v1,v3}
\fmf{gluon,label=$\phantom{x}$,label.side=left}{v3,v2}
\fmf{gluon,left,label=$\phantom{x}$,tension=0.4}{v2,v1}
\fmf{photon,label=$\phantom{x}$}{v2,o1}
\fmfdotn{v}{2}
\fmfv{decor.shape=cross}{v3}
  \end{fmfgraph*}
\end{gathered}\ \ \ \ \ \ \ \Bigg \} \nonumber 
\end{equation}
\begin{equation}
+Z_{20} \Bigg \{\ \ \ \ \ \ \ \  \begin{gathered}
\begin{fmfgraph*}(30,30)
\fmfleft{i1}
\fmflabel{$V^j$}{i1}
\fmfright{o1}
\fmflabel{$V^k$}{o1}
\fmf{photon,label=$\phantom{x}$}{i1,v}
\fmf{photon,label=$\phantom{x}$}{v,o1}
\fmfv{decor.shape=cross}{v}
\end{fmfgraph*}
\end{gathered}\ \ \ \ \ \ \ \ \   + \begin{gathered}
\begin{fmfgraph*}(40,40)
\fmfleft{i1,i2}
\fmflabel{$V^j$}{i1}
\fmfright{o1,o2}
\fmflabel{$V^k$}{o1}
\fmf{phantom,tension=1.5}{i2,v2}
\fmf{phantom,tension=1.5}{o1,v2}
\fmf{photon,label=$\phantom{x}$,tension=1.5,label.side=right}{i1,v1}
\fmf{photon,label=$\phantom{x}$,left,tension=0.05}{v1,v2}
\fmf{photon,label=$\phantom{x}$,left,tension=0.05}{v2,v1}
\fmf{photon,label=$\phantom{x}$,tension=1.5}{v1,o1}
\fmfdotn{v}{1}
\fmfv{decor.shape=cross}{v2}
\fmfv{label=$V^l$,label.angle=90,label.dist=0.3cm}{v2}
\end{fmfgraph*} 
\end{gathered}+\ \ \ \ \ \ \ \ 
\begin{gathered}
\begin{fmfgraph*}(60,40)
\fmfleft{i1}
\fmflabel{$V^j$}{i1}
\fmfright{o1}
\fmflabel{$V^k$}{o1}
\fmf{photon,label=$\phantom{x}$}{i1,v1}
\fmf{photon,right,label=$\phantom{x}$,tension=0.4}{v1,v2}
\fmf{photon,label.side=left,label=$\phantom{x}$}{v1,v3}
\fmf{photon,label.side=left,label=$\phantom{x}$}{v3,v2}
\fmf{photon,label=$\phantom{x}$}{v2,o1}
\fmfdotn{v}{2}
\fmfv{decor.shape=cross}{v3}
\fmfv{label=$V^l$,label.angle=90,label.dist=0.3cm}{v3}
\end{fmfgraph*} 
\end{gathered}\ \ \ \ \ \ \ \ \Bigg \}  \nonumber
\end{equation}
\begin{equation}
+Z_{21} \ \ \ \ \ \ \Bigg \{\ \ \ \ \ \ \begin{gathered}
\begin{fmfgraph*}(30,40)
\fmfleft{i1,i2}
\fmflabel{$V^j$}{i1}
\fmfright{o1,o2}
\fmflabel{$V^k$}{o1}
\fmf{phantom,tension=1.5}{i2,v2}
\fmf{phantom,tension=1.5}{o1,v2}
\fmf{photon,label=$\phantom{x}$,tension=1.5,label.side=right}{i1,v1}
\fmf{ghost,label=$\phantom{x}$,left,tension=0.05}{v1,v2}
\fmf{ghost,label=$\phantom{x}$,left,tension=0.05}{v2,v1}
\fmf{photon,label=$\phantom{x}$,tension=1.5}{v1,o1}
\fmfdotn{v}{1}
\fmfv{decor.shape=cross}{v2}
\fmfv{label=$\omega$,label.angle=90,label.dist=0.3cm}{v2}
\end{fmfgraph*} 
\end{gathered}\ \ \ \ \ \ +\ \ \ \ \ \ \ \ \begin{gathered}
\begin{fmfgraph*}(60,40)
\fmfleft{i1}
\fmflabel{$V^j$}{i1}
\fmfright{o1}
\fmflabel{$V^k$}{o1}
\fmf{photon,label=$\phantom{x}$}{i1,v1}
\fmf{ghost,label=$\phantom{x}$}{v1,v3}
\fmf{ghost,label=$\phantom{x}$}{v3,v2}
\fmf{ghost,left,label=$\phantom{x}$,tension=0.4}{v2,v1}
\fmf{photon,label=$\phantom{x}$}{v2,o1}
\fmfdotn{v}{2}
\fmfv{decor.shape=cross}{v3}
\fmfv{label=$\omega$,label.angle=90,label.dist=0.3cm}{v3}
\end{fmfgraph*} 
\end{gathered} \ \ \ \ \ \ \ \  \Bigg \} \nonumber
\end{equation}
\begin{equation}
+Z_{22} \Bigg  \{ \ \ \ \ \ \ \ \begin{gathered}
\begin{fmfgraph*}(60,40)
\fmfleft{i1}
\fmflabel{$V^j$}{i1}
\fmfright{o1}
\fmflabel{$V^k$}{o1}
\fmf{photon,label=$\phantom{x}$}{i1,v1}
\fmf{ghost,label=$\phantom{x}$}{v1,v3}
\fmf{ghost,label=$\phantom{x}$}{v3,v2}
\fmf{ghost,left,label=$\phantom{x}$,tension=0.4}{v2,v1}
\fmf{photon,label=$\phantom{x}$}{v2,o1}
\fmfdotn{v}{2}
\fmfv{decor.shape=cross}{v3}
\fmfv{label=$C^l$,label.angle=90,label.dist=0.3cm}{v3}
\end{fmfgraph*} 
\end{gathered} \ \ \ \ \ \ \ \Bigg \} +Z_{23}  \{ 0  \} + Z_{24}  \{ 0 \}.   \label{eq10} 
\end{equation}
From this it follows that $Z_{20}$ does not have the tree part.

\begin{equation}
0 \equiv \begin{gathered}
\langle V^j V^k \left[\frac{1}{2}V^\gamma V^\gamma \right]_R \rangle = 
\end{gathered} \nonumber 
\end{equation}
\begin{equation}
Z_{25} \Bigg \{
\begin{gathered}
\begin{fmfgraph*}(40,40)
\fmfleft{i1,i2}
\fmflabel{$V^j$}{i1}
\fmfright{o1,o2}
\fmflabel{$V^k$}{o1}
\fmf{phantom,tension=1.5}{i2,v2}
\fmf{phantom,tension=1.5}{o1,v2}
\fmf{photon,label=$\phantom{x}$,tension=1.5,label.side=right}{i1,v1}
\fmf{gluon,label=$\phantom{x}$,left,tension=0.05}{v1,v2}
\fmf{gluon,label=$\phantom{x}$,left,tension=0.05}{v2,v1}
\fmf{photon,label=$\phantom{x}$,tension=1.5}{v1,o1}
\fmfdotn{v}{1}
\fmfv{decor.shape=cross}{v2}
\end{fmfgraph*} 
\end{gathered} + \ \ \ \ \ \ 
\begin{gathered}
  \begin{fmfgraph*}(100,70)
\fmfleft{i1}
\fmflabel{$V^j$}{i1}
\fmfright{o1}
\fmflabel{$V^k$}{o1}
\fmf{photon,label=$\phantom{x}$}{i1,v1}
\fmf{gluon,label=$\phantom{x}$,label.side=left}{v1,v3}
\fmf{gluon,label=$\phantom{x}$,label.side=left}{v3,v2}
\fmf{gluon,left,label=$\phantom{x}$,tension=0.4}{v2,v1}
\fmf{photon,label=$\phantom{x}$}{v2,o1}
\fmfdotn{v}{2}
\fmfv{decor.shape=cross}{v3}
  \end{fmfgraph*}
\end{gathered}\ \ \ \ \ \ \ \Bigg \} \nonumber 
\end{equation}
\begin{equation}
+Z_{26} \Bigg \{\ \ \ \ \ \ \ \  \begin{gathered}
\begin{fmfgraph*}(30,30)
\fmfleft{i1}
\fmflabel{$V^j$}{i1}
\fmfright{o1}
\fmflabel{$V^k$}{o1}
\fmf{photon,label=$\phantom{x}$}{i1,v}
\fmf{photon,label=$\phantom{x}$}{v,o1}
\fmfv{decor.shape=cross}{v}
\end{fmfgraph*}
\end{gathered}\ \ \ \ \ \ \ \ \  +  \begin{gathered}
\begin{fmfgraph*}(40,40)
\fmfleft{i1,i2}
\fmflabel{$V^j$}{i1}
\fmfright{o1,o2}
\fmflabel{$V^k$}{o1}
\fmf{phantom,tension=1.5}{i2,v2}
\fmf{phantom,tension=1.5}{o1,v2}
\fmf{photon,label=$\phantom{x}$,tension=1.5,label.side=right}{i1,v1}
\fmf{photon,label=$\phantom{x}$,left,tension=0.05}{v1,v2}
\fmf{photon,label=$\phantom{x}$,left,tension=0.05}{v2,v1}
\fmf{photon,label=$\phantom{x}$,tension=1.5}{v1,o1}
\fmfdotn{v}{1}
\fmfv{decor.shape=cross}{v2}
\fmfv{label=$V^l$,label.angle=90,label.dist=0.3cm}{v2}
\end{fmfgraph*} 
\end{gathered}+\ \ \ \ \ \ \ \ 
\begin{gathered}
\begin{fmfgraph*}(60,40)
\fmfleft{i1}
\fmflabel{$V^j$}{i1}
\fmfright{o1}
\fmflabel{$V^k$}{o1}
\fmf{photon,label=$\phantom{x}$}{i1,v1}
\fmf{photon,right,label=$\phantom{x}$,tension=0.4}{v1,v2}
\fmf{photon,label.side=left,label=$\phantom{x}$}{v1,v3}
\fmf{photon,label.side=left,label=$\phantom{x}$}{v3,v2}
\fmf{photon,label=$\phantom{x}$}{v2,o1}
\fmfdotn{v}{2}
\fmfv{decor.shape=cross}{v3}
\fmfv{label=$V^l$,label.angle=90,label.dist=0.3cm}{v3}
\end{fmfgraph*} 
\end{gathered}\ \ \ \ \ \ \ \ \Bigg \}  \nonumber
\end{equation}
\begin{equation}
+Z_{27} \ \ \ \ \ \ \Bigg \{\ \ \ \ \ \ \begin{gathered}
\begin{fmfgraph*}(30,40)
\fmfleft{i1,i2}
\fmflabel{$V^j$}{i1}
\fmfright{o1,o2}
\fmflabel{$V^k$}{o1}
\fmf{phantom,tension=1.5}{i2,v2}
\fmf{phantom,tension=1.5}{o1,v2}
\fmf{photon,label=$\phantom{x}$,tension=1.5,label.side=right}{i1,v1}
\fmf{ghost,label=$\phantom{x}$,left,tension=0.05}{v1,v2}
\fmf{ghost,label=$\phantom{x}$,left,tension=0.05}{v2,v1}
\fmf{photon,label=$\phantom{x}$,tension=1.5}{v1,o1}
\fmfdotn{v}{1}
\fmfv{decor.shape=cross}{v2}
\fmfv{label=$\omega$,label.angle=90,label.dist=0.3cm}{v2}
\end{fmfgraph*} 
\end{gathered}\ \ \ \ \ \ +\ \ \ \ \ \ \ \ \begin{gathered}
\begin{fmfgraph*}(60,40)
\fmfleft{i1}
\fmflabel{$V^j$}{i1}
\fmfright{o1}
\fmflabel{$V^k$}{o1}
\fmf{photon,label=$\phantom{x}$}{i1,v1}
\fmf{ghost,label=$\phantom{x}$}{v1,v3}
\fmf{ghost,label=$\phantom{x}$}{v3,v2}
\fmf{ghost,left,label=$\phantom{x}$,tension=0.4}{v2,v1}
\fmf{photon,label=$\phantom{x}$}{v2,o1}
\fmfdotn{v}{2}
\fmfv{decor.shape=cross}{v3}
\fmfv{label=$\omega$,label.angle=90,label.dist=0.3cm}{v3}
\end{fmfgraph*} 
\end{gathered} \ \ \ \ \ \ \ \  \Bigg \} \nonumber
\end{equation}
\begin{equation}
+Z_{28} \Bigg  \{ \ \ \ \ \ \ \ \begin{gathered}
\begin{fmfgraph*}(60,40)
\fmfleft{i1}
\fmflabel{$V^j$}{i1}
\fmfright{o1}
\fmflabel{$V^k$}{o1}
\fmf{photon,label=$\phantom{x}$}{i1,v1}
\fmf{ghost,label=$\phantom{x}$}{v1,v3}
\fmf{ghost,label=$\phantom{x}$}{v3,v2}
\fmf{ghost,left,label=$\phantom{x}$,tension=0.4}{v2,v1}
\fmf{photon,label=$\phantom{x}$}{v2,o1}
\fmfdotn{v}{2}
\fmfv{decor.shape=cross}{v3}
\fmfv{label=$C^l$,label.angle=90,label.dist=0.3cm}{v3}
\end{fmfgraph*} 
\end{gathered} \ \ \ \ \ \ \ \Bigg \} +Z_{29}  \{ 0  \} + Z_{30}  \{ 0 \}  . \label{eq11} 
\end{equation}
From this it follows that $Z_{26}$ does not have the tree part.

\begin{equation}
0 \equiv \begin{gathered}
\langle V^j V^k \left[iC^\gamma \bar{C}^\gamma \right]_R \rangle = 
\end{gathered} \nonumber 
\end{equation}
\begin{equation}
Z_{31} \Bigg \{
\begin{gathered}
\begin{fmfgraph*}(40,40)
\fmfleft{i1,i2}
\fmflabel{$V^j$}{i1}
\fmfright{o1,o2}
\fmflabel{$V^k$}{o1}
\fmf{phantom,tension=1.5}{i2,v2}
\fmf{phantom,tension=1.5}{o1,v2}
\fmf{photon,label=$\phantom{x}$,tension=1.5,label.side=right}{i1,v1}
\fmf{gluon,label=$\phantom{x}$,left,tension=0.05}{v1,v2}
\fmf{gluon,label=$\phantom{x}$,left,tension=0.05}{v2,v1}
\fmf{photon,label=$\phantom{x}$,tension=1.5}{v1,o1}
\fmfdotn{v}{1}
\fmfv{decor.shape=cross}{v2}
\end{fmfgraph*} 
\end{gathered} + \ \ \ \ \ \ 
\begin{gathered}
  \begin{fmfgraph*}(100,70)
\fmfleft{i1}
\fmflabel{$V^j$}{i1}
\fmfright{o1}
\fmflabel{$V^k$}{o1}
\fmf{photon,label=$\phantom{x}$}{i1,v1}
\fmf{gluon,label=$\phantom{x}$,label.side=left}{v1,v3}
\fmf{gluon,label=$\phantom{x}$,label.side=left}{v3,v2}
\fmf{gluon,left,label=$\phantom{x}$,tension=0.4}{v2,v1}
\fmf{photon,label=$\phantom{x}$}{v2,o1}
\fmfdotn{v}{2}
\fmfv{decor.shape=cross}{v3}
  \end{fmfgraph*}
\end{gathered}\ \ \ \ \ \ \ \Bigg \} \nonumber 
\end{equation}
\begin{equation}
+Z_{32} \Bigg \{\ \ \ \ \ \ \ \   \begin{gathered}
\begin{fmfgraph*}(30,30)
\fmfleft{i1}
\fmflabel{$V^j$}{i1}
\fmfright{o1}
\fmflabel{$V^k$}{o1}
\fmf{photon,label=$\phantom{x}$}{i1,v}
\fmf{photon,label=$\phantom{x}$}{v,o1}
\fmfv{decor.shape=cross}{v}
\end{fmfgraph*}
\end{gathered}\ \ \ \ \ \ \ \ \  + \begin{gathered}
\begin{fmfgraph*}(40,40)
\fmfleft{i1,i2}
\fmflabel{$V^j$}{i1}
\fmfright{o1,o2}
\fmflabel{$V^k$}{o1}
\fmf{phantom,tension=1.5}{i2,v2}
\fmf{phantom,tension=1.5}{o1,v2}
\fmf{photon,label=$\phantom{x}$,tension=1.5,label.side=right}{i1,v1}
\fmf{photon,label=$\phantom{x}$,left,tension=0.05}{v1,v2}
\fmf{photon,label=$\phantom{x}$,left,tension=0.05}{v2,v1}
\fmf{photon,label=$\phantom{x}$,tension=1.5}{v1,o1}
\fmfdotn{v}{1}
\fmfv{decor.shape=cross}{v2}
\fmfv{label=$V^l$,label.angle=90,label.dist=0.3cm}{v2}
\end{fmfgraph*} 
\end{gathered}+\ \ \ \ \ \ \ \ 
\begin{gathered}
\begin{fmfgraph*}(60,40)
\fmfleft{i1}
\fmflabel{$V^j$}{i1}
\fmfright{o1}
\fmflabel{$V^k$}{o1}
\fmf{photon,label=$\phantom{x}$}{i1,v1}
\fmf{photon,right,label=$\phantom{x}$,tension=0.4}{v1,v2}
\fmf{photon,label.side=left,label=$\phantom{x}$}{v1,v3}
\fmf{photon,label.side=left,label=$\phantom{x}$}{v3,v2}
\fmf{photon,label=$\phantom{x}$}{v2,o1}
\fmfdotn{v}{2}
\fmfv{decor.shape=cross}{v3}
\fmfv{label=$V^l$,label.angle=90,label.dist=0.3cm}{v3}
\end{fmfgraph*} 
\end{gathered}\ \ \ \ \ \ \ \ \Bigg \}  \nonumber
\end{equation}
\begin{equation}
+Z_{33} \ \ \ \ \ \ \Bigg \{\ \ \ \ \ \ \begin{gathered}
\begin{fmfgraph*}(40,40)
\fmfleft{i1,i2}
\fmflabel{$V^j$}{i1}
\fmfright{o1,o2}
\fmflabel{$V^k$}{o1}
\fmf{phantom,tension=1.5}{i2,v2}
\fmf{phantom,tension=1.5}{o1,v2}
\fmf{photon,label=$\phantom{x}$,tension=1.5,label.side=right}{i1,v1}
\fmf{ghost,label=$\phantom{x}$,left,tension=0.05}{v1,v2}
\fmf{ghost,label=$\phantom{x}$,left,tension=0.05}{v2,v1}
\fmf{photon,label=$\phantom{x}$,tension=1.5}{v1,o1}
\fmfdotn{v}{1}
\fmfv{decor.shape=cross}{v2}
\fmfv{label=$\omega$,label.angle=90,label.dist=0.3cm}{v2}
\end{fmfgraph*} 
\end{gathered}\ \ \ \ \ \ +\ \ \ \ \ \ \ \ \begin{gathered}
\begin{fmfgraph*}(60,40)
\fmfleft{i1}
\fmflabel{$V^j$}{i1}
\fmfright{o1}
\fmflabel{$V^k$}{o1}
\fmf{photon,label=$\phantom{x}$}{i1,v1}
\fmf{ghost,label=$\phantom{x}$}{v1,v3}
\fmf{ghost,label=$\phantom{x}$}{v3,v2}
\fmf{ghost,left,label=$\phantom{x}$,tension=0.4}{v2,v1}
\fmf{photon,label=$\phantom{x}$}{v2,o1}
\fmfdotn{v}{2}
\fmfv{decor.shape=cross}{v3}
\fmfv{label=$\omega$,label.angle=90,label.dist=0.3cm}{v3}
\end{fmfgraph*} 
\end{gathered} \ \ \ \ \ \ \ \  \Bigg \} \nonumber
\end{equation}
\begin{equation}
+Z_{34} \Bigg  \{ \ \ \ \ \ \ \ \begin{gathered}
\begin{fmfgraph*}(60,40)
\fmfleft{i1}
\fmflabel{$V^j$}{i1}
\fmfright{o1}
\fmflabel{$V^k$}{o1}
\fmf{photon,label=$\phantom{x}$}{i1,v1}
\fmf{ghost,label=$\phantom{x}$}{v1,v3}
\fmf{ghost,label=$\phantom{x}$}{v3,v2}
\fmf{ghost,left,label=$\phantom{x}$,tension=0.4}{v2,v1}
\fmf{photon,label=$\phantom{x}$}{v2,o1}
\fmfdotn{v}{2}
\fmfv{decor.shape=cross}{v3}
\fmfv{label=$C^l$,label.angle=90,label.dist=0.3cm}{v3}
\end{fmfgraph*} 
\end{gathered} \ \ \ \ \ \ \ \Bigg \} +Z_{35}  \{ 0  \} + Z_{36}  \{ 0 \}  . \label{eq12} 
\end{equation}
From this it follows that $Z_{32}$ does not have the tree part.

\subsubsection{Insertion into $\langle \omega \bar{\omega} \rangle$}


\begin{equation}
0 \equiv \begin{gathered}
\langle \omega \bar{\omega} \left[\frac{1}{2} X X \right]_R \rangle = 
\end{gathered} \nonumber
\end{equation}
\begin{equation}
Z_1 \Bigg \{
\begin{gathered}
\begin{fmfgraph*}(40,40)
\fmfleft{i1,i2}
\fmflabel{$\omega$}{i1}
\fmfright{o1,o2}
\fmflabel{$\bar{\omega}$}{o1}
\fmf{phantom,tension=1.5}{i2,v2}
\fmf{phantom,tension=1.5}{o1,v2}
\fmf{ghost,label=$\phantom{x}$,tension=1.5,label.side=right}{i1,v1}
\fmf{gluon,label=$\phantom{x}$,left,tension=0.05}{v1,v2}
\fmf{gluon,label=$\phantom{x}$,left,tension=0.05}{v2,v1}
\fmf{ghost,label=$\phantom{x}$,tension=1.5}{v1,o1}
\fmfdotn{v}{1}
\fmfv{decor.shape=cross}{v2}
\end{fmfgraph*}
\end{gathered} \Bigg \} +Z_2 \Bigg \{\ \ \ \begin{gathered}
\begin{fmfgraph*}(40,40)
\fmfleft{i1,i2}
\fmflabel{$\omega$}{i1}
\fmfright{o1,o2}
\fmflabel{$\bar{\omega}$}{o1}
\fmf{phantom,tension=1.5}{i2,v2}
\fmf{phantom,tension=1.5}{o1,v2}
\fmf{ghost,label=$\phantom{x}$,tension=1.5,label.side=right}{i1,v1}
\fmf{photon,label=$\phantom{x}$,left,tension=0.05}{v1,v2}
\fmf{photon,label=$\phantom{x}$,left,tension=0.05}{v2,v1}
\fmf{ghost,label=$\phantom{x}$,tension=1.5}{v1,o1}
\fmfdotn{v}{1}
\fmfv{decor.shape=cross}{v2}
\fmfv{label=$V^j$,label.angle=90,label.dist=0.3cm}{v2}
\end{fmfgraph*} 
\end{gathered}+\ \ \ \ \ 
\begin{gathered}
\begin{fmfgraph*}(70,40)
\fmfleft{i1}
\fmflabel{$\omega$}{i1}
\fmfright{o1}
\fmflabel{$\bar{\omega}$}{o1}
\fmf{ghost,label=$\phantom{x}$}{i1,v1}
\fmf{photon,label.side=left,label=$\phantom{x}$}{v1,v3}
\fmf{photon,label.side=left,label=$\phantom{x}$}{v3,v2}
\fmf{ghost,right,tension=0.4,label=$\phantom{x}$}{v1,v2}
\fmf{ghost,label=$\phantom{x}$}{v2,o1}
\fmfdotn{v}{2}
\fmfv{decor.shape=cross}{v3}
\fmfv{label=$V^j$,label.angle=90,label.dist=0.3cm}{v3}
\end{fmfgraph*} 
\end{gathered}\ \ \ \ \  \Bigg \} \nonumber 
\end{equation}
\begin{equation}
+Z_3 \Bigg \{\ \ \ \ \ \ \begin{fmfgraph*}(30,30)
\fmfleft{i1}
\fmflabel{$\omega$}{i1}
\fmfright{o1}
\fmflabel{$\bar{\omega}$}{o1}
\fmf{ghost,label=$\phantom{x}$}{i1,v}
\fmf{ghost,label=$\phantom{x}$}{v,o1}
\fmfv{decor.shape=cross}{v}
\end{fmfgraph*}\ \ \ \ \  + \ \ \ \ \  \begin{gathered}
\begin{fmfgraph*}(50,40)
\fmfleft{i1}
\fmflabel{$\omega$}{i1}
\fmfright{o1}
\fmflabel{$\bar{\omega}$}{o1}
\fmf{ghost,label=$\phantom{x}$}{i1,v1}
\fmf{photon,left,tension=0.4,label=$V^j+V^\gamma$}{v1,v2}
\fmf{ghost,label=$\phantom{x}$}{v3,v2}
\fmf{ghost,label=$\phantom{x}$}{v1,v3}
\fmf{ghost,label=$\phantom{x}$}{v2,o1}
\fmfdotn{v}{2}
\fmfv{decor.shape=cross}{v3}
\end{fmfgraph*} 
\end{gathered}\ \ \ \ \ \ +\ \ \ \ \ \ \ \ \begin{gathered}
\begin{fmfgraph*}(40,40)
\fmfleft{i1,i2}
\fmflabel{$\omega$}{i1}
\fmfright{o1,o2}
\fmflabel{$\bar{\omega}$}{o1}
\fmf{phantom,tension=1.5}{i2,v2}
\fmf{phantom,tension=1.5}{o1,v2}
\fmf{ghost,label=$\phantom{x}$,tension=1.5,label.side=right}{i1,v1}
\fmf{ghost,label=$\phantom{x}$,left,tension=0.05}{v1,v2}
\fmf{ghost,label=$\phantom{x}$,left,tension=0.05}{v2,v1}
\fmf{ghost,label=$\phantom{x}$,tension=1.5}{v1,o1}
\fmfdotn{v}{1}
\fmfv{decor.shape=cross}{v2}
\fmfv{label=$\omega$,label.angle=90,label.dist=0.3cm}{v2}
\end{fmfgraph*} 
\end{gathered} \ \ \ \ \ \ \ \  \Bigg \} \nonumber
\end{equation}
\begin{equation}
+Z_4  \{ 0 \} + Z_5 \Bigg \{\ \ \ \begin{gathered}
\begin{fmfgraph*}(40,40)
\fmfleft{i1,i2}
\fmflabel{$\omega$}{i1}
\fmfright{o1,o2}
\fmflabel{$\bar{\omega}$}{o1}
\fmf{phantom,tension=1.5}{i2,v2}
\fmf{phantom,tension=1.5}{o1,v2}
\fmf{ghost,label=$\phantom{x}$,tension=1.5,label.side=right}{i1,v1}
\fmf{photon,label=$\phantom{x}$,left,tension=0.05}{v1,v2}
\fmf{photon,label=$\phantom{x}$,left,tension=0.05}{v2,v1}
\fmf{ghost,label=$\phantom{x}$,tension=1.5}{v1,o1}
\fmfdotn{v}{1}
\fmfv{decor.shape=cross}{v2}
\fmfv{label=$V^\gamma$,label.angle=90,label.dist=0.3cm}{v2}
\end{fmfgraph*} 
\end{gathered}+\ \ \ \ \ 
\begin{gathered}
\begin{fmfgraph*}(70,40)
\fmfleft{i1}
\fmflabel{$\omega$}{i1}
\fmfright{o1}
\fmflabel{$\bar{\omega}$}{o1}
\fmf{ghost,label=$\phantom{x}$}{i1,v1}
\fmf{photon,label.side=left,label=$\phantom{x}$}{v1,v3}
\fmf{photon,label.side=left,label=$\phantom{x}$}{v3,v2}
\fmf{ghost,right,tension=0.4,label=$\phantom{x}$}{v1,v2}
\fmf{ghost,label=$\phantom{x}$}{v2,o1}
\fmfdotn{v}{2}
\fmfv{decor.shape=cross}{v3}
\fmfv{label=$V^\gamma$,label.angle=90,label.dist=0.3cm}{v3}
\end{fmfgraph*} 
\end{gathered}\ \ \ \ \  \Bigg \} + Z_6  \{ 0 \}.   \label{eq13} 
\end{equation}
From this it follows that $Z_3$ does not have the tree part.

\begin{equation}
0 \equiv \begin{gathered}
\langle \omega \bar{\omega} \left[\frac{1}{2} V^j V^j \right]_R \rangle = 
\end{gathered} \nonumber
\end{equation}
\begin{equation}
Z_7 \Bigg \{
\begin{gathered}
\begin{fmfgraph*}(40,40)
\fmfleft{i1,i2}
\fmflabel{$\omega$}{i1}
\fmfright{o1,o2}
\fmflabel{$\bar{\omega}$}{o1}
\fmf{phantom,tension=1.5}{i2,v2}
\fmf{phantom,tension=1.5}{o1,v2}
\fmf{ghost,label=$\phantom{x}$,tension=1.5,label.side=right}{i1,v1}
\fmf{gluon,label=$\phantom{x}$,left,tension=0.05}{v1,v2}
\fmf{gluon,label=$\phantom{x}$,left,tension=0.05}{v2,v1}
\fmf{ghost,label=$\phantom{x}$,tension=1.5}{v1,o1}
\fmfdotn{v}{1}
\fmfv{decor.shape=cross}{v2}
\end{fmfgraph*}
\end{gathered} \Bigg \} +Z_8 \Bigg \{\ \ \ \begin{gathered}
\begin{fmfgraph*}(40,40)
\fmfleft{i1,i2}
\fmflabel{$\omega$}{i1}
\fmfright{o1,o2}
\fmflabel{$\bar{\omega}$}{o1}
\fmf{phantom,tension=1.5}{i2,v2}
\fmf{phantom,tension=1.5}{o1,v2}
\fmf{ghost,label=$\phantom{x}$,tension=1.5,label.side=right}{i1,v1}
\fmf{photon,label=$\phantom{x}$,left,tension=0.05}{v1,v2}
\fmf{photon,label=$\phantom{x}$,left,tension=0.05}{v2,v1}
\fmf{ghost,label=$\phantom{x}$,tension=1.5}{v1,o1}
\fmfdotn{v}{1}
\fmfv{decor.shape=cross}{v2}
\fmfv{label=$V^j$,label.angle=90,label.dist=0.3cm}{v2}
\end{fmfgraph*} 
\end{gathered}+\ \ \ \ \ 
\begin{gathered}
\begin{fmfgraph*}(70,40)
\fmfleft{i1}
\fmflabel{$\omega$}{i1}
\fmfright{o1}
\fmflabel{$\bar{\omega}$}{o1}
\fmf{ghost,label=$\phantom{x}$}{i1,v1}
\fmf{photon,label.side=left,label=$\phantom{x}$}{v1,v3}
\fmf{photon,label.side=left,label=$\phantom{x}$}{v3,v2}
\fmf{ghost,right,tension=0.4,label=$\phantom{x}$}{v1,v2}
\fmf{ghost,label=$\phantom{x}$}{v2,o1}
\fmfdotn{v}{2}
\fmfv{decor.shape=cross}{v3}
\fmfv{label=$V^j$,label.angle=90,label.dist=0.3cm}{v3}
\end{fmfgraph*} 
\end{gathered}\ \ \ \ \  \Bigg \} \nonumber 
\end{equation}
\begin{equation}
+Z_9 \Bigg \{\ \ \ \ \ \ \begin{fmfgraph*}(30,30)
\fmfleft{i1}
\fmflabel{$\omega$}{i1}
\fmfright{o1}
\fmflabel{$\bar{\omega}$}{o1}
\fmf{ghost,label=$\phantom{x}$}{i1,v}
\fmf{ghost,label=$\phantom{x}$}{v,o1}
\fmfv{decor.shape=cross}{v}
\end{fmfgraph*}\ \ \ \ \  + \ \ \ \ \  \begin{gathered}
\begin{fmfgraph*}(50,40)
\fmfleft{i1}
\fmflabel{$\omega$}{i1}
\fmfright{o1}
\fmflabel{$\bar{\omega}$}{o1}
\fmf{ghost,label=$\phantom{x}$}{i1,v1}
\fmf{photon,left,tension=0.4,label=$V^j+V^\gamma$}{v1,v2}
\fmf{ghost,label=$\phantom{x}$}{v3,v2}
\fmf{ghost,label=$\phantom{x}$}{v1,v3}
\fmf{ghost,label=$\phantom{x}$}{v2,o1}
\fmfdotn{v}{2}
\fmfv{decor.shape=cross}{v3}
\end{fmfgraph*} 
\end{gathered}\ \ \ \ \ \ +\ \ \ \ \ \ \ \ \begin{gathered}
\begin{fmfgraph*}(40,40)
\fmfleft{i1,i2}
\fmflabel{$\omega$}{i1}
\fmfright{o1,o2}
\fmflabel{$\bar{\omega}$}{o1}
\fmf{phantom,tension=1.5}{i2,v2}
\fmf{phantom,tension=1.5}{o1,v2}
\fmf{ghost,label=$\phantom{x}$,tension=1.5,label.side=right}{i1,v1}
\fmf{ghost,label=$\phantom{x}$,left,tension=0.05}{v1,v2}
\fmf{ghost,label=$\phantom{x}$,left,tension=0.05}{v2,v1}
\fmf{ghost,label=$\phantom{x}$,tension=1.5}{v1,o1}
\fmfdotn{v}{1}
\fmfv{decor.shape=cross}{v2}
\fmfv{label=$\omega$,label.angle=90,label.dist=0.3cm}{v2}
\end{fmfgraph*} 
\end{gathered} \ \ \ \ \ \ \ \  \Bigg \} \nonumber
\end{equation}
\begin{equation}
+Z_{10} \{ 0 \} + Z_{11} \Bigg \{\ \ \ \begin{gathered}
\begin{fmfgraph*}(40,40)
\fmfleft{i1,i2}
\fmflabel{$\omega$}{i1}
\fmfright{o1,o2}
\fmflabel{$\bar{\omega}$}{o1}
\fmf{phantom,tension=1.5}{i2,v2}
\fmf{phantom,tension=1.5}{o1,v2}
\fmf{ghost,label=$\phantom{x}$,tension=1.5,label.side=right}{i1,v1}
\fmf{photon,label=$\phantom{x}$,left,tension=0.05}{v1,v2}
\fmf{photon,label=$\phantom{x}$,left,tension=0.05}{v2,v1}
\fmf{ghost,label=$\phantom{x}$,tension=1.5}{v1,o1}
\fmfdotn{v}{1}
\fmfv{decor.shape=cross}{v2}
\fmfv{label=$V^\gamma$,label.angle=90,label.dist=0.3cm}{v2}
\end{fmfgraph*} 
\end{gathered}+\ \ \ \ \ 
\begin{gathered}
\begin{fmfgraph*}(70,40)
\fmfleft{i1}
\fmflabel{$\omega$}{i1}
\fmfright{o1}
\fmflabel{$\bar{\omega}$}{o1}
\fmf{ghost,label=$\phantom{x}$}{i1,v1}
\fmf{photon,label.side=left,label=$\phantom{x}$}{v1,v3}
\fmf{photon,label.side=left,label=$\phantom{x}$}{v3,v2}
\fmf{ghost,right,tension=0.4,label=$\phantom{x}$}{v1,v2}
\fmf{ghost,label=$\phantom{x}$}{v2,o1}
\fmfdotn{v}{2}
\fmfv{decor.shape=cross}{v3}
\fmfv{label=$V^\gamma$,label.angle=90,label.dist=0.3cm}{v3}
\end{fmfgraph*} 
\end{gathered}\ \ \ \ \  \Bigg \} + Z_{12}  \{ 0 \}.   \label{eq14} 
\end{equation}
From this it follows that $Z_9$ does not have the tree part.

\begin{equation}
\begin{gathered}\begin{fmfgraph*}(60,30)
\fmfleft{i1}
\fmflabel{$\omega$}{i1}
\fmfright{o1}
\fmflabel{$\bar{\omega}$}{o1}
\fmf{ghost,label=$\phantom{x}$}{i1,v}
\fmf{ghost,label=$\phantom{x}$}{v,o1}
\fmfv{decor.shape=cross}{v}
\end{fmfgraph*}\end{gathered} \ \ \ \ \ \ \ \ \  \equiv \ \ \ \begin{gathered}
\langle \omega \bar{\omega} \left[i\omega \bar{\omega} \right]_R \rangle = 
\end{gathered} \nonumber
\end{equation}
\begin{equation}
Z_{13} \Bigg \{
\begin{gathered}
\begin{fmfgraph*}(40,40)
\fmfleft{i1,i2}
\fmflabel{$\omega$}{i1}
\fmfright{o1,o2}
\fmflabel{$\bar{\omega}$}{o1}
\fmf{phantom,tension=1.5}{i2,v2}
\fmf{phantom,tension=1.5}{o1,v2}
\fmf{ghost,label=$\phantom{x}$,tension=1.5,label.side=right}{i1,v1}
\fmf{gluon,label=$\phantom{x}$,left,tension=0.05}{v1,v2}
\fmf{gluon,label=$\phantom{x}$,left,tension=0.05}{v2,v1}
\fmf{ghost,label=$\phantom{x}$,tension=1.5}{v1,o1}
\fmfdotn{v}{1}
\fmfv{decor.shape=cross}{v2}
\end{fmfgraph*}
\end{gathered} \Bigg \} +Z_{14} \Bigg \{\ \ \ \begin{gathered}
\begin{fmfgraph*}(40,40)
\fmfleft{i1,i2}
\fmflabel{$\omega$}{i1}
\fmfright{o1,o2}
\fmflabel{$\bar{\omega}$}{o1}
\fmf{phantom,tension=1.5}{i2,v2}
\fmf{phantom,tension=1.5}{o1,v2}
\fmf{ghost,label=$\phantom{x}$,tension=1.5,label.side=right}{i1,v1}
\fmf{photon,label=$\phantom{x}$,left,tension=0.05}{v1,v2}
\fmf{photon,label=$\phantom{x}$,left,tension=0.05}{v2,v1}
\fmf{ghost,label=$\phantom{x}$,tension=1.5}{v1,o1}
\fmfdotn{v}{1}
\fmfv{decor.shape=cross}{v2}
\fmfv{label=$V^j$,label.angle=90,label.dist=0.3cm}{v2}
\end{fmfgraph*} 
\end{gathered}+\ \ \ \ \ 
\begin{gathered}
\begin{fmfgraph*}(70,40)
\fmfleft{i1}
\fmflabel{$\omega$}{i1}
\fmfright{o1}
\fmflabel{$\bar{\omega}$}{o1}
\fmf{ghost,label=$\phantom{x}$}{i1,v1}
\fmf{photon,label.side=left,label=$\phantom{x}$}{v1,v3}
\fmf{photon,label.side=left,label=$\phantom{x}$}{v3,v2}
\fmf{ghost,right,tension=0.4,label=$\phantom{x}$}{v1,v2}
\fmf{ghost,label=$\phantom{x}$}{v2,o1}
\fmfdotn{v}{2}
\fmfv{decor.shape=cross}{v3}
\fmfv{label=$V^j$,label.angle=90,label.dist=0.3cm}{v3}
\end{fmfgraph*} 
\end{gathered}\ \ \ \ \  \Bigg \} \nonumber 
\end{equation}
\begin{equation}
+Z_{15} \Bigg \{\ \ \ \ \ \ \begin{fmfgraph*}(30,30)
\fmfleft{i1}
\fmflabel{$\omega$}{i1}
\fmfright{o1}
\fmflabel{$\bar{\omega}$}{o1}
\fmf{ghost,label=$\phantom{x}$}{i1,v}
\fmf{ghost,label=$\phantom{x}$}{v,o1}
\fmfv{decor.shape=cross}{v}
\end{fmfgraph*}\ \ \ \ \  + \ \ \ \ \  \begin{gathered}
\begin{fmfgraph*}(50,40)
\fmfleft{i1}
\fmflabel{$\omega$}{i1}
\fmfright{o1}
\fmflabel{$\bar{\omega}$}{o1}
\fmf{ghost,label=$\phantom{x}$}{i1,v1}
\fmf{photon,left,tension=0.4,label=$V^j+V^\gamma$}{v1,v2}
\fmf{ghost,label=$\phantom{x}$}{v3,v2}
\fmf{ghost,label=$\phantom{x}$}{v1,v3}
\fmf{ghost,label=$\phantom{x}$}{v2,o1}
\fmfdotn{v}{2}
\fmfv{decor.shape=cross}{v3}
\end{fmfgraph*} 
\end{gathered}\ \ \ \ \ \ +\ \ \ \ \ \ \ \ \begin{gathered}
\begin{fmfgraph*}(40,40)
\fmfleft{i1,i2}
\fmflabel{$\omega$}{i1}
\fmfright{o1,o2}
\fmflabel{$\bar{\omega}$}{o1}
\fmf{phantom,tension=1.5}{i2,v2}
\fmf{phantom,tension=1.5}{o1,v2}
\fmf{ghost,label=$\phantom{x}$,tension=1.5,label.side=right}{i1,v1}
\fmf{ghost,label=$\phantom{x}$,left,tension=0.05}{v1,v2}
\fmf{ghost,label=$\phantom{x}$,left,tension=0.05}{v2,v1}
\fmf{ghost,label=$\phantom{x}$,tension=1.5}{v1,o1}
\fmfdotn{v}{1}
\fmfv{decor.shape=cross}{v2}
\fmfv{label=$\omega$,label.angle=90,label.dist=0.3cm}{v2}
\end{fmfgraph*} 
\end{gathered} \ \ \ \ \ \ \ \  \Bigg \} \nonumber
\end{equation}
\begin{equation}
+Z_{16}  \{ 0 \} + Z_{17} \Bigg \{\ \ \ \begin{gathered}
\begin{fmfgraph*}(40,40)
\fmfleft{i1,i2}
\fmflabel{$\omega$}{i1}
\fmfright{o1,o2}
\fmflabel{$\bar{\omega}$}{o1}
\fmf{phantom,tension=1.5}{i2,v2}
\fmf{phantom,tension=1.5}{o1,v2}
\fmf{ghost,label=$\phantom{x}$,tension=1.5,label.side=right}{i1,v1}
\fmf{photon,label=$\phantom{x}$,left,tension=0.05}{v1,v2}
\fmf{photon,label=$\phantom{x}$,left,tension=0.05}{v2,v1}
\fmf{ghost,label=$\phantom{x}$,tension=1.5}{v1,o1}
\fmfdotn{v}{1}
\fmfv{decor.shape=cross}{v2}
\fmfv{label=$V^\gamma$,label.angle=90,label.dist=0.3cm}{v2}
\end{fmfgraph*} 
\end{gathered}+\ \ \ \ \ 
\begin{gathered}
\begin{fmfgraph*}(70,40)
\fmfleft{i1}
\fmflabel{$\omega$}{i1}
\fmfright{o1}
\fmflabel{$\bar{\omega}$}{o1}
\fmf{ghost,label=$\phantom{x}$}{i1,v1}
\fmf{photon,label.side=left,label=$\phantom{x}$}{v1,v3}
\fmf{photon,label.side=left,label=$\phantom{x}$}{v3,v2}
\fmf{ghost,right,tension=0.4,label=$\phantom{x}$}{v1,v2}
\fmf{ghost,label=$\phantom{x}$}{v2,o1}
\fmfdotn{v}{2}
\fmfv{decor.shape=cross}{v3}
\fmfv{label=$V^\gamma$,label.angle=90,label.dist=0.3cm}{v3}
\end{fmfgraph*} 
\end{gathered}\ \ \ \ \  \Bigg \} + Z_{18}  \{ 0 \}.   \label{eq15} 
\end{equation}
From this it follows that $Z_{15}$ does have the tree part.

\begin{equation}
0 \equiv \begin{gathered}
\langle \omega \bar{\omega} \left[iC^j \bar{C}^j\right]_R \rangle = 
\end{gathered} \nonumber
\end{equation}
\begin{equation}
Z_{19} \Bigg \{
\begin{gathered}
\begin{fmfgraph*}(40,40)
\fmfleft{i1,i2}
\fmflabel{$\omega$}{i1}
\fmfright{o1,o2}
\fmflabel{$\bar{\omega}$}{o1}
\fmf{phantom,tension=1.5}{i2,v2}
\fmf{phantom,tension=1.5}{o1,v2}
\fmf{ghost,label=$\phantom{x}$,tension=1.5,label.side=right}{i1,v1}
\fmf{gluon,label=$\phantom{x}$,left,tension=0.05}{v1,v2}
\fmf{gluon,label=$\phantom{x}$,left,tension=0.05}{v2,v1}
\fmf{ghost,label=$\phantom{x}$,tension=1.5}{v1,o1}
\fmfdotn{v}{1}
\fmfv{decor.shape=cross}{v2}
\end{fmfgraph*}
\end{gathered} \Bigg \} +Z_{20} \Bigg \{\ \ \ \begin{gathered}
\begin{fmfgraph*}(40,40)
\fmfleft{i1,i2}
\fmflabel{$\omega$}{i1}
\fmfright{o1,o2}
\fmflabel{$\bar{\omega}$}{o1}
\fmf{phantom,tension=1.5}{i2,v2}
\fmf{phantom,tension=1.5}{o1,v2}
\fmf{ghost,label=$\phantom{x}$,tension=1.5,label.side=right}{i1,v1}
\fmf{photon,label=$\phantom{x}$,left,tension=0.05}{v1,v2}
\fmf{photon,label=$\phantom{x}$,left,tension=0.05}{v2,v1}
\fmf{ghost,label=$\phantom{x}$,tension=1.5}{v1,o1}
\fmfdotn{v}{1}
\fmfv{decor.shape=cross}{v2}
\fmfv{label=$V^j$,label.angle=90,label.dist=0.3cm}{v2}
\end{fmfgraph*} 
\end{gathered}+\ \ \ \ \ 
\begin{gathered}
\begin{fmfgraph*}(70,40)
\fmfleft{i1}
\fmflabel{$\omega$}{i1}
\fmfright{o1}
\fmflabel{$\bar{\omega}$}{o1}
\fmf{ghost,label=$\phantom{x}$}{i1,v1}
\fmf{photon,label.side=left,label=$\phantom{x}$}{v1,v3}
\fmf{photon,label.side=left,label=$\phantom{x}$}{v3,v2}
\fmf{ghost,right,tension=0.4,label=$\phantom{x}$}{v1,v2}
\fmf{ghost,label=$\phantom{x}$}{v2,o1}
\fmfdotn{v}{2}
\fmfv{decor.shape=cross}{v3}
\fmfv{label=$V^j$,label.angle=90,label.dist=0.3cm}{v3}
\end{fmfgraph*} 
\end{gathered}\ \ \ \ \  \Bigg \} \nonumber 
\end{equation}
\begin{equation}
+Z_{21} \Bigg \{\ \ \ \ \ \ \begin{fmfgraph*}(30,30)
\fmfleft{i1}
\fmflabel{$\omega$}{i1}
\fmfright{o1}
\fmflabel{$\bar{\omega}$}{o1}
\fmf{ghost,label=$\phantom{x}$}{i1,v}
\fmf{ghost,label=$\phantom{x}$}{v,o1}
\fmfv{decor.shape=cross}{v}
\end{fmfgraph*}\ \ \ \ \  + \ \ \ \ \  \begin{gathered}
\begin{fmfgraph*}(50,40)
\fmfleft{i1}
\fmflabel{$\omega$}{i1}
\fmfright{o1}
\fmflabel{$\bar{\omega}$}{o1}
\fmf{ghost,label=$\phantom{x}$}{i1,v1}
\fmf{photon,left,tension=0.4,label=$V^j+V^\gamma$}{v1,v2}
\fmf{ghost,label=$\phantom{x}$}{v3,v2}
\fmf{ghost,label=$\phantom{x}$}{v1,v3}
\fmf{ghost,label=$\phantom{x}$}{v2,o1}
\fmfdotn{v}{2}
\fmfv{decor.shape=cross}{v3}
\end{fmfgraph*} 
\end{gathered}\ \ \ \ \ \ +\ \ \ \ \ \ \ \ \begin{gathered}
\begin{fmfgraph*}(40,40)
\fmfleft{i1,i2}
\fmflabel{$\omega$}{i1}
\fmfright{o1,o2}
\fmflabel{$\bar{\omega}$}{o1}
\fmf{phantom,tension=1.5}{i2,v2}
\fmf{phantom,tension=1.5}{o1,v2}
\fmf{ghost,label=$\phantom{x}$,tension=1.5,label.side=right}{i1,v1}
\fmf{ghost,label=$\phantom{x}$,left,tension=0.05}{v1,v2}
\fmf{ghost,label=$\phantom{x}$,left,tension=0.05}{v2,v1}
\fmf{ghost,label=$\phantom{x}$,tension=1.5}{v1,o1}
\fmfdotn{v}{1}
\fmfv{decor.shape=cross}{v2}
\fmfv{label=$\omega$,label.angle=90,label.dist=0.3cm}{v2}
\end{fmfgraph*} 
\end{gathered} \ \ \ \ \ \ \ \  \Bigg \} \nonumber
\end{equation}
\begin{equation}
+Z_{22} \{ 0 \} + Z_{23} \Bigg \{\ \ \ \begin{gathered}
\begin{fmfgraph*}(40,40)
\fmfleft{i1,i2}
\fmflabel{$\omega$}{i1}
\fmfright{o1,o2}
\fmflabel{$\bar{\omega}$}{o1}
\fmf{phantom,tension=1.5}{i2,v2}
\fmf{phantom,tension=1.5}{o1,v2}
\fmf{ghost,label=$\phantom{x}$,tension=1.5,label.side=right}{i1,v1}
\fmf{photon,label=$\phantom{x}$,left,tension=0.05}{v1,v2}
\fmf{photon,label=$\phantom{x}$,left,tension=0.05}{v2,v1}
\fmf{ghost,label=$\phantom{x}$,tension=1.5}{v1,o1}
\fmfdotn{v}{1}
\fmfv{decor.shape=cross}{v2}
\fmfv{label=$V^\gamma$,label.angle=90,label.dist=0.3cm}{v2}
\end{fmfgraph*} 
\end{gathered}+\ \ \ \ \ 
\begin{gathered}
\begin{fmfgraph*}(70,40)
\fmfleft{i1}
\fmflabel{$\omega$}{i1}
\fmfright{o1}
\fmflabel{$\bar{\omega}$}{o1}
\fmf{ghost,label=$\phantom{x}$}{i1,v1}
\fmf{photon,label.side=left,label=$\phantom{x}$}{v1,v3}
\fmf{photon,label.side=left,label=$\phantom{x}$}{v3,v2}
\fmf{ghost,right,tension=0.4,label=$\phantom{x}$}{v1,v2}
\fmf{ghost,label=$\phantom{x}$}{v2,o1}
\fmfdotn{v}{2}
\fmfv{decor.shape=cross}{v3}
\fmfv{label=$V^\gamma$,label.angle=90,label.dist=0.3cm}{v3}
\end{fmfgraph*} 
\end{gathered}\ \ \ \ \  \Bigg \} + Z_{24}  \{ 0 \}.   \label{eq16} 
\end{equation}
From this it follows that $Z_{21}$ does not have the tree part.

\begin{equation}
0 \equiv \begin{gathered}
\langle \omega \bar{\omega} \left[\frac{1}{2}V^\gamma V^\gamma \right]_R \rangle = 
\end{gathered} \nonumber
\end{equation}
\begin{equation}
Z_{25} \Bigg \{
\begin{gathered}
\begin{fmfgraph*}(40,40)
\fmfleft{i1,i2}
\fmflabel{$\omega$}{i1}
\fmfright{o1,o2}
\fmflabel{$\bar{\omega}$}{o1}
\fmf{phantom,tension=1.5}{i2,v2}
\fmf{phantom,tension=1.5}{o1,v2}
\fmf{ghost,label=$\phantom{x}$,tension=1.5,label.side=right}{i1,v1}
\fmf{gluon,label=$\phantom{x}$,left,tension=0.05}{v1,v2}
\fmf{gluon,label=$\phantom{x}$,left,tension=0.05}{v2,v1}
\fmf{ghost,label=$\phantom{x}$,tension=1.5}{v1,o1}
\fmfdotn{v}{1}
\fmfv{decor.shape=cross}{v2}
\end{fmfgraph*}
\end{gathered} \Bigg \} +Z_{26} \Bigg \{\ \ \ \begin{gathered}
\begin{fmfgraph*}(40,40)
\fmfleft{i1,i2}
\fmflabel{$\omega$}{i1}
\fmfright{o1,o2}
\fmflabel{$\bar{\omega}$}{o1}
\fmf{phantom,tension=1.5}{i2,v2}
\fmf{phantom,tension=1.5}{o1,v2}
\fmf{ghost,label=$\phantom{x}$,tension=1.5,label.side=right}{i1,v1}
\fmf{photon,label=$\phantom{x}$,left,tension=0.05}{v1,v2}
\fmf{photon,label=$\phantom{x}$,left,tension=0.05}{v2,v1}
\fmf{ghost,label=$\phantom{x}$,tension=1.5}{v1,o1}
\fmfdotn{v}{1}
\fmfv{decor.shape=cross}{v2}
\fmfv{label=$V^j$,label.angle=90,label.dist=0.3cm}{v2}
\end{fmfgraph*} 
\end{gathered}+\ \ \ \ \ 
\begin{gathered}
\begin{fmfgraph*}(70,40)
\fmfleft{i1}
\fmflabel{$\omega$}{i1}
\fmfright{o1}
\fmflabel{$\bar{\omega}$}{o1}
\fmf{ghost,label=$\phantom{x}$}{i1,v1}
\fmf{photon,label.side=left,label=$\phantom{x}$}{v1,v3}
\fmf{photon,label.side=left,label=$\phantom{x}$}{v3,v2}
\fmf{ghost,right,tension=0.4,label=$\phantom{x}$}{v1,v2}
\fmf{ghost,label=$\phantom{x}$}{v2,o1}
\fmfdotn{v}{2}
\fmfv{decor.shape=cross}{v3}
\fmfv{label=$V^j$,label.angle=90,label.dist=0.3cm}{v3}
\end{fmfgraph*} 
\end{gathered}\ \ \ \ \  \Bigg \} \nonumber 
\end{equation}
\begin{equation}
+Z_{27} \Bigg \{\ \ \ \ \ \ \begin{fmfgraph*}(30,30)
\fmfleft{i1}
\fmflabel{$\omega$}{i1}
\fmfright{o1}
\fmflabel{$\bar{\omega}$}{o1}
\fmf{ghost,label=$\phantom{x}$}{i1,v}
\fmf{ghost,label=$\phantom{x}$}{v,o1}
\fmfv{decor.shape=cross}{v}
\end{fmfgraph*}\ \ \ \ \  + \ \ \ \ \  \begin{gathered}
\begin{fmfgraph*}(50,40)
\fmfleft{i1}
\fmflabel{$\omega$}{i1}
\fmfright{o1}
\fmflabel{$\bar{\omega}$}{o1}
\fmf{ghost,label=$\phantom{x}$}{i1,v1}
\fmf{photon,left,tension=0.4,label=$V^j+V^\gamma$}{v1,v2}
\fmf{ghost,label=$\phantom{x}$}{v3,v2}
\fmf{ghost,label=$\phantom{x}$}{v1,v3}
\fmf{ghost,label=$\phantom{x}$}{v2,o1}
\fmfdotn{v}{2}
\fmfv{decor.shape=cross}{v3}
\end{fmfgraph*} 
\end{gathered}\ \ \ \ \ \ +\ \ \ \ \ \ \ \ \begin{gathered}
\begin{fmfgraph*}(40,40)
\fmfleft{i1,i2}
\fmflabel{$\omega$}{i1}
\fmfright{o1,o2}
\fmflabel{$\bar{\omega}$}{o1}
\fmf{phantom,tension=1.5}{i2,v2}
\fmf{phantom,tension=1.5}{o1,v2}
\fmf{ghost,label=$\phantom{x}$,tension=1.5,label.side=right}{i1,v1}
\fmf{ghost,label=$\phantom{x}$,left,tension=0.05}{v1,v2}
\fmf{ghost,label=$\phantom{x}$,left,tension=0.05}{v2,v1}
\fmf{ghost,label=$\phantom{x}$,tension=1.5}{v1,o1}
\fmfdotn{v}{1}
\fmfv{decor.shape=cross}{v2}
\fmfv{label=$\omega$,label.angle=90,label.dist=0.3cm}{v2}
\end{fmfgraph*} 
\end{gathered} \ \ \ \ \ \ \ \  \Bigg \} \nonumber
\end{equation}
\begin{equation}
+Z_{28} \{ 0 \} + Z_{29} \Bigg \{\ \ \ \begin{gathered}
\begin{fmfgraph*}(40,40)
\fmfleft{i1,i2}
\fmflabel{$\omega$}{i1}
\fmfright{o1,o2}
\fmflabel{$\bar{\omega}$}{o1}
\fmf{phantom,tension=1.5}{i2,v2}
\fmf{phantom,tension=1.5}{o1,v2}
\fmf{ghost,label=$\phantom{x}$,tension=1.5,label.side=right}{i1,v1}
\fmf{photon,label=$\phantom{x}$,left,tension=0.05}{v1,v2}
\fmf{photon,label=$\phantom{x}$,left,tension=0.05}{v2,v1}
\fmf{ghost,label=$\phantom{x}$,tension=1.5}{v1,o1}
\fmfdotn{v}{1}
\fmfv{decor.shape=cross}{v2}
\fmfv{label=$V^\gamma$,label.angle=90,label.dist=0.3cm}{v2}
\end{fmfgraph*} 
\end{gathered}+\ \ \ \ \ 
\begin{gathered}
\begin{fmfgraph*}(70,40)
\fmfleft{i1}
\fmflabel{$\omega$}{i1}
\fmfright{o1}
\fmflabel{$\bar{\omega}$}{o1}
\fmf{ghost,label=$\phantom{x}$}{i1,v1}
\fmf{photon,label.side=left,label=$\phantom{x}$}{v1,v3}
\fmf{photon,label.side=left,label=$\phantom{x}$}{v3,v2}
\fmf{ghost,right,tension=0.4,label=$\phantom{x}$}{v1,v2}
\fmf{ghost,label=$\phantom{x}$}{v2,o1}
\fmfdotn{v}{2}
\fmfv{decor.shape=cross}{v3}
\fmfv{label=$V^\gamma$,label.angle=90,label.dist=0.3cm}{v3}
\end{fmfgraph*} 
\end{gathered}\ \ \ \ \  \Bigg \} + Z_{30}  \{ 0 \} .  \label{eq17} 
\end{equation}
From this it follows that $Z_{27}$ does not have the tree part.

\begin{equation}
0 \equiv \begin{gathered}
\langle \omega \bar{\omega} \left[iC^\gamma \bar{C}^\gamma\right]_R \rangle = 
\end{gathered} \nonumber
\end{equation}
\begin{equation}
Z_{31} \Bigg \{
\begin{gathered}
\begin{fmfgraph*}(40,40)
\fmfleft{i1,i2}
\fmflabel{$\omega$}{i1}
\fmfright{o1,o2}
\fmflabel{$\bar{\omega}$}{o1}
\fmf{phantom,tension=1.5}{i2,v2}
\fmf{phantom,tension=1.5}{o1,v2}
\fmf{ghost,label=$\phantom{x}$,tension=1.5,label.side=right}{i1,v1}
\fmf{gluon,label=$\phantom{x}$,left,tension=0.05}{v1,v2}
\fmf{gluon,label=$\phantom{x}$,left,tension=0.05}{v2,v1}
\fmf{ghost,label=$\phantom{x}$,tension=1.5}{v1,o1}
\fmfdotn{v}{1}
\fmfv{decor.shape=cross}{v2}
\end{fmfgraph*}
\end{gathered} \Bigg \} +Z_{32} \Bigg \{\ \ \ \begin{gathered}
\begin{fmfgraph*}(40,40)
\fmfleft{i1,i2}
\fmflabel{$\omega$}{i1}
\fmfright{o1,o2}
\fmflabel{$\bar{\omega}$}{o1}
\fmf{phantom,tension=1.5}{i2,v2}
\fmf{phantom,tension=1.5}{o1,v2}
\fmf{ghost,label=$\phantom{x}$,tension=1.5,label.side=right}{i1,v1}
\fmf{photon,label=$\phantom{x}$,left,tension=0.05}{v1,v2}
\fmf{photon,label=$\phantom{x}$,left,tension=0.05}{v2,v1}
\fmf{ghost,label=$\phantom{x}$,tension=1.5}{v1,o1}
\fmfdotn{v}{1}
\fmfv{decor.shape=cross}{v2}
\fmfv{label=$V^j$,label.angle=90,label.dist=0.3cm}{v2}
\end{fmfgraph*} 
\end{gathered}+\ \ \ \ \ 
\begin{gathered}
\begin{fmfgraph*}(70,40)
\fmfleft{i1}
\fmflabel{$\omega$}{i1}
\fmfright{o1}
\fmflabel{$\bar{\omega}$}{o1}
\fmf{ghost,label=$\phantom{x}$}{i1,v1}
\fmf{photon,label.side=left,label=$\phantom{x}$}{v1,v3}
\fmf{photon,label.side=left,label=$\phantom{x}$}{v3,v2}
\fmf{ghost,right,tension=0.4,label=$\phantom{x}$}{v1,v2}
\fmf{ghost,label=$\phantom{x}$}{v2,o1}
\fmfdotn{v}{2}
\fmfv{decor.shape=cross}{v3}
\fmfv{label=$V^j$,label.angle=90,label.dist=0.3cm}{v3}
\end{fmfgraph*} 
\end{gathered}\ \ \ \ \  \Bigg \} \nonumber 
\end{equation}
\begin{equation}
+Z_{33} \Bigg \{\ \ \ \ \ \ \begin{fmfgraph*}(30,30)
\fmfleft{i1}
\fmflabel{$\omega$}{i1}
\fmfright{o1}
\fmflabel{$\bar{\omega}$}{o1}
\fmf{ghost,label=$\phantom{x}$}{i1,v}
\fmf{ghost,label=$\phantom{x}$}{v,o1}
\fmfv{decor.shape=cross}{v}
\end{fmfgraph*}\ \ \ \ \  + \ \ \ \ \  \begin{gathered}
\begin{fmfgraph*}(50,40)
\fmfleft{i1}
\fmflabel{$\omega$}{i1}
\fmfright{o1}
\fmflabel{$\bar{\omega}$}{o1}
\fmf{ghost,label=$\phantom{x}$}{i1,v1}
\fmf{photon,left,tension=0.4,label=$V^j+V^\gamma$}{v1,v2}
\fmf{ghost,label=$\phantom{x}$}{v3,v2}
\fmf{ghost,label=$\phantom{x}$}{v1,v3}
\fmf{ghost,label=$\phantom{x}$}{v2,o1}
\fmfdotn{v}{2}
\fmfv{decor.shape=cross}{v3}
\end{fmfgraph*} 
\end{gathered}\ \ \ \ \ \ +\ \ \ \ \ \ \ \ \begin{gathered}
\begin{fmfgraph*}(40,40)
\fmfleft{i1,i2}
\fmflabel{$\omega$}{i1}
\fmfright{o1,o2}
\fmflabel{$\bar{\omega}$}{o1}
\fmf{phantom,tension=1.5}{i2,v2}
\fmf{phantom,tension=1.5}{o1,v2}
\fmf{ghost,label=$\phantom{x}$,tension=1.5,label.side=right}{i1,v1}
\fmf{ghost,label=$\phantom{x}$,left,tension=0.05}{v1,v2}
\fmf{ghost,label=$\phantom{x}$,left,tension=0.05}{v2,v1}
\fmf{ghost,label=$\phantom{x}$,tension=1.5}{v1,o1}
\fmfdotn{v}{1}
\fmfv{decor.shape=cross}{v2}
\fmfv{label=$\omega$,label.angle=90,label.dist=0.3cm}{v2}
\end{fmfgraph*} 
\end{gathered} \ \ \ \ \ \ \ \  \Bigg \} \nonumber
\end{equation}
\begin{equation}
+Z_{34} \{ 0 \} + Z_{35} \Bigg \{\ \ \ \begin{gathered}
\begin{fmfgraph*}(40,40)
\fmfleft{i1,i2}
\fmflabel{$\omega$}{i1}
\fmfright{o1,o2}
\fmflabel{$\bar{\omega}$}{o1}
\fmf{phantom,tension=1.5}{i2,v2}
\fmf{phantom,tension=1.5}{o1,v2}
\fmf{ghost,label=$\phantom{x}$,tension=1.5,label.side=right}{i1,v1}
\fmf{photon,label=$\phantom{x}$,left,tension=0.05}{v1,v2}
\fmf{photon,label=$\phantom{x}$,left,tension=0.05}{v2,v1}
\fmf{ghost,label=$\phantom{x}$,tension=1.5}{v1,o1}
\fmfdotn{v}{1}
\fmfv{decor.shape=cross}{v2}
\fmfv{label=$V^\gamma$,label.angle=90,label.dist=0.3cm}{v2}
\end{fmfgraph*} 
\end{gathered}+\ \ \ \ \ 
\begin{gathered}
\begin{fmfgraph*}(70,40)
\fmfleft{i1}
\fmflabel{$\omega$}{i1}
\fmfright{o1}
\fmflabel{$\bar{\omega}$}{o1}
\fmf{ghost,label=$\phantom{x}$}{i1,v1}
\fmf{photon,label.side=left,label=$\phantom{x}$}{v1,v3}
\fmf{photon,label.side=left,label=$\phantom{x}$}{v3,v2}
\fmf{ghost,right,tension=0.4,label=$\phantom{x}$}{v1,v2}
\fmf{ghost,label=$\phantom{x}$}{v2,o1}
\fmfdotn{v}{2}
\fmfv{decor.shape=cross}{v3}
\fmfv{label=$V^\gamma$,label.angle=90,label.dist=0.3cm}{v3}
\end{fmfgraph*} 
\end{gathered}\ \ \ \ \  \Bigg \} + Z_{36}  \{ 0 \}.   \label{eq18} 
\end{equation}
From this it follows that $Z_{33}$ does not have the tree part.

\subsubsection{Insertion into $\langle C^j \bar{C}^k \rangle$}


\begin{equation}
\begin{gathered}
0 \equiv \langle C^j \bar{C}^k \left[\frac{1}{2} X X \right]_R \rangle = 
\end{gathered} \nonumber
\end{equation}
\begin{equation}
Z_1 \{ 0\} +Z_2 \Bigg \{\ \ \ \ \ \begin{gathered}
\begin{fmfgraph*}(80,40)
\fmfleft{i1}
\fmflabel{$C^j$}{i1}
\fmfright{o1}
\fmflabel{$\bar{C}^k$}{o1}
\fmf{ghost,label=$\phantom{x}$}{i1,v1}
\fmf{photon,label.side=left,label=$\phantom{x}$}{v1,v3}
\fmf{photon,label.side=left,label=$\phantom{x}$}{v3,v2}
\fmf{ghost,right,label=$\phantom{x}$,tension=0.4}{v1,v2}
\fmf{ghost,label=$\phantom{x}$}{v2,o1}
\fmfdotn{v}{2}
\fmfv{decor.shape=cross}{v3}
\fmfv{label=$V^l$,label.angle=90,label.dist=0.3cm}{v3}
\end{fmfgraph*} 
\end{gathered}\ \ \ \ \ \  \Bigg \} +Z_3 \{0\} \nonumber 
\end{equation}
\begin{equation}
+Z_4 \Bigg \{\ \ \ \ \ \ \   \begin{gathered}\begin{fmfgraph*}(40,40)
\fmfleft{i1}
\fmflabel{$C^j$}{i1}
\fmfright{o1}
\fmflabel{$\bar{C}^k$}{o1}
\fmf{ghost,label=$\phantom{x}$}{i1,v}
\fmf{ghost,label=$\phantom{x}$}{v,o1}
\fmfv{decor.shape=cross}{v}
\end{fmfgraph*} \end{gathered}\ \ \ \ \ \ \ \ \ \ \   + \ \ \ \ \ \ \ \ \begin{gathered}  \begin{fmfgraph*}(60,40)
\fmfleft{i1}
\fmflabel{$C^j$}{i1}
\fmfright{o1}
\fmflabel{$\bar{C}^k$}{o1}
\fmf{ghost,label=$\phantom{x}$}{i1,v1}
\fmf{photon,label=$\phantom{x}$,right,tension=0.4}{v1,v2}
\fmf{ghost,label=$\phantom{x}$}{v1,v3}
\fmf{ghost,label=$\phantom{x}$}{v3,v2}
\fmf{ghost,label=$\phantom{x}$}{v2,o1}
\fmfdotn{v}{2}
\fmfv{decor.shape=cross}{v3}
\fmfv{label=$C^l$,label.angle=90,label.dist=0.3cm}{v3}
\end{fmfgraph*} \end{gathered} \ \ \ \ \ \ \   \Bigg \}\nonumber
\end{equation}
\begin{equation}
+Z_5  \{0\} + Z_6 \{ 0 \}. \label{eq19} 
\end{equation}
From this it follows that $Z_4$ does not have the tree part.

\begin{equation}
\begin{gathered}
0 \equiv \langle C^j \bar{C}^k \left[\frac{1}{2} V^l V^l \right]_R \rangle = 
\end{gathered} \nonumber
\end{equation}
\begin{equation}
Z_7 \{ 0\} +Z_8 \Bigg \{\ \ \ \ \ \begin{gathered}
\begin{fmfgraph*}(80,40)
\fmfleft{i1}
\fmflabel{$C^j$}{i1}
\fmfright{o1}
\fmflabel{$\bar{C}^k$}{o1}
\fmf{ghost,label=$\phantom{x}$}{i1,v1}
\fmf{photon,label.side=left,label=$\phantom{x}$}{v1,v3}
\fmf{photon,label.side=left,label=$\phantom{x}$}{v3,v2}
\fmf{ghost,right,label=$\phantom{x}$,tension=0.4}{v1,v2}
\fmf{ghost,label=$\phantom{x}$}{v2,o1}
\fmfdotn{v}{2}
\fmfv{decor.shape=cross}{v3}
\fmfv{label=$V^l$,label.angle=90,label.dist=0.3cm}{v3}
\end{fmfgraph*} 
\end{gathered}\ \ \ \ \ \  \Bigg \} +Z_9 \{0\} \nonumber 
\end{equation}
\begin{equation}
+Z_{10} \Bigg \{\ \ \ \ \ \ \   \begin{gathered}\begin{fmfgraph*}(40,40)
\fmfleft{i1}
\fmflabel{$C^j$}{i1}
\fmfright{o1}
\fmflabel{$\bar{C}^k$}{o1}
\fmf{ghost,label=$\phantom{x}$}{i1,v}
\fmf{ghost,label=$\phantom{x}$}{v,o1}
\fmfv{decor.shape=cross}{v}
\end{fmfgraph*} \end{gathered}\ \ \ \ \ \ \ \ \ \ \   + \ \ \ \ \ \ \ \ \begin{gathered}  \begin{fmfgraph*}(60,40)
\fmfleft{i1}
\fmflabel{$C^j$}{i1}
\fmfright{o1}
\fmflabel{$\bar{C}^k$}{o1}
\fmf{ghost,label=$\phantom{x}$}{i1,v1}
\fmf{photon,label=$\phantom{x}$,right,tension=0.4}{v1,v2}
\fmf{ghost,label=$\phantom{x}$}{v1,v3}
\fmf{ghost,label=$\phantom{x}$}{v3,v2}
\fmf{ghost,label=$\phantom{x}$}{v2,o1}
\fmfdotn{v}{2}
\fmfv{decor.shape=cross}{v3}
\fmfv{label=$C^l$,label.angle=90,label.dist=0.3cm}{v3}
\end{fmfgraph*} \end{gathered} \ \ \ \ \ \ \   \Bigg \}\nonumber
\end{equation}
\begin{equation}
+Z_{11}  \{0\} + Z_{12} \{ 0 \}. \label{eq20} 
\end{equation}
From this it follows that $Z_{10}$ does not have the tree part.

\begin{equation}
\begin{gathered}
0 \equiv \langle C^j \bar{C}^k \left[i\omega \bar{\omega} \right]_R \rangle = 
\end{gathered} \nonumber
\end{equation}
\begin{equation}
Z_{13} \{ 0\} +Z_{14} \Bigg \{\ \ \ \ \ \begin{gathered}
\begin{fmfgraph*}(80,40)
\fmfleft{i1}
\fmflabel{$C^j$}{i1}
\fmfright{o1}
\fmflabel{$\bar{C}^k$}{o1}
\fmf{ghost,label=$\phantom{x}$}{i1,v1}
\fmf{photon,label.side=left,label=$\phantom{x}$}{v1,v3}
\fmf{photon,label.side=left,label=$\phantom{x}$}{v3,v2}
\fmf{ghost,right,label=$\phantom{x}$,tension=0.4}{v1,v2}
\fmf{ghost,label=$\phantom{x}$}{v2,o1}
\fmfdotn{v}{2}
\fmfv{decor.shape=cross}{v3}
\fmfv{label=$V^l$,label.angle=90,label.dist=0.3cm}{v3}
\end{fmfgraph*} 
\end{gathered}\ \ \ \ \ \  \Bigg \} +Z_{15} \{0\} \nonumber 
\end{equation}
\begin{equation}
+Z_{16} \Bigg \{\ \ \ \ \ \ \   \begin{gathered}\begin{fmfgraph*}(40,40)
\fmfleft{i1}
\fmflabel{$C^j$}{i1}
\fmfright{o1}
\fmflabel{$\bar{C}^k$}{o1}
\fmf{ghost,label=$\phantom{x}$}{i1,v}
\fmf{ghost,label=$\phantom{x}$}{v,o1}
\fmfv{decor.shape=cross}{v}
\end{fmfgraph*} \end{gathered}\ \ \ \ \ \ \ \ \ \ \   + \ \ \ \ \ \ \ \ \begin{gathered}  \begin{fmfgraph*}(60,40)
\fmfleft{i1}
\fmflabel{$C^j$}{i1}
\fmfright{o1}
\fmflabel{$\bar{C}^k$}{o1}
\fmf{ghost,label=$\phantom{x}$}{i1,v1}
\fmf{photon,label=$\phantom{x}$,right,tension=0.4}{v1,v2}
\fmf{ghost,label=$\phantom{x}$}{v1,v3}
\fmf{ghost,label=$\phantom{x}$}{v3,v2}
\fmf{ghost,label=$\phantom{x}$}{v2,o1}
\fmfdotn{v}{2}
\fmfv{decor.shape=cross}{v3}
\fmfv{label=$C^l$,label.angle=90,label.dist=0.3cm}{v3}
\end{fmfgraph*} \end{gathered} \ \ \ \ \ \ \   \Bigg \}\nonumber
\end{equation}
\begin{equation}
+Z_{17}  \{0\} + Z_{18} \{ 0 \}. \label{eq21} 
\end{equation}
From this it follows that $Z_{16}$ does not have the tree part.

\begin{equation}
\begin{gathered}
\begin{gathered}\begin{fmfgraph*}(40,40)
\fmfleft{i1}
\fmflabel{$C^j$}{i1}
\fmfright{o1}
\fmflabel{$\bar{C}^k$}{o1}
\fmf{ghost,label=$\phantom{x}$}{i1,v}
\fmf{ghost,label=$\phantom{x}$}{v,o1}
\fmfv{decor.shape=cross}{v}
\end{fmfgraph*} \end{gathered}\ \ \ \ \ \ \  \equiv \ \ \  \langle C^j \bar{C}^k \left[iC^l\bar{C}^l \right]_R \rangle = 
\end{gathered} \nonumber
\end{equation}
\begin{equation}
Z_{19} \{ 0\} +Z_{20} \Bigg \{\ \ \ \ \ \begin{gathered}
\begin{fmfgraph*}(80,40)
\fmfleft{i1}
\fmflabel{$C^j$}{i1}
\fmfright{o1}
\fmflabel{$\bar{C}^k$}{o1}
\fmf{ghost,label=$\phantom{x}$}{i1,v1}
\fmf{photon,label.side=left,label=$\phantom{x}$}{v1,v3}
\fmf{photon,label.side=left,label=$\phantom{x}$}{v3,v2}
\fmf{ghost,right,label=$\phantom{x}$,tension=0.4}{v1,v2}
\fmf{ghost,label=$\phantom{x}$}{v2,o1}
\fmfdotn{v}{2}
\fmfv{decor.shape=cross}{v3}
\fmfv{label=$V^l$,label.angle=90,label.dist=0.3cm}{v3}
\end{fmfgraph*} 
\end{gathered}\ \ \ \ \ \  \Bigg \} +Z_{21} \{0\} \nonumber 
\end{equation}
\begin{equation}
+Z_{22} \Bigg \{\ \ \ \ \ \ \   \begin{gathered}\begin{fmfgraph*}(40,40)
\fmfleft{i1}
\fmflabel{$C^j$}{i1}
\fmfright{o1}
\fmflabel{$\bar{C}^k$}{o1}
\fmf{ghost,label=$\phantom{x}$}{i1,v}
\fmf{ghost,label=$\phantom{x}$}{v,o1}
\fmfv{decor.shape=cross}{v}
\end{fmfgraph*} \end{gathered}\ \ \ \ \ \ \ \ \ \ \   + \ \ \ \ \ \ \ \ \begin{gathered}  \begin{fmfgraph*}(60,40)
\fmfleft{i1}
\fmflabel{$C^j$}{i1}
\fmfright{o1}
\fmflabel{$\bar{C}^k$}{o1}
\fmf{ghost,label=$\phantom{x}$}{i1,v1}
\fmf{photon,label=$\phantom{x}$,right,tension=0.4}{v1,v2}
\fmf{ghost,label=$\phantom{x}$}{v1,v3}
\fmf{ghost,label=$\phantom{x}$}{v3,v2}
\fmf{ghost,label=$\phantom{x}$}{v2,o1}
\fmfdotn{v}{2}
\fmfv{decor.shape=cross}{v3}
\fmfv{label=$C^l$,label.angle=90,label.dist=0.3cm}{v3}
\end{fmfgraph*} \end{gathered} \ \ \ \ \ \ \   \Bigg \}\nonumber
\end{equation}
\begin{equation}
+Z_{23}  \{0\} + Z_{24} \{ 0 \}. \label{eq22} 
\end{equation}
From this it follows that $Z_{22}$ does have the tree part.

\begin{equation}
\begin{gathered}
0 \equiv \langle C^j \bar{C}^k \left[\frac{1}{2} V^\gamma V^\gamma \right]_R \rangle = 
\end{gathered} \nonumber
\end{equation}
\begin{equation}
Z_{25} \{ 0\} +Z_{26} \Bigg \{\ \ \ \ \ \begin{gathered}
\begin{fmfgraph*}(80,40)
\fmfleft{i1}
\fmflabel{$C^j$}{i1}
\fmfright{o1}
\fmflabel{$\bar{C}^k$}{o1}
\fmf{ghost,label=$\phantom{x}$}{i1,v1}
\fmf{photon,label.side=left,label=$\phantom{x}$}{v1,v3}
\fmf{photon,label.side=left,label=$\phantom{x}$}{v3,v2}
\fmf{ghost,right,label=$\phantom{x}$,tension=0.4}{v1,v2}
\fmf{ghost,label=$\phantom{x}$}{v2,o1}
\fmfdotn{v}{2}
\fmfv{decor.shape=cross}{v3}
\fmfv{label=$V^l$,label.angle=90,label.dist=0.3cm}{v3}
\end{fmfgraph*} 
\end{gathered}\ \ \ \ \ \  \Bigg \} +Z_{27} \{0\} \nonumber 
\end{equation}
\begin{equation}
+Z_{28} \Bigg \{\ \ \ \ \ \ \   \begin{gathered}\begin{fmfgraph*}(40,40)
\fmfleft{i1}
\fmflabel{$C^j$}{i1}
\fmfright{o1}
\fmflabel{$\bar{C}^k$}{o1}
\fmf{ghost,label=$\phantom{x}$}{i1,v}
\fmf{ghost,label=$\phantom{x}$}{v,o1}
\fmfv{decor.shape=cross}{v}
\end{fmfgraph*} \end{gathered}\ \ \ \ \ \ \ \ \ \ \   + \ \ \ \ \ \ \ \ \begin{gathered}  \begin{fmfgraph*}(60,40)
\fmfleft{i1}
\fmflabel{$C^j$}{i1}
\fmfright{o1}
\fmflabel{$\bar{C}^k$}{o1}
\fmf{ghost,label=$\phantom{x}$}{i1,v1}
\fmf{photon,label=$\phantom{x}$,right,tension=0.4}{v1,v2}
\fmf{ghost,label=$\phantom{x}$}{v1,v3}
\fmf{ghost,label=$\phantom{x}$}{v3,v2}
\fmf{ghost,label=$\phantom{x}$}{v2,o1}
\fmfdotn{v}{2}
\fmfv{decor.shape=cross}{v3}
\fmfv{label=$C^l$,label.angle=90,label.dist=0.3cm}{v3}
\end{fmfgraph*} \end{gathered} \ \ \ \ \ \ \   \Bigg \}\nonumber
\end{equation}
\begin{equation}
+Z_{29}  \{0\} + Z_{30} \{ 0 \}. \label{eq23} 
\end{equation}
From this it follows that $Z_{28}$ does not have the tree part.

\begin{equation}
\begin{gathered}
0 \equiv \langle C^j \bar{C}^k \left[iC^\gamma \bar{C}^\gamma \right]_R \rangle = 
\end{gathered} \nonumber
\end{equation}
\begin{equation}
Z_{31} \{ 0\} +Z_{32} \Bigg \{\ \ \ \ \ \begin{gathered}
\begin{fmfgraph*}(80,40)
\fmfleft{i1}
\fmflabel{$C^j$}{i1}
\fmfright{o1}
\fmflabel{$\bar{C}^k$}{o1}
\fmf{ghost,label=$\phantom{x}$}{i1,v1}
\fmf{photon,label.side=left,label=$\phantom{x}$}{v1,v3}
\fmf{photon,label.side=left,label=$\phantom{x}$}{v3,v2}
\fmf{ghost,right,label=$\phantom{x}$,tension=0.4}{v1,v2}
\fmf{ghost,label=$\phantom{x}$}{v2,o1}
\fmfdotn{v}{2}
\fmfv{decor.shape=cross}{v3}
\fmfv{label=$V^l$,label.angle=90,label.dist=0.3cm}{v3}
\end{fmfgraph*} 
\end{gathered}\ \ \ \ \ \  \Bigg \} +Z_{33} \{0\} \nonumber 
\end{equation}
\begin{equation}
+Z_{34} \Bigg \{\ \ \ \ \ \ \   \begin{gathered}\begin{fmfgraph*}(40,40)
\fmfleft{i1}
\fmflabel{$C^j$}{i1}
\fmfright{o1}
\fmflabel{$\bar{C}^k$}{o1}
\fmf{ghost,label=$\phantom{x}$}{i1,v}
\fmf{ghost,label=$\phantom{x}$}{v,o1}
\fmfv{decor.shape=cross}{v}
\end{fmfgraph*} \end{gathered}\ \ \ \ \ \ \ \ \ \ \   + \ \ \ \ \ \ \ \ \begin{gathered}  \begin{fmfgraph*}(60,40)
\fmfleft{i1}
\fmflabel{$C^j$}{i1}
\fmfright{o1}
\fmflabel{$\bar{C}^k$}{o1}
\fmf{ghost,label=$\phantom{x}$}{i1,v1}
\fmf{photon,label=$\phantom{x}$,right,tension=0.4}{v1,v2}
\fmf{ghost,label=$\phantom{x}$}{v1,v3}
\fmf{ghost,label=$\phantom{x}$}{v3,v2}
\fmf{ghost,label=$\phantom{x}$}{v2,o1}
\fmfdotn{v}{2}
\fmfv{decor.shape=cross}{v3}
\fmfv{label=$C^l$,label.angle=90,label.dist=0.3cm}{v3}
\end{fmfgraph*} \end{gathered} \ \ \ \ \ \ \   \Bigg \}\nonumber
\end{equation}
\begin{equation}
+Z_{35}  \{0\} + Z_{36} \{ 0 \}. \label{eq24} 
\end{equation}
From this it follows that $Z_{34}$ does not have the tree part.

\subsubsection{Insertion into $\langle V^\gamma V^\gamma \rangle$}


\begin{equation}
0 \equiv\begin{gathered}
\langle V^\gamma V^\gamma \left[\frac{1}{2} X X \right]_R \rangle = 
\end{gathered}\nonumber 
\end{equation}
\begin{equation} 
Z_1 \Bigg \{
\begin{gathered}
\begin{fmfgraph*}(50,50)
\fmfleft{i1,i2}
\fmflabel{$V^\gamma$}{i1}
\fmfright{o1,o2}
\fmflabel{$V^\gamma$}{o1}
\fmf{phantom,tension=1.5}{i2,v2}
\fmf{phantom,tension=1.5}{o1,v2}
\fmf{photon,label=$\phantom{x}$,tension=1.5,label.side=right}{i1,v1}
\fmf{gluon,label=$\phantom{x}$,left,tension=0.05}{v1,v2}
\fmf{gluon,label=$\phantom{x}$,left,tension=0.05}{v2,v1}
\fmf{photon,label=$\phantom{x}$,tension=1.5}{v1,o1}
\fmfdotn{v}{1}
\fmfv{decor.shape=cross}{v2}
\end{fmfgraph*} 
\end{gathered} + \ \ \ \ \ \ 
\begin{gathered}
  \begin{fmfgraph*}(80,40)
\fmfleft{i1}
\fmflabel{$V^\gamma$}{i1}
\fmfright{o1}
\fmflabel{$V^\gamma$}{o1}
\fmf{photon,label=$\phantom{x}$}{i1,v1}
\fmf{gluon,label=$\phantom{x}$,label.side=left}{v1,v3}
\fmf{gluon,label=$\phantom{x}$,label.side=left}{v3,v2}
\fmf{gluon,left,label=$\phantom{x}$,tension=0.4}{v2,v1}
\fmf{photon,label=$\phantom{x}$}{v2,o1}
\fmfdotn{v}{2}
\fmfv{decor.shape=cross}{v3}
  \end{fmfgraph*}
\end{gathered}\ \ \ \ \ \ \ \Bigg \} +Z_2  \{0  \}   \nonumber 
\end{equation}
\begin{equation}
+Z_3 \Bigg \{\ \ \ \ \ \ \begin{gathered}
\begin{fmfgraph*}(50,50)
\fmfleft{i1,i2}
\fmflabel{$V^\gamma$}{i1}
\fmfv{label=$\omega$,label.angle=90,label.dist=0.3cm}{v2}
\fmfright{o1,o2}
\fmflabel{$V^\gamma$}{o1}
\fmf{phantom,tension=1.5}{i2,v2}
\fmf{phantom,tension=1.5}{o1,v2}
\fmf{photon,label=$\phantom{x}$,tension=1.5,label.side=right}{i1,v1}
\fmf{ghost,label=$\phantom{x}$,left,tension=0.05}{v1,v2}
\fmf{ghost,label=$\phantom{x}$,left,tension=0.05}{v2,v1}
\fmf{photon,label=$\phantom{x}$,tension=1.5}{v1,o1}
\fmfdotn{v}{1}
\fmfv{decor.shape=cross}{v2}
\end{fmfgraph*} 
\end{gathered}\ \ \ \ \ \ +\ \ \ \ \ \ \ \ \begin{gathered}
\begin{fmfgraph*}(70,40)
\fmfleft{i1}
\fmflabel{$V^\gamma$}{i1}
\fmfright{o1}
\fmflabel{$V^\gamma$}{o1}
\fmf{photon,label=$\phantom{x}$}{i1,v1}
\fmf{ghost,label=$\phantom{x}$}{v1,v3}
\fmf{ghost,label=$\phantom{x}$}{v3,v2}
\fmf{ghost,left,label=$\phantom{x}$,tension=0.4}{v2,v1}
\fmf{photon,label=$\phantom{x}$}{v2,o1}
\fmfdotn{v}{2}
\fmfv{decor.shape=cross}{v3}
\fmfv{label=$\omega$,label.angle=90,label.dist=0.3cm}{v3}
\end{fmfgraph*} 
\end{gathered} \ \ \ \ \ \ \ \  \Bigg \} \nonumber
\end{equation}
\begin{equation}
+Z_4   \{ 0\} +Z_5 \Bigg \{\ \ \ \ \ \ \  \begin{gathered}\begin{fmfgraph*}(40,40)
\fmfleft{i1}
\fmflabel{$V^\gamma$}{i1}
\fmfright{o1}
\fmflabel{$V^\gamma$}{o1}
\fmf{photon,label=$\phantom{x}$}{i1,v}
\fmf{photon,label=$\phantom{x}$}{v,o1}
\fmfv{decor.shape=cross}{v}
\end{fmfgraph*} \end{gathered} \ \ \ \ \ \ \  \Bigg \} + Z_6  \{ 0 \}  . \label{eq25} 
\end{equation}
From this it follows that $Z_5$ does not have the tree part.

\begin{equation}
0 \equiv\begin{gathered}
\langle V^\gamma V^\gamma \left[\frac{1}{2} V^j V^j \right]_R \rangle = 
\end{gathered}\nonumber 
\end{equation}
\begin{equation} 
Z_7 \Bigg \{
\begin{gathered}
\begin{fmfgraph*}(50,50)
\fmfleft{i1,i2}
\fmflabel{$V^\gamma$}{i1}
\fmfright{o1,o2}
\fmflabel{$V^\gamma$}{o1}
\fmf{phantom,tension=1.5}{i2,v2}
\fmf{phantom,tension=1.5}{o1,v2}
\fmf{photon,label=$\phantom{x}$,tension=1.5,label.side=right}{i1,v1}
\fmf{gluon,label=$\phantom{x}$,left,tension=0.05}{v1,v2}
\fmf{gluon,label=$\phantom{x}$,left,tension=0.05}{v2,v1}
\fmf{photon,label=$\phantom{x}$,tension=1.5}{v1,o1}
\fmfdotn{v}{1}
\fmfv{decor.shape=cross}{v2}
\end{fmfgraph*} 
\end{gathered} + \ \ \ \ \ \ 
\begin{gathered}
  \begin{fmfgraph*}(80,40)
\fmfleft{i1}
\fmflabel{$V^\gamma$}{i1}
\fmfright{o1}
\fmflabel{$V^\gamma$}{o1}
\fmf{photon,label=$\phantom{x}$}{i1,v1}
\fmf{gluon,label=$\phantom{x}$,label.side=left}{v1,v3}
\fmf{gluon,label=$\phantom{x}$,label.side=left}{v3,v2}
\fmf{gluon,left,label=$\phantom{x}$,tension=0.4}{v2,v1}
\fmf{photon,label=$\phantom{x}$}{v2,o1}
\fmfdotn{v}{2}
\fmfv{decor.shape=cross}{v3}
  \end{fmfgraph*}
\end{gathered}\ \ \ \ \ \ \ \Bigg \} +Z_8  \{0  \}   \nonumber 
\end{equation}
\begin{equation}
+Z_9 \Bigg \{\ \ \ \ \ \ \begin{gathered}
\begin{fmfgraph*}(50,50)
\fmfleft{i1,i2}
\fmflabel{$V^\gamma$}{i1}
\fmfv{label=$\omega$,label.angle=90,label.dist=0.3cm}{v2}
\fmfright{o1,o2}
\fmflabel{$V^\gamma$}{o1}
\fmf{phantom,tension=1.5}{i2,v2}
\fmf{phantom,tension=1.5}{o1,v2}
\fmf{photon,label=$\phantom{x}$,tension=1.5,label.side=right}{i1,v1}
\fmf{ghost,label=$\phantom{x}$,left,tension=0.05}{v1,v2}
\fmf{ghost,label=$\phantom{x}$,left,tension=0.05}{v2,v1}
\fmf{photon,label=$\phantom{x}$,tension=1.5}{v1,o1}
\fmfdotn{v}{1}
\fmfv{decor.shape=cross}{v2}
\end{fmfgraph*} 
\end{gathered}\ \ \ \ \ \ +\ \ \ \ \ \ \ \ \begin{gathered}
\begin{fmfgraph*}(70,40)
\fmfleft{i1}
\fmflabel{$V^\gamma$}{i1}
\fmfright{o1}
\fmflabel{$V^\gamma$}{o1}
\fmf{photon,label=$\phantom{x}$}{i1,v1}
\fmf{ghost,label=$\phantom{x}$}{v1,v3}
\fmf{ghost,label=$\phantom{x}$}{v3,v2}
\fmf{ghost,left,label=$\phantom{x}$,tension=0.4}{v2,v1}
\fmf{photon,label=$\phantom{x}$}{v2,o1}
\fmfdotn{v}{2}
\fmfv{decor.shape=cross}{v3}
\fmfv{label=$\omega$,label.angle=90,label.dist=0.3cm}{v3}
\end{fmfgraph*} 
\end{gathered} \ \ \ \ \ \ \ \  \Bigg \} \nonumber
\end{equation}
\begin{equation}
+Z_{10}  \{ 0\} +Z_{11} \Bigg \{\ \ \ \ \ \ \  \begin{gathered}\begin{fmfgraph*}(40,40)
\fmfleft{i1}
\fmflabel{$V^\gamma$}{i1}
\fmfright{o1}
\fmflabel{$V^\gamma$}{o1}
\fmf{photon,label=$\phantom{x}$}{i1,v}
\fmf{photon,label=$\phantom{x}$}{v,o1}
\fmfv{decor.shape=cross}{v}
\end{fmfgraph*} \end{gathered} \ \ \ \ \ \ \  \Bigg \} + Z_{12}  \{ 0 \}  . \label{eq26} 
\end{equation}
From this it follows that $Z_{11}$ does not have the tree part.

\begin{equation}
0 \equiv\begin{gathered}
\langle V^\gamma V^\gamma \left[i\omega \bar{\omega} \right]_R \rangle = 
\end{gathered}\nonumber 
\end{equation}
\begin{equation} 
Z_{13} \Bigg \{
\begin{gathered}
\begin{fmfgraph*}(50,50)
\fmfleft{i1,i2}
\fmflabel{$V^\gamma$}{i1}
\fmfright{o1,o2}
\fmflabel{$V^\gamma$}{o1}
\fmf{phantom,tension=1.5}{i2,v2}
\fmf{phantom,tension=1.5}{o1,v2}
\fmf{photon,label=$\phantom{x}$,tension=1.5,label.side=right}{i1,v1}
\fmf{gluon,label=$\phantom{x}$,left,tension=0.05}{v1,v2}
\fmf{gluon,label=$\phantom{x}$,left,tension=0.05}{v2,v1}
\fmf{photon,label=$\phantom{x}$,tension=1.5}{v1,o1}
\fmfdotn{v}{1}
\fmfv{decor.shape=cross}{v2}
\end{fmfgraph*} 
\end{gathered} + \ \ \ \ \ \ 
\begin{gathered}
  \begin{fmfgraph*}(80,40)
\fmfleft{i1}
\fmflabel{$V^\gamma$}{i1}
\fmfright{o1}
\fmflabel{$V^\gamma$}{o1}
\fmf{photon,label=$\phantom{x}$}{i1,v1}
\fmf{gluon,label=$\phantom{x}$,label.side=left}{v1,v3}
\fmf{gluon,label=$\phantom{x}$,label.side=left}{v3,v2}
\fmf{gluon,left,label=$\phantom{x}$,tension=0.4}{v2,v1}
\fmf{photon,label=$\phantom{x}$}{v2,o1}
\fmfdotn{v}{2}
\fmfv{decor.shape=cross}{v3}
  \end{fmfgraph*}
\end{gathered}\ \ \ \ \ \ \ \Bigg \} +Z_{14}  \{0  \}   \nonumber 
\end{equation}
\begin{equation}
+Z_{15} \Bigg \{\ \ \ \ \ \ \begin{gathered}
\begin{fmfgraph*}(50,50)
\fmfleft{i1,i2}
\fmflabel{$V^\gamma$}{i1}
\fmfv{label=$\omega$,label.angle=90,label.dist=0.3cm}{v2}
\fmfright{o1,o2}
\fmflabel{$V^\gamma$}{o1}
\fmf{phantom,tension=1.5}{i2,v2}
\fmf{phantom,tension=1.5}{o1,v2}
\fmf{photon,label=$\phantom{x}$,tension=1.5,label.side=right}{i1,v1}
\fmf{ghost,label=$\phantom{x}$,left,tension=0.05}{v1,v2}
\fmf{ghost,label=$\phantom{x}$,left,tension=0.05}{v2,v1}
\fmf{photon,label=$\phantom{x}$,tension=1.5}{v1,o1}
\fmfdotn{v}{1}
\fmfv{decor.shape=cross}{v2}
\end{fmfgraph*} 
\end{gathered}\ \ \ \ \ \ +\ \ \ \ \ \ \ \ \begin{gathered}
\begin{fmfgraph*}(70,40)
\fmfleft{i1}
\fmflabel{$V^\gamma$}{i1}
\fmfright{o1}
\fmflabel{$V^\gamma$}{o1}
\fmf{photon,label=$\phantom{x}$}{i1,v1}
\fmf{ghost,label=$\phantom{x}$}{v1,v3}
\fmf{ghost,label=$\phantom{x}$}{v3,v2}
\fmf{ghost,left,label=$\phantom{x}$,tension=0.4}{v2,v1}
\fmf{photon,label=$\phantom{x}$}{v2,o1}
\fmfdotn{v}{2}
\fmfv{decor.shape=cross}{v3}
\fmfv{label=$\omega$,label.angle=90,label.dist=0.3cm}{v3}
\end{fmfgraph*} 
\end{gathered} \ \ \ \ \ \ \ \  \Bigg \} \nonumber
\end{equation}
\begin{equation}
+Z_{16}   \{ 0\} +Z_{17} \Bigg \{\ \ \ \ \ \ \  \begin{gathered}\begin{fmfgraph*}(40,40)
\fmfleft{i1}
\fmflabel{$V^\gamma$}{i1}
\fmfright{o1}
\fmflabel{$V^\gamma$}{o1}
\fmf{photon,label=$\phantom{x}$}{i1,v}
\fmf{photon,label=$\phantom{x}$}{v,o1}
\fmfv{decor.shape=cross}{v}
\end{fmfgraph*} \end{gathered} \ \ \ \ \ \ \  \Bigg \} + Z_{18}  \{ 0 \}  . \label{eq27} 
\end{equation}
From this it follows that $Z_{17}$ does not have the tree part.

\end{fmffile}

\begin{fmffile}{2ndPart}

\begin{equation}
0 \equiv\begin{gathered}
\langle V^\gamma V^\gamma \left[iC^j \bar{C}^j \right]_R \rangle = 
\end{gathered}\nonumber 
\end{equation}
\begin{equation} 
Z_{19} \Bigg \{
\begin{gathered}
\begin{fmfgraph*}(50,50)
\fmfleft{i1,i2}
\fmflabel{$V^\gamma$}{i1}
\fmfright{o1,o2}
\fmflabel{$V^\gamma$}{o1}
\fmf{phantom,tension=1.5}{i2,v2}
\fmf{phantom,tension=1.5}{o1,v2}
\fmf{photon,label=$\phantom{x}$,tension=1.5,label.side=right}{i1,v1}
\fmf{gluon,label=$\phantom{x}$,left,tension=0.05}{v1,v2}
\fmf{gluon,label=$\phantom{x}$,left,tension=0.05}{v2,v1}
\fmf{photon,label=$\phantom{x}$,tension=1.5}{v1,o1}
\fmfdotn{v}{1}
\fmfv{decor.shape=cross}{v2}
\end{fmfgraph*} 
\end{gathered} + \ \ \ \ \ \ 
\begin{gathered}
  \begin{fmfgraph*}(80,40)
\fmfleft{i1}
\fmflabel{$V^\gamma$}{i1}
\fmfright{o1}
\fmflabel{$V^\gamma$}{o1}
\fmf{photon,label=$\phantom{x}$}{i1,v1}
\fmf{gluon,label=$\phantom{x}$,label.side=left}{v1,v3}
\fmf{gluon,label=$\phantom{x}$,label.side=left}{v3,v2}
\fmf{gluon,left,label=$\phantom{x}$,tension=0.4}{v2,v1}
\fmf{photon,label=$\phantom{x}$}{v2,o1}
\fmfdotn{v}{2}
\fmfv{decor.shape=cross}{v3}
  \end{fmfgraph*}
\end{gathered}\ \ \ \ \ \ \ \Bigg \} +Z_{20}  \{0  \}   \nonumber 
\end{equation}
\begin{equation}
+Z_{21} \Bigg \{\ \ \ \ \ \ \begin{gathered}
\begin{fmfgraph*}(50,50)
\fmfleft{i1,i2}
\fmflabel{$V^\gamma$}{i1}
\fmfv{label=$\omega$,label.angle=90,label.dist=0.3cm}{v2}
\fmfright{o1,o2}
\fmflabel{$V^\gamma$}{o1}
\fmf{phantom,tension=1.5}{i2,v2}
\fmf{phantom,tension=1.5}{o1,v2}
\fmf{photon,label=$\phantom{x}$,tension=1.5,label.side=right}{i1,v1}
\fmf{ghost,label=$\phantom{x}$,left,tension=0.05}{v1,v2}
\fmf{ghost,label=$\phantom{x}$,left,tension=0.05}{v2,v1}
\fmf{photon,label=$\phantom{x}$,tension=1.5}{v1,o1}
\fmfdotn{v}{1}
\fmfv{decor.shape=cross}{v2}
\end{fmfgraph*} 
\end{gathered}\ \ \ \ \ \ +\ \ \ \ \ \ \ \ \begin{gathered}
\begin{fmfgraph*}(70,40)
\fmfleft{i1}
\fmflabel{$V^\gamma$}{i1}
\fmfright{o1}
\fmflabel{$V^\gamma$}{o1}
\fmf{photon,label=$\phantom{x}$}{i1,v1}
\fmf{ghost,label=$\phantom{x}$}{v1,v3}
\fmf{ghost,label=$\phantom{x}$}{v3,v2}
\fmf{ghost,left,label=$\phantom{x}$,tension=0.4}{v2,v1}
\fmf{photon,label=$\phantom{x}$}{v2,o1}
\fmfdotn{v}{2}
\fmfv{decor.shape=cross}{v3}
\fmfv{label=$\omega$,label.angle=90,label.dist=0.3cm}{v3}
\end{fmfgraph*} 
\end{gathered} \ \ \ \ \ \ \ \  \Bigg \} \nonumber
\end{equation}
\begin{equation}
+Z_{22}   \{ 0\} +Z_{23} \Bigg \{\ \ \ \ \ \ \  \begin{gathered}\begin{fmfgraph*}(40,40)
\fmfleft{i1}
\fmflabel{$V^\gamma$}{i1}
\fmfright{o1}
\fmflabel{$V^\gamma$}{o1}
\fmf{photon,label=$\phantom{x}$}{i1,v}
\fmf{photon,label=$\phantom{x}$}{v,o1}
\fmfv{decor.shape=cross}{v}
\end{fmfgraph*} \end{gathered} \ \ \ \ \ \ \  \Bigg \} + Z_{24}  \{ 0 \}  . \label{eq28} 
\end{equation}
From this it follows that $Z_{23}$ does not have the tree part.

\begin{equation}
\begin{gathered}\begin{fmfgraph*}(40,40)
\fmfleft{i1}
\fmflabel{$V^\gamma$}{i1}
\fmfright{o1}
\fmflabel{$V^\gamma$}{o1}
\fmf{photon,label=$\phantom{x}$}{i1,v}
\fmf{photon,label=$\phantom{x}$}{v,o1}
\fmfv{decor.shape=cross}{v}
\end{fmfgraph*} \end{gathered} \ \ \ \ \ \ \ \  \equiv \ \ \begin{gathered}
\langle V^\gamma V^\gamma \left[\frac{1}{2} V^\gamma V^\gamma \right]_R \rangle = 
\end{gathered}\nonumber 
\end{equation}
\begin{equation} 
Z_{25} \Bigg \{
\begin{gathered}
\begin{fmfgraph*}(50,50)
\fmfleft{i1,i2}
\fmflabel{$V^\gamma$}{i1}
\fmfright{o1,o2}
\fmflabel{$V^\gamma$}{o1}
\fmf{phantom,tension=1.5}{i2,v2}
\fmf{phantom,tension=1.5}{o1,v2}
\fmf{photon,label=$\phantom{x}$,tension=1.5,label.side=right}{i1,v1}
\fmf{gluon,label=$\phantom{x}$,left,tension=0.05}{v1,v2}
\fmf{gluon,label=$\phantom{x}$,left,tension=0.05}{v2,v1}
\fmf{photon,label=$\phantom{x}$,tension=1.5}{v1,o1}
\fmfdotn{v}{1}
\fmfv{decor.shape=cross}{v2}
\end{fmfgraph*} 
\end{gathered} + \ \ \ \ \ \ 
\begin{gathered}
  \begin{fmfgraph*}(80,40)
\fmfleft{i1}
\fmflabel{$V^\gamma$}{i1}
\fmfright{o1}
\fmflabel{$V^\gamma$}{o1}
\fmf{photon,label=$\phantom{x}$}{i1,v1}
\fmf{gluon,label=$\phantom{x}$,label.side=left}{v1,v3}
\fmf{gluon,label=$\phantom{x}$,label.side=left}{v3,v2}
\fmf{gluon,left,label=$\phantom{x}$,tension=0.4}{v2,v1}
\fmf{photon,label=$\phantom{x}$}{v2,o1}
\fmfdotn{v}{2}
\fmfv{decor.shape=cross}{v3}
  \end{fmfgraph*}
\end{gathered}\ \ \ \ \ \ \ \Bigg \} +Z_{26}  \{0  \}   \nonumber 
\end{equation}
\begin{equation}
+Z_{27} \Bigg \{\ \ \ \ \ \ \begin{gathered}
\begin{fmfgraph*}(50,50)
\fmfleft{i1,i2}
\fmflabel{$V^\gamma$}{i1}
\fmfv{label=$\omega$,label.angle=90,label.dist=0.3cm}{v2}
\fmfright{o1,o2}
\fmflabel{$V^\gamma$}{o1}
\fmf{phantom,tension=1.5}{i2,v2}
\fmf{phantom,tension=1.5}{o1,v2}
\fmf{photon,label=$\phantom{x}$,tension=1.5,label.side=right}{i1,v1}
\fmf{ghost,label=$\phantom{x}$,left,tension=0.05}{v1,v2}
\fmf{ghost,label=$\phantom{x}$,left,tension=0.05}{v2,v1}
\fmf{photon,label=$\phantom{x}$,tension=1.5}{v1,o1}
\fmfdotn{v}{1}
\fmfv{decor.shape=cross}{v2}
\end{fmfgraph*} 
\end{gathered}\ \ \ \ \ \ +\ \ \ \ \ \ \ \ \begin{gathered}
\begin{fmfgraph*}(70,40)
\fmfleft{i1}
\fmflabel{$V^\gamma$}{i1}
\fmfright{o1}
\fmflabel{$V^\gamma$}{o1}
\fmf{photon,label=$\phantom{x}$}{i1,v1}
\fmf{ghost,label=$\phantom{x}$}{v1,v3}
\fmf{ghost,label=$\phantom{x}$}{v3,v2}
\fmf{ghost,left,label=$\phantom{x}$,tension=0.4}{v2,v1}
\fmf{photon,label=$\phantom{x}$}{v2,o1}
\fmfdotn{v}{2}
\fmfv{decor.shape=cross}{v3}
\fmfv{label=$\omega$,label.angle=90,label.dist=0.3cm}{v3}
\end{fmfgraph*} 
\end{gathered} \ \ \ \ \ \ \ \  \Bigg \} \nonumber
\end{equation}
\begin{equation}
+Z_{28}   \{ 0\} +Z_{29} \Bigg \{\ \ \ \ \ \ \  \begin{gathered}\begin{fmfgraph*}(40,40)
\fmfleft{i1}
\fmflabel{$V^\gamma$}{i1}
\fmfright{o1}
\fmflabel{$V^\gamma$}{o1}
\fmf{photon,label=$\phantom{x}$}{i1,v}
\fmf{photon,label=$\phantom{x}$}{v,o1}
\fmfv{decor.shape=cross}{v}
\end{fmfgraph*} \end{gathered} \ \ \ \ \ \ \  \Bigg \} + Z_{30}  \{ 0 \}.   \label{eq29} 
\end{equation}
From this it follows that $Z_{29}$ does have the tree part.

\begin{equation}
0 \equiv\begin{gathered}
\langle V^\gamma V^\gamma \left[iC^\gamma \bar{C}^\gamma \right]_R \rangle = 
\end{gathered}\nonumber 
\end{equation}
\begin{equation} 
Z_{31} \Bigg \{
\begin{gathered}
\begin{fmfgraph*}(50,50)
\fmfleft{i1,i2}
\fmflabel{$V^\gamma$}{i1}
\fmfright{o1,o2}
\fmflabel{$V^\gamma$}{o1}
\fmf{phantom,tension=1.5}{i2,v2}
\fmf{phantom,tension=1.5}{o1,v2}
\fmf{photon,label=$\phantom{x}$,tension=1.5,label.side=right}{i1,v1}
\fmf{gluon,label=$\phantom{x}$,left,tension=0.05}{v1,v2}
\fmf{gluon,label=$\phantom{x}$,left,tension=0.05}{v2,v1}
\fmf{photon,label=$\phantom{x}$,tension=1.5}{v1,o1}
\fmfdotn{v}{1}
\fmfv{decor.shape=cross}{v2}
\end{fmfgraph*} 
\end{gathered} + \ \ \ \ \ \ 
\begin{gathered}
  \begin{fmfgraph*}(80,40)
\fmfleft{i1}
\fmflabel{$V^\gamma$}{i1}
\fmfright{o1}
\fmflabel{$V^\gamma$}{o1}
\fmf{photon,label=$\phantom{x}$}{i1,v1}
\fmf{gluon,label=$\phantom{x}$,label.side=left}{v1,v3}
\fmf{gluon,label=$\phantom{x}$,label.side=left}{v3,v2}
\fmf{gluon,left,label=$\phantom{x}$,tension=0.4}{v2,v1}
\fmf{photon,label=$\phantom{x}$}{v2,o1}
\fmfdotn{v}{2}
\fmfv{decor.shape=cross}{v3}
  \end{fmfgraph*}
\end{gathered}\ \ \ \ \ \ \ \Bigg \} +Z_{32}  \{0  \}   \nonumber 
\end{equation}
\begin{equation}
+Z_{33} \Bigg \{\ \ \ \ \ \ \begin{gathered}
\begin{fmfgraph*}(50,50)
\fmfleft{i1,i2}
\fmflabel{$V^\gamma$}{i1}
\fmfv{label=$\omega$,label.angle=90,label.dist=0.3cm}{v2}
\fmfright{o1,o2}
\fmflabel{$V^\gamma$}{o1}
\fmf{phantom,tension=1.5}{i2,v2}
\fmf{phantom,tension=1.5}{o1,v2}
\fmf{photon,label=$\phantom{x}$,tension=1.5,label.side=right}{i1,v1}
\fmf{ghost,label=$\phantom{x}$,left,tension=0.05}{v1,v2}
\fmf{ghost,label=$\phantom{x}$,left,tension=0.05}{v2,v1}
\fmf{photon,label=$\phantom{x}$,tension=1.5}{v1,o1}
\fmfdotn{v}{1}
\fmfv{decor.shape=cross}{v2}
\end{fmfgraph*} 
\end{gathered}\ \ \ \ \ \ +\ \ \ \ \ \ \ \ \begin{gathered}
\begin{fmfgraph*}(70,40)
\fmfleft{i1}
\fmflabel{$V^\gamma$}{i1}
\fmfright{o1}
\fmflabel{$V^\gamma$}{o1}
\fmf{photon,label=$\phantom{x}$}{i1,v1}
\fmf{ghost,label=$\phantom{x}$}{v1,v3}
\fmf{ghost,label=$\phantom{x}$}{v3,v2}
\fmf{ghost,left,label=$\phantom{x}$,tension=0.4}{v2,v1}
\fmf{photon,label=$\phantom{x}$}{v2,o1}
\fmfdotn{v}{2}
\fmfv{decor.shape=cross}{v3}
\fmfv{label=$\omega$,label.angle=90,label.dist=0.3cm}{v3}
\end{fmfgraph*} 
\end{gathered} \ \ \ \ \ \ \ \  \Bigg \} \nonumber
\end{equation}
\begin{equation}
+Z_{34}   \{ 0\} +Z_{35} \Bigg \{\ \ \ \ \ \ \  \begin{gathered}\begin{fmfgraph*}(40,40)
\fmfleft{i1}
\fmflabel{$V^\gamma$}{i1}
\fmfright{o1}
\fmflabel{$V^\gamma$}{o1}
\fmf{photon,label=$\phantom{x}$}{i1,v}
\fmf{photon,label=$\phantom{x}$}{v,o1}
\fmfv{decor.shape=cross}{v}
\end{fmfgraph*} \end{gathered} \ \ \ \ \ \ \  \Bigg \} + Z_{36}  \{ 0 \}.   \label{eq30} 
\end{equation}
From this it follows that $Z_{35}$ does not have the tree part.

\subsubsection{Insertion into $\langle C^\gamma \bar{C}^\gamma \rangle$}


\begin{equation}
\begin{gathered}
\langle C^\gamma \bar{C}^\gamma \left[\frac{1}{2} X X \right]_R \rangle = 
\end{gathered} 
Z_1 \{ 0\} +Z_2 \{ 0\} +Z_3 \{0\} +Z_4\{0\} \nonumber 
\end{equation}
\begin{equation}
 + Z_5\{ 0\} 
+Z_6 \Bigg \{\ \ \ \ \ \ \   \begin{gathered}\begin{fmfgraph*}(40,40)
\fmfleft{i1}
\fmflabel{$C^\gamma$}{i1}
\fmfright{o1}
\fmflabel{$\bar{C}^\gamma$}{o1}
\fmf{ghost,label=$\phantom{x}$}{i1,v}
\fmf{ghost,label=$\phantom{x}$}{v,o1}
\fmfv{decor.shape=cross}{v}
\end{fmfgraph*} \end{gathered}\ \ \ \ \ \  \Bigg \} \equiv 0.  \label{eq31} 
\end{equation}
From this it follows that $Z_6$ does not have the tree part.

\begin{equation}
\begin{gathered}
\langle C^\gamma \bar{C}^\gamma \left[\frac{1}{2} V^j V^j \right]_R \rangle = 
\end{gathered} 
Z_7 \{ 0\} +Z_8 \{ 0\} +Z_9 \{0\} +Z_{10}\{0\} \nonumber 
\end{equation}
\begin{equation}
 + Z_{11}\{ 0\} 
+Z_{12} \Bigg \{\ \ \ \ \ \ \   \begin{gathered}\begin{fmfgraph*}(40,40)
\fmfleft{i1}
\fmflabel{$C^\gamma$}{i1}
\fmfright{o1}
\fmflabel{$\bar{C}^\gamma$}{o1}
\fmf{ghost,label=$\phantom{x}$}{i1,v}
\fmf{ghost,label=$\phantom{x}$}{v,o1}
\fmfv{decor.shape=cross}{v}
\end{fmfgraph*} \end{gathered}\ \ \ \ \ \  \Bigg \} \equiv 0 . \label{eq32} 
\end{equation}
From this it follows that $Z_{12}$ does not have the tree part.

\begin{equation}
\begin{gathered}
\langle C^\gamma \bar{C}^\gamma \left[i\omega \bar{\omega} \right]_R \rangle = 
\end{gathered} 
Z_{13} \{ 0\} +Z_{14} \{ 0\} +Z_{15} \{0\} +Z_{16}\{0\} \nonumber 
\end{equation}
\begin{equation}
 + Z_{17}\{ 0\} 
+Z_{18} \Bigg \{\ \ \ \ \ \ \   \begin{gathered}\begin{fmfgraph*}(40,40)
\fmfleft{i1}
\fmflabel{$C^\gamma$}{i1}
\fmfright{o1}
\fmflabel{$\bar{C}^\gamma$}{o1}
\fmf{ghost,label=$\phantom{x}$}{i1,v}
\fmf{ghost,label=$\phantom{x}$}{v,o1}
\fmfv{decor.shape=cross}{v}
\end{fmfgraph*} \end{gathered}\ \ \ \ \ \  \Bigg \} \equiv 0.  \label{eq33} 
\end{equation}
From this it follows that $Z_{18}$ does not have the tree part.

\begin{equation}
\begin{gathered}
\langle C^\gamma \bar{C}^\gamma \left[iC^j \bar{C}^j \right]_R \rangle = 
\end{gathered} 
Z_{19} \{ 0\} +Z_{20} \{ 0\} +Z_{21} \{0\} +Z_{22}\{0\} \nonumber 
\end{equation}
\begin{equation}
 + Z_{23}\{ 0\} 
+Z_{24} \Bigg \{\ \ \ \ \ \ \   \begin{gathered}\begin{fmfgraph*}(40,40)
\fmfleft{i1}
\fmflabel{$C^\gamma$}{i1}
\fmfright{o1}
\fmflabel{$\bar{C}^\gamma$}{o1}
\fmf{ghost,label=$\phantom{x}$}{i1,v}
\fmf{ghost,label=$\phantom{x}$}{v,o1}
\fmfv{decor.shape=cross}{v}
\end{fmfgraph*} \end{gathered}\ \ \ \ \ \  \Bigg \} \equiv 0.  \label{eq34} 
\end{equation}
From this it follows that $Z_{24}$ does not have the tree part.

\begin{equation}
\begin{gathered}
\langle C^\gamma \bar{C}^\gamma \left[\frac{1}{2}V^\gamma V^\gamma \right]_R \rangle = 
\end{gathered} 
Z_{25} \{ 0\} +Z_{26} \{ 0\} +Z_{27} \{0\} +Z_{28}\{0\} \nonumber 
\end{equation}
\begin{equation}
 + Z_{29}\{ 0\} 
+Z_{30} \Bigg \{\ \ \ \ \ \ \   \begin{gathered}\begin{fmfgraph*}(40,40)
\fmfleft{i1}
\fmflabel{$C^\gamma$}{i1}
\fmfright{o1}
\fmflabel{$\bar{C}^\gamma$}{o1}
\fmf{ghost,label=$\phantom{x}$}{i1,v}
\fmf{ghost,label=$\phantom{x}$}{v,o1}
\fmfv{decor.shape=cross}{v}
\end{fmfgraph*} \end{gathered}\ \ \ \ \ \  \Bigg \} \equiv 0.  \label{eq35} 
\end{equation}
From this it follows that $Z_{30}$ does not have the tree part.

\begin{equation}
\begin{gathered}\begin{fmfgraph*}(40,40)
\fmfleft{i1}
\fmflabel{$C^\gamma$}{i1}
\fmfright{o1}
\fmflabel{$\bar{C}^\gamma$}{o1}
\fmf{ghost,label=$\phantom{x}$}{i1,v}
\fmf{ghost,label=$\phantom{x}$}{v,o1}
\fmfv{decor.shape=cross}{v}
\end{fmfgraph*} \end{gathered} \ \ \ \ \ \ \ \ \equiv \ \ \begin{gathered}
\langle C^\gamma \bar{C}^\gamma \left[iC^\gamma \bar{C}^\gamma \right]_R \rangle =\end{gathered} \nonumber
\end{equation}
\begin{equation} 
Z_{31} \{ 0\} +Z_{32} \{ 0\} +Z_{33} \{0\} +Z_{34}\{0\} \nonumber 
\end{equation}
\begin{equation}
 + Z_{35}\{ 0\} 
+Z_{36} \Bigg \{\ \ \ \ \ \ \   \begin{gathered}\begin{fmfgraph*}(40,40)
\fmfleft{i1}
\fmflabel{$C^\gamma$}{i1}
\fmfright{o1}
\fmflabel{$\bar{C}^\gamma$}{o1}
\fmf{ghost,label=$\phantom{x}$}{i1,v}
\fmf{ghost,label=$\phantom{x}$}{v,o1}
\fmfv{decor.shape=cross}{v}
\end{fmfgraph*} \end{gathered}\ \ \ \ \ \  \Bigg \} .  \label{eq36} 
\end{equation}
From this it follows that $Z_{36}$ does have the tree part.

At this stage we have proven the following form of the renormalization matrix,
\muu{
\mathcal{Z}= \mathbbm{1} + \mathcal{Z}^{(1)}. \nonumber
}

We therefore can reconsider the equations \eqref{eq1} - \eqref{eq36} to obtain the diagrammatic equations for the elements of $\mathcal{Z}^{(1)}$. From equations \eqref{eq1}-\eqref{eq6} we find $Z_{19}^{(1)}=Z_{31}^{(1)}=0$ as well as
\begin{equation}
Z_1^{(1)}=-\ \Bigg \{\begin{gathered}
\begin{fmfgraph*}(40,40)
   \fmfleft{i1,i2}
\fmflabel{$\phantom{x}$}{i1}
\fmfright{o1,o2}
\fmflabel{$\phantom{x}$}{o1}
\fmf{phantom,tension=1.5}{i2,v2}
\fmf{phantom,tension=1.5}{o1,v2}
\fmf{gluon,label=$\phantom{x}$,tension=1.5,label.side=right}{i1,v1}
\fmf{gluon,label=$\phantom{x}$,left,tension=0.05}{v1,v2}
\fmf{gluon,label=$\phantom{x}$,left,tension=0.05}{v2,v1}
\fmf{gluon,label=$\phantom{x}$,tension=1.5}{v1,o1}
\fmfdotn{v}{1}
\fmfv{decor.shape=cross}{v2}
  \end{fmfgraph*}
\end{gathered} + 
\begin{gathered}
\begin{fmfgraph*}(70,40)
\fmfleft{i1}
\fmflabel{$\phantom{x}$}{i1}
\fmfright{o1}
\fmflabel{$\phantom{x}$}{o1}
\fmf{gluon,label=$\phantom{x}$}{v1,i1}
\fmf{photon,left,tension=0.4,label=$V^j+V^\gamma$}{v1,v2}
\fmf{gluon,label=$\phantom{x}$}{v3,v1}
\fmf{gluon,label=$\phantom{x}$}{v2,v3}
\fmfv{decor.shape=cross}{v3}
\fmf{gluon,label=$\phantom{x}$}{o1,v2}
\fmfdotn{v}{2}
  \end{fmfgraph*}
\end{gathered}\Bigg \}.
\end{equation}
\begin{equation}
Z_7^{(1)}= - \Bigg \{ \begin{gathered}
\begin{fmfgraph*}(40,40)
\fmfleft{i1,i2}
\fmflabel{$\phantom{x}$}{i1}
\fmfright{o1,o2}
\fmflabel{$\phantom{x}$}{o1}
\fmflabel{$\phantom{x}$}{v2}
\fmf{phantom,tension=1.5}{i2,v2}
\fmf{phantom,tension=1.5}{o1,v2}
\fmf{gluon,label=$\phantom{x}$,tension=1.5,label.side=right}{i1,v1}
\fmf{photon,label=$\phantom{x}$,left,tension=0.05}{v1,v2}
\fmf{photon,label=$\phantom{x}$,left,tension=0.05}{v2,v1}
\fmf{gluon,label=$\phantom{x}$,tension=1.5}{v1,o1}
\fmfv{label=$V^j$,label.angle=90,label.dist=0.3cm}{v2}
\fmfdotn{v}{1}
\fmfv{decor.shape=cross}{v2}
\end{fmfgraph*} 
\end{gathered}+\begin{gathered}
\begin{fmfgraph*}(80,40)
\fmfleft{i1}
\fmflabel{$\phantom{x}$}{i1}
\fmfright{o1}
\fmflabel{$\phantom{x}$}{o1}
\fmf{gluon,label=$\phantom{x}$}{v1,i1}
\fmf{photon,label.side=left,label=$\phantom{x}$}{v1,v3}
\fmf{photon,label.side=left,label=$\phantom{x}$}{v3,v2}
\fmf{gluon,left,tension=0.4,label=$\phantom{x}$}{v2,v1}
\fmfv{decor.shape=cross}{v3}
\fmfv{label=$V^j$,label.angle=90,label.dist=0.3cm}{v3}
\fmf{gluon,label=$\phantom{x}$}{o1,v2}
\fmfdotn{v}{2}
\end{fmfgraph*} 
\end{gathered} \Bigg \}. 
\end{equation}
\begin{equation}
Z_{13}^{(1)}=- \Bigg \{\begin{gathered}
\begin{fmfgraph*}(40,40)
\fmfleft{i1,i2}
\fmflabel{$\phantom{x}$}{i1}
\fmfright{o1,o2}
\fmflabel{$\phantom{x}$}{o1}
\fmf{phantom,tension=1.5}{i2,v2}
\fmf{phantom,tension=1.5}{o1,v2}
\fmf{gluon,label=$\phantom{x}$,tension=1.5,label.side=right}{i1,v1}
\fmf{ghost,label=$\phantom{x}$,left,tension=0.05}{v1,v2}
\fmf{ghost,label=$\phantom{x}$,left,tension=0.05}{v2,v1}
\fmf{gluon,label=$\phantom{x}$,tension=1.5}{v1,o1}
\fmfdotn{v}{1}
\fmfv{decor.shape=cross}{v2}
\fmfv{label=$\omega$,label.angle=90,label.dist=0.3cm}{v2}
\end{fmfgraph*} 
\end{gathered}
\Bigg \}.
\end{equation}
\begin{equation}
Z_{25}^{(1)}=-\Bigg \{\begin{gathered}
\begin{fmfgraph*}(40,40)
\fmfleft{i1,i2}
\fmflabel{$\phantom{x}$}{i1}
\fmfright{o1,o2}
\fmflabel{$\phantom{x}$}{o1}
\fmflabel{$\phantom{x}$}{v2}
\fmf{phantom,tension=1.5}{i2,v2}
\fmf{phantom,tension=1.5}{o1,v2}
\fmf{gluon,label=$\phantom{x}$,tension=1.5,label.side=right}{i1,v1}
\fmf{photon,label=$\phantom{x}$,left,tension=0.05}{v1,v2}
\fmf{photon,label=$\phantom{x}$,left,tension=0.05}{v2,v1}
\fmf{gluon,label=$\phantom{x}$,tension=1.5}{v1,o1}
\fmfdotn{v}{1}
\fmfv{decor.shape=cross}{v2}
\fmfv{label=$V^\gamma$,label.angle=90,label.dist=0.3cm}{v2}
\end{fmfgraph*} 
\end{gathered}+\begin{gathered}
\begin{fmfgraph*}(80,40)
\fmfleft{i1}
\fmflabel{$\phantom{x}$}{i1}
\fmfright{o1}
\fmflabel{$\phantom{x}$}{o1}
\fmf{gluon,label=$\phantom{x}$}{v1,i1}
\fmf{photon,label.side=left,label=$\phantom{x}$}{v1,v3}
\fmf{photon,label.side=left,label=$\phantom{x}$}{v3,v2}
\fmf{gluon,left,tension=0.4,label=$\phantom{x}$}{v2,v1}
\fmfv{decor.shape=cross}{v3}
\fmfv{label=$V^\gamma$,label.angle=90,label.dist=0.3cm}{v3}
\fmf{gluon,label=$\phantom{x}$}{o1,v2}
\fmfdotn{v}{2}
\end{fmfgraph*} 
\end{gathered} \Bigg \}.
\end{equation}
From equations \eqref{eq7}-\eqref{eq12} we find $Z_{26}^{(1)}=Z_{32}^{(1)}=0$ as well as
\begin{equation}
Z_{14}^{(1)}=-\Bigg\{
\ \ \ \ \ \ \begin{gathered}
\begin{fmfgraph*}(30,40)
\fmfleft{i1,i2}
\fmflabel{$V^j$}{i1}
\fmfright{o1,o2}
\fmflabel{$V^k$}{o1}
\fmf{phantom,tension=1.5}{i2,v2}
\fmf{phantom,tension=1.5}{o1,v2}
\fmf{photon,label=$\phantom{x}$,tension=1.5,label.side=right}{i1,v1}
\fmf{ghost,label=$\phantom{x}$,left,tension=0.05}{v1,v2}
\fmf{ghost,label=$\phantom{x}$,left,tension=0.05}{v2,v1}
\fmf{photon,label=$\phantom{x}$,tension=1.5}{v1,o1}
\fmfdotn{v}{1}
\fmfv{decor.shape=cross}{v2}
\fmfv{label=$\omega$,label.angle=90,label.dist=0.3cm}{v2}
\end{fmfgraph*} 
\end{gathered}\ \ \ \ \ \ +\ \ \ \ \ \ \ \ \begin{gathered}
\begin{fmfgraph*}(60,40)
\fmfleft{i1}
\fmflabel{$V^j$}{i1}
\fmfright{o1}
\fmflabel{$V^k$}{o1}
\fmf{photon,label=$\phantom{x}$}{i1,v1}
\fmf{ghost,label=$\phantom{x}$}{v1,v3}
\fmf{ghost,label=$\phantom{x}$}{v3,v2}
\fmf{ghost,left,label=$\phantom{x}$,tension=0.4}{v2,v1}
\fmf{photon,label=$\phantom{x}$}{v2,o1}
\fmfdotn{v}{2}
\fmfv{decor.shape=cross}{v3}
\fmfv{label=$\omega$,label.angle=90,label.dist=0.3cm}{v3}
\end{fmfgraph*} 
\end{gathered} \ \ \ \ \ \ \ \ 
\Bigg \}.\label{z14}
\end{equation}

\begin{equation}
Z_2^{(1)}=- \Bigg \{
\begin{gathered}
\begin{fmfgraph*}(40,40)
\fmfleft{i1,i2}
\fmflabel{$V^j$}{i1}
\fmfright{o1,o2}
\fmflabel{$V^k$}{o1}
\fmf{phantom,tension=1.5}{i2,v2}
\fmf{phantom,tension=1.5}{o1,v2}
\fmf{photon,label=$\phantom{x}$,tension=1.5,label.side=right}{i1,v1}
\fmf{gluon,label=$\phantom{x}$,left,tension=0.05}{v1,v2}
\fmf{gluon,label=$\phantom{x}$,left,tension=0.05}{v2,v1}
\fmf{photon,label=$\phantom{x}$,tension=1.5}{v1,o1}
\fmfdotn{v}{1}
\fmfv{decor.shape=cross}{v2}
\end{fmfgraph*} 
\end{gathered} + \ \ \ \ \ \ \ \ 
\begin{gathered}
  \begin{fmfgraph*}(80,40)
\fmfleft{i1}
\fmflabel{$V^j$}{i1}
\fmfright{o1}
\fmflabel{$V^k$}{o1}
\fmf{photon,label=$\phantom{x}$}{i1,v1}
\fmf{gluon,label=$\phantom{x}$,label.side=left}{v1,v3}
\fmf{gluon,label=$\phantom{x}$,label.side=left}{v3,v2}
\fmf{gluon,left,label=$\phantom{x}$,tension=0.4}{v2,v1}
\fmf{photon,label=$\phantom{x}$}{v2,o1}
\fmfdotn{v}{2}
\fmfv{decor.shape=cross}{v3}
  \end{fmfgraph*}
\end{gathered}\ \ \ \ \ \ \ \Bigg \}.\label{z2}
\end{equation}

\begin{equation}
Z_8^{(1)}=- \ \Bigg \{ \begin{gathered}
\begin{fmfgraph*}(40,40)
\fmfleft{i1,i2}
\fmflabel{$V^j$}{i1}
\fmfright{o1,o2}
\fmflabel{$V^k$}{o1}
\fmf{phantom,tension=1.5}{i2,v2}
\fmf{phantom,tension=1.5}{o1,v2}
\fmf{photon,label=$\phantom{x}$,tension=1.5,label.side=right}{i1,v1}
\fmf{photon,label=$\phantom{x}$,left,tension=0.05}{v1,v2}
\fmf{photon,label=$\phantom{x}$,left,tension=0.05}{v2,v1}
\fmf{photon,label=$\phantom{x}$,tension=1.5}{v1,o1}
\fmfdotn{v}{1}
\fmfv{decor.shape=cross}{v2}
\fmfv{label=$V^l$,label.angle=90,label.dist=0.3cm}{v2}
\end{fmfgraph*} 
\end{gathered}+\ \ \ \ \ \ \ \ 
\begin{gathered}
\begin{fmfgraph*}(80,40)
\fmfleft{i1}
\fmflabel{$V^j$}{i1}
\fmfright{o1}
\fmflabel{$V^k$}{o1}
\fmf{photon,label=$\phantom{x}$}{i1,v1}
\fmf{photon,right,label=$\phantom{x}$,tension=0.4}{v1,v2}
\fmf{photon,label.side=left,label=$\phantom{x}$}{v1,v3}
\fmf{photon,label.side=left,label=$\phantom{x}$}{v3,v2}
\fmf{photon,label=$\phantom{x}$}{v2,o1}
\fmfdotn{v}{2}
\fmfv{decor.shape=cross}{v3}
\fmfv{label=$V^l$,label.angle=90,label.dist=0.3cm}{v3}
\end{fmfgraph*} 
\end{gathered}\ \ \ \ \ \ \ \ \Bigg \}.
\end{equation}

\begin{equation}
Z_{20}^{(1)}=- \Bigg \{ \ \ \ \ \ \ \ \begin{gathered}
\begin{fmfgraph*}(80,40)
\fmfleft{i1}
\fmflabel{$V^j$}{i1}
\fmfright{o1}
\fmflabel{$V^k$}{o1}
\fmf{photon,label=$\phantom{x}$}{i1,v1}
\fmf{ghost,label=$\phantom{x}$}{v1,v3}
\fmf{ghost,label=$\phantom{x}$}{v3,v2}
\fmf{ghost,left,label=$\phantom{x}$,tension=0.4}{v2,v1}
\fmf{photon,label=$\phantom{x}$}{v2,o1}
\fmfdotn{v}{2}
\fmfv{decor.shape=cross}{v3}
\fmfv{label=$C^l$,label.angle=90,label.dist=0.3cm}{v3}
\end{fmfgraph*} 
\end{gathered} \ \ \ \ \ \ \ \Bigg \}.
\end{equation}

From equations \eqref{eq13} - \eqref{eq18} we find $Z_{21}^{(1)}=Z_{33}^{(1)}=0$ as well as
\begin{equation}
Z_3^{(1)}=-\ \Bigg \{
\begin{gathered}
\begin{fmfgraph*}(50,50)
\fmfleft{i1,i2}
\fmflabel{$\omega$}{i1}
\fmfright{o1,o2}
\fmflabel{$\bar{\omega}$}{o1}
\fmf{phantom,tension=1.5}{i2,v2}
\fmf{phantom,tension=1.5}{o1,v2}
\fmf{ghost,label=$\phantom{x}$,tension=1.5,label.side=right}{i1,v1}
\fmf{gluon,label=$\phantom{x}$,left,tension=0.05}{v1,v2}
\fmf{gluon,label=$\phantom{x}$,left,tension=0.05}{v2,v1}
\fmf{ghost,label=$\phantom{x}$,tension=1.5}{v1,o1}
\fmfdotn{v}{1}
\fmfv{decor.shape=cross}{v2}
\end{fmfgraph*}
\end{gathered} \Bigg \}.
\end{equation}
\begin{equation}
Z_9^{(1)}=-\ \Bigg \{\ \ \ \begin{gathered}
\begin{fmfgraph*}(60,60)
\fmfleft{i1,i2}
\fmflabel{$\omega$}{i1}
\fmfright{o1,o2}
\fmflabel{$\bar{\omega}$}{o1}
\fmf{phantom,tension=1.5}{i2,v2}
\fmf{phantom,tension=1.5}{o1,v2}
\fmf{ghost,label=$\phantom{x}$,tension=1.5,label.side=right}{i1,v1}
\fmf{photon,label=$\phantom{x}$,left,tension=0.05}{v1,v2}
\fmf{photon,label=$\phantom{x}$,left,tension=0.05}{v2,v1}
\fmf{ghost,label=$\phantom{x}$,tension=1.5}{v1,o1}
\fmfdotn{v}{1}
\fmfv{decor.shape=cross}{v2}
\fmfv{label=$V^j$,label.angle=90,label.dist=0.3cm}{v2}
\end{fmfgraph*} 
\end{gathered}+\ \ \ \ \ 
\begin{gathered}
\begin{fmfgraph*}(80,40)
\fmfleft{i1}
\fmflabel{$\omega$}{i1}
\fmfright{o1}
\fmflabel{$\bar{\omega}$}{o1}
\fmf{ghost,label=$\phantom{x}$}{i1,v1}
\fmf{photon,label.side=left,label=$\phantom{x}$}{v1,v3}
\fmf{photon,label.side=left,label=$\phantom{x}$}{v3,v2}
\fmf{ghost,right,tension=0.4,label=$\phantom{x}$}{v1,v2}
\fmf{ghost,label=$\phantom{x}$}{v2,o1}
\fmfdotn{v}{2}
\fmfv{decor.shape=cross}{v3}
\fmfv{label=$V^j$,label.angle=90,label.dist=0.3cm}{v3}
\end{fmfgraph*} 
\end{gathered}\ \ \ \ \  \Bigg \}.
\end{equation}

\begin{equation}
Z_{15}^{(1)}=-\ \Bigg \{\ \ \ \  \begin{gathered}
\begin{fmfgraph*}(50,50)
\fmfleft{i1,i2}
\fmflabel{$\omega$}{i1}
\fmfright{o1,o2}
\fmflabel{$\bar{\omega}$}{o1}
\fmfv{label=$\omega$,label.angle=90,label.dist=0.3cm}{v2}
\fmf{phantom,tension=1.5}{i2,v2}
\fmf{phantom,tension=1.5}{o1,v2}
\fmf{ghost,label=$\phantom{x}$,tension=1.5,label.side=right}{i1,v1}
\fmf{ghost,label=$\phantom{x}$,left,tension=0.05}{v1,v2}
\fmf{ghost,label=$\phantom{x}$,left,tension=0.05}{v2,v1}
\fmf{ghost,label=$\phantom{x}$,tension=1.5}{v1,o1}
\fmfdotn{v}{1}
\fmfv{decor.shape=cross}{v2}
\end{fmfgraph*} 
\end{gathered}  + \ \ \ \ \  \begin{gathered}
\begin{fmfgraph*}(80,40)
\fmfleft{i1}
\fmflabel{$\omega$}{i1}
\fmfright{o1}
\fmflabel{$\bar{\omega}$}{o1}
\fmf{ghost,label=$\phantom{x}$}{i1,v1}
\fmf{photon,left,tension=0.4,label=$V^j+V^\gamma$}{v1,v2}
\fmf{ghost,label=$\phantom{x}$}{v3,v2}
\fmf{ghost,label=$\phantom{x}$}{v1,v3}
\fmf{ghost,label=$\phantom{x}$}{v2,o1}
\fmfdotn{v}{2}
\fmfv{decor.shape=cross}{v3}
\end{fmfgraph*} 
\end{gathered} \ \ \ \ \  \Bigg \}.
\end{equation}

\begin{equation}
Z_{27}^{(1)}=-\ \Bigg \{\ \ \ \begin{gathered}
\begin{fmfgraph*}(50,50)
\fmfleft{i1,i2}
\fmflabel{$\omega$}{i1}
\fmfright{o1,o2}
\fmflabel{$\bar{\omega}$}{o1}
\fmf{phantom,tension=1.5}{i2,v2}
\fmf{phantom,tension=1.5}{o1,v2}
\fmf{ghost,label=$\phantom{x}$,tension=1.5,label.side=right}{i1,v1}
\fmf{photon,label=$\phantom{x}$,left,tension=0.05}{v1,v2}
\fmf{photon,label=$\phantom{x}$,left,tension=0.05}{v2,v1}
\fmf{ghost,label=$\phantom{x}$,tension=1.5}{v1,o1}
\fmfdotn{v}{1}
\fmfv{decor.shape=cross}{v2}
\fmfv{label=$V^\gamma$,label.angle=90,label.dist=0.3cm}{v2}
\end{fmfgraph*} 
\end{gathered}+\ \ \ \ \ 
\begin{gathered}
\begin{fmfgraph*}(80,40)
\fmfleft{i1}
\fmflabel{$\omega$}{i1}
\fmfright{o1}
\fmflabel{$\bar{\omega}$}{o1}
\fmf{ghost,label=$\phantom{x}$}{i1,v1}
\fmf{photon,label.side=left,label=$\phantom{x}$}{v1,v3}
\fmf{photon,label.side=left,label=$\phantom{x}$}{v3,v2}
\fmf{ghost,right,tension=0.4,label=$\phantom{x}$}{v1,v2}
\fmf{ghost,label=$\phantom{x}$}{v2,o1}
\fmfdotn{v}{2}
\fmfv{decor.shape=cross}{v3}
\fmfv{label=$V^\gamma$,label.angle=90,label.dist=0.3cm}{v3}
\end{fmfgraph*} 
\end{gathered}\ \ \ \ \  \Bigg \}.
\end{equation}

Considering equations \eqref{eq19} - \eqref{eq24} we obtain $Z_4^{(1)}=Z_{16}^{(1)}=Z_{28}^{(1)}=Z_{34}^{(1)}=0$ and 
\begin{equation}
Z_{10}^{(1)}=-\ \Bigg \{\ \ \ \ \ \begin{gathered}
\begin{fmfgraph*}(80,40)
\fmfleft{i1}
\fmflabel{$C^j$}{i1}
\fmfright{o1}
\fmflabel{$\bar{C}^k$}{o1}
\fmf{ghost,label=$\phantom{x}$}{i1,v1}
\fmf{photon,label.side=left,label=$\phantom{x}$}{v1,v3}
\fmf{photon,label.side=left,label=$\phantom{x}$}{v3,v2}
\fmf{ghost,right,label=$\phantom{x}$,tension=0.4}{v1,v2}
\fmf{ghost,label=$\phantom{x}$}{v2,o1}
\fmfdotn{v}{2}
\fmfv{decor.shape=cross}{v3}
\fmfv{label=$V^j$,label.angle=90,label.dist=0.3cm}{v3}
\end{fmfgraph*} 
\end{gathered}\ \ \ \ \ \  \Bigg \}.\label{z10}
\end{equation}
\begin{equation}
Z_{22}^{(1)}=-\ \Bigg \{\ \ \ \ \ \  \begin{gathered}  \begin{fmfgraph*}(80,40)
\fmfleft{i1}
\fmflabel{$C^j$}{i1}
\fmfright{o1}
\fmflabel{$\bar{C}^k$}{o1}
\fmf{ghost,label=$\phantom{x}$}{i1,v1}
\fmf{photon,label=$\phantom{x}$,right,tension=0.4}{v1,v2}
\fmf{ghost,label=$\phantom{x}$}{v1,v3}
\fmf{ghost,label=$\phantom{x}$}{v3,v2}
\fmf{ghost,label=$\phantom{x}$}{v2,o1}
\fmfdotn{v}{2}
\fmfv{decor.shape=cross}{v3}
\fmfv{label=$C^l$,label.angle=90,label.dist=0.2cm}{v3}
\end{fmfgraph*} \end{gathered} \ \ \ \ \ \ \   \Bigg \}.\label{z22}
\end{equation}

The equations \eqref{eq25} - \eqref{eq30} imply $Z_{11}^{(1)}=Z_{23}^{(1)}=Z_{29}^{(1)}=Z_{35}^{(1)}=0$ and 
\begin{equation}
Z_{5}^{(1)}=-\ \Bigg \{
\begin{gathered}
\begin{fmfgraph*}(50,50)
\fmfleft{i1,i2}
\fmflabel{$V^\gamma$}{i1}
\fmfright{o1,o2}
\fmflabel{$V^\gamma$}{o1}
\fmf{phantom,tension=1.5}{i2,v2}
\fmf{phantom,tension=1.5}{o1,v2}
\fmf{photon,label=$\phantom{x}$,tension=1.5,label.side=right}{i1,v1}
\fmf{gluon,label=$\phantom{x}$,left,tension=0.05}{v1,v2}
\fmf{gluon,label=$\phantom{x}$,left,tension=0.05}{v2,v1}
\fmf{photon,label=$\phantom{x}$,tension=1.5}{v1,o1}
\fmfdotn{v}{1}
\fmfv{decor.shape=cross}{v2}
\end{fmfgraph*} 
\end{gathered} + \ \ \ \ \ \ 
\begin{gathered}
  \begin{fmfgraph*}(80,40)
\fmfleft{i1}
\fmflabel{$V^\gamma$}{i1}
\fmfright{o1}
\fmflabel{$V^\gamma$}{o1}
\fmf{photon,label=$\phantom{x}$}{i1,v1}
\fmf{gluon,label=$\phantom{x}$,label.side=left}{v1,v3}
\fmf{gluon,label=$\phantom{x}$,label.side=left}{v3,v2}
\fmf{gluon,left,label=$\phantom{x}$,tension=0.4}{v2,v1}
\fmf{photon,label=$\phantom{x}$}{v2,o1}
\fmfdotn{v}{2}
\fmfv{decor.shape=cross}{v3}
  \end{fmfgraph*}
\end{gathered}\ \ \ \ \ \ \ \Bigg \}.\label{z5}
\end{equation}
\begin{equation}
Z_{17}^{(1)}=- \ \Bigg \{\ \ \ \ \ \ \begin{gathered}
\begin{fmfgraph*}(50,50)
\fmfleft{i1,i2}
\fmflabel{$V^\gamma$}{i1}
\fmfv{label=$\omega$,label.angle=90,label.dist=0.2cm}{v2}
\fmfright{o1,o2}
\fmflabel{$V^\gamma$}{o1}
\fmf{phantom,tension=1.5}{i2,v2}
\fmf{phantom,tension=1.5}{o1,v2}
\fmf{photon,label=$\phantom{x}$,tension=1.5,label.side=right}{i1,v1}
\fmf{ghost,label=$\phantom{x}$,left,tension=0.05}{v1,v2}
\fmf{ghost,label=$\phantom{x}$,left,tension=0.05}{v2,v1}
\fmf{photon,label=$\phantom{x}$,tension=1.5}{v1,o1}
\fmfdotn{v}{1}
\fmfv{decor.shape=cross}{v2}
\end{fmfgraph*} 
\end{gathered} +\ \ \ \ \ \ \ \ \begin{gathered}
\begin{fmfgraph*}(80,40)
\fmfleft{i1}
\fmflabel{$V^\gamma$}{i1}
\fmfright{o1}
\fmflabel{$V^\gamma$}{o1}
\fmf{photon,label=$\phantom{x}$}{i1,v1}
\fmf{ghost,label=$\phantom{x}$}{v1,v3}
\fmf{ghost,label=$\phantom{x}$}{v3,v2}
\fmf{ghost,left,label=$\phantom{x}$,tension=0.4}{v2,v1}
\fmf{photon,label=$\phantom{x}$}{v2,o1}
\fmfdotn{v}{2}
\fmfv{decor.shape=cross}{v3}
\fmfv{label=$\omega$,label.angle=90,label.dist=0.2cm}{v3}
\end{fmfgraph*} 
\end{gathered} \ \ \ \ \ \ \ \  \Bigg \}.\label{z17}
\end{equation}
Finally, from the equations \eqref{eq31} - \eqref{eq36} we find \\
$Z_6^{(1)}=Z_{12}^{(1)}=Z_{18}^{(1)}=Z_{24}^{(1)}=Z_{30}^{(1)}=Z_{36}^{(1)}=0$. This leaves us with a priori 16 non-vanishing renormalization factors. The divergent parts of the diagrams necessary to obtain these renormalization factors are calculated as
$\;$ \\
\begin{equation}
\begin{gathered}
\begin{fmfgraph*}(80,40)
\fmfleft{i1}
\fmflabel{$\phantom{x}$}{i1}
\fmfright{o1}
\fmflabel{$\phantom{x}$}{o1}
\fmf{gluon,label=$\phantom{x}$}{v1,i1}
\fmf{photon,left,tension=0.4,label=$V^j+V^\gamma$}{v1,v2}
\fmf{gluon,label=$\phantom{x}$}{v3,v1}
\fmf{gluon,label=$\phantom{x}$}{v2,v3}
\fmfv{decor.shape=cross}{v3}
\fmf{gluon,label=$\phantom{x}$}{o1,v2}
\fmfdotn{v}{2}
\end{fmfgraph*}
\end{gathered}\nonumber
\end{equation}
\muu{
=  i\frac{g^2\mu^{-2\epsilon}}{(4\pi)^2 \epsilon}  g^{\mu \nu}\delta^{ab} \frac{N}{2} \left[\frac{3}{4}\left(5+\xi (\xi+2)\right]+\frac{\alpha+(N-2)\lambda}{N-1}\right).
}

\begin{equation}
\begin{gathered}
\begin{fmfgraph*}(80,40)
\fmfleft{i1}
\fmflabel{$\phantom{x}$}{i1}
\fmfright{o1}
\fmflabel{$\phantom{x}$}{o1}
\fmf{gluon,label=$\phantom{x}$}{v1,i1}
\fmf{photon,label=$\phantom{x}$}{v1,v3}
\fmf{photon,label=$\phantom{x}$}{v3,v2}
\fmf{gluon,left,tension=0.4,label=$\phantom{x}$}{v2,v1}
\fmfv{decor.shape=cross}{v3}
\fmfv{label=$V^j$,label.angle=90,label.dist=0.2cm}{v3}
\fmf{gluon,label=$\phantom{x}$}{o1,v2}
\fmfdotn{v}{2}
\end{fmfgraph*}
\end{gathered}\nonumber
\end{equation}
\muu{
 =i\frac{g^2\mu^{-2\epsilon}}{(4\pi)^2 \epsilon}g^{\mu \nu} \delta^{ab} \frac{N(N-2)}{2(N-1)} \left(\frac{3(\lambda^2+6)\xi+\lambda^2+3\xi^2+3}{4\xi} \right).
}

\begin{equation}
\begin{gathered}
\begin{fmfgraph*}(80,40)
\fmfleft{i1}
\fmflabel{$\phantom{x}$}{i1}
\fmfright{o1}
\fmflabel{$\phantom{x}$}{o1}
\fmf{gluon,label=$\phantom{x}$}{v1,i1}
\fmf{photon,label=$\phantom{x}$}{v1,v3}
\fmf{photon,label=$\phantom{x}$}{v3,v2}
\fmf{gluon,left,tension=0.4,label=$\phantom{x}$}{v2,v1}
\fmfv{decor.shape=cross}{v3}
\fmfv{label=$V^\gamma$,label.angle=90,label.dist=0.2cm}{v3}
\fmf{gluon,label=$\phantom{x}$}{o1,v2}
\fmfdotn{v}{2}
\end{fmfgraph*}
\end{gathered} \nonumber
\end{equation}
\muu{
 =i\frac{g^2\mu^{-2\epsilon}}{(4\pi)^2 \epsilon}g^{\mu \nu} \delta^{ab} \frac{N}{2(N-1)} \left(\frac{3(\alpha^2+6)\xi+\alpha^2+3\xi^2+3}{4\xi} \right).
}

\begin{equation}
\begin{gathered}
\begin{fmfgraph*}(70,40)
\fmfleft{i1}
\fmflabel{$\omega$}{i1}
\fmfright{o1}
\fmflabel{$\bar{\omega}$}{o1}
\fmf{ghost,label=$\phantom{x}$}{i1,v1}
\fmf{photon,label=$\phantom{x}$}{v1,v3}
\fmf{photon,label=$\phantom{x}$}{v3,v2}
\fmf{ghost,right,tension=0.4,label=$\phantom{x}$}{v1,v2}
\fmf{ghost,label=$\phantom{x}$}{v2,o1}
\fmfdotn{v}{2}
\fmfv{decor.shape=cross}{v3}
\fmfv{label=$V^j$,label.angle=90,label.dist=0.2cm}{v3}
\end{fmfgraph*}
\end{gathered}\nonumber
\end{equation}
\muu{
=-\frac{g^2 \mu^{-2\epsilon}}{(4\pi)^2\epsilon}\frac{N(N-2)}{2(N-1)}\lambda^2 \delta^{ab}.
}

\begin{equation}
\begin{gathered}
\begin{fmfgraph*}(100,40)
\fmfleft{i1}
\fmflabel{$\omega$}{i1}
\fmfright{o1}
\fmflabel{$\bar{\omega}$}{o1}
\fmf{ghost,label=$\phantom{x}$}{i1,v1}
\fmf{photon,label=$\phantom{x}$}{v1,v3}
\fmf{photon,label=$\phantom{x}$}{v3,v2}
\fmf{ghost,right,tension=0.4,label=$\phantom{x}$}{v1,v2}
\fmf{ghost,label=$\phantom{x}$}{v2,o1}
\fmfdotn{v}{2}
\fmfv{decor.shape=cross}{v3}
\fmfv{label=$V^\gamma$,label.angle=90,label.dist=0.2cm}{v3}
\end{fmfgraph*}
\end{gathered}\nonumber
\end{equation}
\muu{
=-\frac{g^2 \mu^{-2\epsilon}}{(4\pi)^2\epsilon}\frac{N}{2(N-1)}\alpha^2 \delta^{ab}.
}
$\;$ \\
$\;$ \\
\begin{equation}
\begin{gathered}
\begin{fmfgraph*}(100,40)
\fmfleft{i1}
\fmflabel{$\omega$}{i1}
\fmfright{o1}
\fmflabel{$\bar{\omega}$}{o1}
\fmf{ghost,label=$\phantom{x}$}{i1,v1}
\fmf{ghost,label=$\phantom{x}$}{v1,v3}
\fmf{ghost,label=$\phantom{x}$}{v3,v2}
\fmf{photon,label=$V^j$,left,label.side=left,tension=0.4}{v1,v2}
\fmf{ghost,label=$\phantom{x}$}{v2,o1}
\fmfdotn{v}{2}
\fmfv{decor.shape=cross}{v3}
\end{fmfgraph*}
\end{gathered}\ \ \ \ \ \ \
=0.
\end{equation}
$\;$ \\
$\;$ \\
\begin{equation}
\begin{gathered}
\begin{fmfgraph*}(100,40)
\fmfleft{i1}
\fmflabel{$\omega$}{i1}
\fmfright{o1}
\fmflabel{$\bar{\omega}$}{o1}
\fmf{ghost,label=$\phantom{x}$}{i1,v1}
\fmf{ghost,label=$\phantom{x}$}{v1,v3}
\fmf{ghost,label=$\phantom{x}$}{v3,v2}
\fmf{photon,label=$V^\gamma$,left,label.side=left,tension=0.4}{v1,v2}
\fmf{ghost,label=$\phantom{x}$}{v2,o1}
\fmfdotn{v}{2}
\fmfv{decor.shape=cross}{v3}
\end{fmfgraph*}
\end{gathered}\ \ \ \ \ \ \
=0.
\end{equation}

\begin{equation}
\begin{gathered}
\begin{fmfgraph*}(80,50)
\fmfleft{i1}
\fmflabel{$V^j$}{i1}
\fmfright{o1}
\fmflabel{$V^k$}{o1}
\fmf{photon,label=$\phantom{x}$}{i1,v1}
\fmf{photon,label=$\phantom{x}$}{v1,v3}
\fmf{photon,label=$\phantom{x}$}{v3,v2}
\fmf{photon,right,tension=0.4,label=$\phantom{x}$}{v1,v2}
\fmf{photon,label=$\phantom{x}$}{v2,o1}
\fmfdotn{v}{2}
\fmfv{decor.shape=cross}{v3}
\fmfv{label=$V^l$,label.angle=90,label.dist=0.2cm}{v3}
\end{fmfgraph*}
\end{gathered}\nonumber
\end{equation}
\muu{
=i\frac{g^2 \mu^{-2\epsilon}}{(4\pi)^2 \epsilon}g^{\mu \nu} \delta^{jk} (N-1)\frac{3}{4}(\lambda^2+\lambda+4).
}

\begin{equation}
\begin{gathered}
\begin{fmfgraph*}(70,40)
\fmfleft{i1}
\fmflabel{$V^j$}{i1}
\fmfright{o1}
\fmflabel{$V^k$}{o1}
\fmf{photon,label=$\phantom{x}$}{v1,i1}
\fmf{gluon,label=$\phantom{x}$}{v1,v3}
\fmf{gluon,label=$\phantom{x}$}{v3,v2}
\fmf{gluon,left,tension=0.4,label=$\phantom{x}$}{v2,v1}
\fmf{photon,label=$\phantom{x}$}{v2,o1}
\fmfdotn{v}{2}
\fmfv{decor.shape=cross}{v3}
\end{fmfgraph*}
\end{gathered}\ \ \ \ \ \
=i\frac{g^2 \mu^{-2\epsilon}}{(4\pi)^2 \epsilon}g^{\mu \nu} \delta^{jk} \frac{3+\xi^2}{4}\left(3+\frac{1}{\xi}\right).
\end{equation}

\begin{equation}
\begin{gathered}
\begin{fmfgraph*}(70,40)
\fmfleft{i1}
\fmflabel{$V^\gamma$}{i1}
\fmfright{o1}
\fmflabel{$V^\gamma$}{o1}
\fmf{photon,label=$\phantom{x}$}{v1,i1}
\fmf{gluon,label=$\phantom{x}$}{v1,v3}
\fmf{gluon,label=$\phantom{x}$}{v3,v2}
\fmf{gluon,left,tension=0.4,label=$\phantom{x}$}{v2,v1}
\fmf{photon,label=$\phantom{x}$}{v2,o1}
\fmfdotn{v}{2}
\fmfv{decor.shape=cross}{v3}
\end{fmfgraph*}
\end{gathered}\ \ \ \ \ \
=iN\frac{g^2 \mu^{-2\epsilon}}{(4\pi)^2 \epsilon}g^{\mu \nu}  \frac{3+\xi^2}{4}\left(3+\frac{1}{\xi}\right).
\end{equation}

\begin{equation}
\begin{gathered}
\begin{fmfgraph*}(70,40)
\fmfleft{i1}
\fmflabel{$V^j$}{i1}
\fmfright{o1}
\fmflabel{$V^k$}{o1}
\fmf{photon,label=$\phantom{x}$}{v1,i1}
\fmf{ghost,label=$\phantom{x}$}{v1,v3}
\fmf{ghost,label=$\phantom{x}$}{v3,v2}
\fmf{ghost,left,tension=0.4,label=$\phantom{x}$}{v2,v1}
\fmf{photon,label=$\phantom{x}$}{v2,o1}
\fmfdotn{v}{2}
\fmfv{decor.shape=cross}{v3}
\fmfv{label=$C^l$,label.angle=90,label.dist=0.2cm}{v3}
\end{fmfgraph*}
\end{gathered}\ \ \ \ \ \
=i\frac{g^2 \mu^{-2\epsilon}}{(4\pi)^2 \epsilon}g^{\mu \nu} \delta^{jk} \frac{(N-1)}{2}.
\end{equation}

\begin{equation}
\begin{gathered}
\begin{fmfgraph*}(70,40)
\fmfleft{i1}
\fmflabel{$V^j$}{i1}
\fmfright{o1}
\fmflabel{$V^k$}{o1}
\fmf{photon,label=$\phantom{x}$}{v1,i1}
\fmf{ghost,label=$\phantom{x}$}{v1,v3}
\fmf{ghost,label=$\phantom{x}$}{v3,v2}
\fmf{ghost,left,tension=0.4,label=$\phantom{x}$}{v2,v1}
\fmf{photon,label=$\phantom{x}$}{v2,o1}
\fmfdotn{v}{2}
\fmfv{decor.shape=cross}{v3}
\fmfv{label=$\omega$,label.angle=90,label.dist=0.2cm}{v3}
\end{fmfgraph*}
\end{gathered}\ \ \ \ \ \
=2i\frac{g^2 \mu^{-2\epsilon}}{(4\pi)^2 \epsilon}g^{\mu \nu} \delta^{jk}. 
\end{equation}

\begin{equation}
\begin{gathered}
\begin{fmfgraph*}(70,40)
\fmfleft{i1}
\fmflabel{$V^\gamma$}{i1}
\fmfright{o1}
\fmflabel{$V^\gamma$}{o1}
\fmf{photon,label=$\phantom{x}$}{v1,i1}
\fmf{ghost,label=$\phantom{x}$}{v1,v3}
\fmf{ghost,label=$\phantom{x}$}{v3,v2}
\fmf{ghost,left,tension=0.4,label=$\phantom{x}$}{v2,v1}
\fmf{photon,label=$\phantom{x}$}{v2,o1}
\fmfdotn{v}{2}
\fmfv{decor.shape=cross}{v3}
\fmfv{label=$\omega$,label.angle=90,label.dist=0.2cm}{v3}
\end{fmfgraph*}
\end{gathered}\ \ \ \ \ \
=i\frac{g^2 \mu^{-2\epsilon}}{(4\pi)^2 \epsilon}g^{\mu \nu} \delta^{jk} 2N.
\end{equation}

\begin{equation}
\begin{gathered}
\begin{fmfgraph*}(70,40)
\fmfleft{i1}
\fmflabel{$C^j$}{i1}
\fmfright{o1}
\fmflabel{$\bar{C}^k$}{o1}
\fmf{ghost,label=$\phantom{x}$}{i1,v1}
\fmf{photon,label=$\phantom{x}$}{v1,v3}
\fmf{photon,label=$\phantom{x}$}{v3,v2}
\fmf{ghost,right,tension=0.4,label=$\phantom{x}$}{v1,v2}
\fmf{ghost,label=$\phantom{x}$}{v2,o1}
\fmfdotn{v}{2}
\fmfv{decor.shape=cross}{v3}
\fmfv{label=$V^l$,label.angle=90,label.dist=0.2cm}{v3}
\end{fmfgraph*}
\end{gathered}\ \ \ \ \ \
=0.
\end{equation}

\begin{equation}
\begin{gathered}
\begin{fmfgraph*}(70,40)
\fmfleft{i1}
\fmflabel{$C^j$}{i1}
\fmfright{o1}
\fmflabel{$\bar{C}^k$}{o1}
\fmf{ghost,label=$\phantom{x}$}{i1,v1}
\fmf{ghost,label=$\phantom{x}$}{v1,v3}
\fmf{ghost,label=$\phantom{x}$}{v3,v2}
\fmf{photon,right,tension=0.4,label=$V^l$}{v1,v2}
\fmf{ghost,label=$\phantom{x}$}{v2,o1}
\fmfdotn{v}{2}
\fmfv{decor.shape=cross}{v3}
\end{fmfgraph*}
\end{gathered}\ \ \ \ \ \
=0.
\end{equation}

\begin{equation}
\begin{gathered}
\begin{fmfgraph*}(60,60)
\fmfleft{i1,i2}
\fmflabel{$\phantom{x}$}{i1}
\fmfright{o1,o2}
\fmflabel{$\phantom{x}$}{o1}
\fmf{phantom,tension=1.5}{i2,v2}
\fmf{phantom,tension=1.5}{o1,v2}
\fmf{gluon,label=$\phantom{x}$,tension=1.5,label.side=right}{i1,v1}
\fmf{gluon,label=$\phantom{x}$,left,tension=0.05}{v1,v2}
\fmf{gluon,label=$\phantom{x}$,left,tension=0.05}{v2,v1}
\fmf{gluon,label=$\phantom{x}$,tension=1.5}{v1,o1}
\fmfdotn{v}{1}
\fmfv{decor.shape=cross}{v2}
\fmfv{label=$\phantom{x}$,label.angle=90,label.dist=0.2cm}{v2}
\end{fmfgraph*}
\end{gathered} \ \ 
= -i 3N\frac{g^2\mu^{-2\epsilon}}{(4\pi)^2\epsilon} \delta^{ab} g^{\mu \nu} \frac{3+\xi^2}{8}.
\end{equation}

\begin{equation}
\begin{gathered}
\begin{fmfgraph*}(60,60)
\fmfleft{i1,i2}
\fmflabel{$\phantom{x}$}{i1}
\fmfright{o1,o2}
\fmflabel{$\phantom{x}$}{o1}
\fmf{phantom,tension=1.5}{i2,v2}
\fmf{phantom,tension=1.5}{o1,v2}
\fmf{gluon,label=$\phantom{x}$,tension=1.5,label.side=right}{i1,v1}
\fmf{ghost,label=$\phantom{x}$,left,tension=0.05}{v1,v2}
\fmf{ghost,label=$\phantom{x}$,left,tension=0.05}{v2,v1}
\fmf{gluon,label=$\phantom{x}$,tension=1.5}{v1,o1}
\fmfdotn{v}{1}
\fmfv{decor.shape=cross}{v2}
\fmfv{label=$\omega$,label.angle=90,label.dist=0.2cm}{v2}
\end{fmfgraph*}
\end{gathered} \ \ 
= i N\frac{g^2\mu^{-2\epsilon}}{(4\pi)^2\epsilon} \delta^{ab} g^{\mu \nu}.
\end{equation}

\begin{equation}
\begin{gathered}
\begin{fmfgraph*}(60,60)
\fmfleft{i1,i2}
\fmflabel{$\phantom{x}$}{i1}
\fmfright{o1,o2}
\fmflabel{$\phantom{x}$}{o1}
\fmf{phantom,tension=1.5}{i2,v2}
\fmf{phantom,tension=1.5}{o1,v2}
\fmf{ghost,label=$\omega$,tension=1.5,label.side=right}{i1,v1}
\fmf{gluon,label=$\phantom{x}$,left,tension=0.05}{v1,v2}
\fmf{gluon,label=$\phantom{x}$,left,tension=0.05}{v2,v1}
\fmf{ghost,label=$\bar{\omega}$,tension=1.5}{v1,o1}
\fmfdotn{v}{1}
\fmfv{decor.shape=cross}{v2}
\fmfv{label=$\phantom{x}$,label.angle=90,label.dist=0.2cm}{v2}
\end{fmfgraph*}
\end{gathered} \ \ 
= -N \frac{g^2\mu^{-2\epsilon}}{(4\pi)^2\epsilon} \delta^{ab}  \frac{3+\xi^2}{2}.
\end{equation}

\begin{equation}
\begin{gathered}
\begin{fmfgraph*}(60,60)
\fmfleft{i1,i2}
\fmflabel{$\phantom{x}$}{i1}
\fmfright{o1,o2}
\fmflabel{$\phantom{x}$}{o1}
\fmf{phantom,tension=1.5}{i2,v2}
\fmf{phantom,tension=1.5}{o1,v2}
\fmf{gluon,label=$\phantom{x}$,tension=1.5,label.side=right}{i1,v1}
\fmf{photon,label=$\phantom{x}$,left,tension=0.05}{v1,v2}
\fmf{photon,label=$\phantom{x}$,left,tension=0.05}{v2,v1}
\fmf{gluon,label=$\phantom{x}$,tension=1.5}{v1,o1}
\fmfdotn{v}{1}
\fmfv{decor.shape=cross}{v2}
\fmfv{label=$V^j$,label.angle=90,label.dist=0.2cm}{v2}
\end{fmfgraph*}
\end{gathered}  
= -i \frac{g^2\mu^{-2\epsilon}}{(4\pi)^2\epsilon} \delta^{ab} g^{\mu \nu} \frac{N(N-2)}{2(N-1)}(3+\lambda^2)\left[\frac{3}{4}+\frac{1}{4\xi} \right].
\end{equation}

\begin{equation}
\begin{gathered}
\begin{fmfgraph*}(60,60)
\fmfleft{i1,i2}
\fmflabel{$\phantom{x}$}{i1}
\fmfright{o1,o2}
\fmflabel{$\phantom{x}$}{o1}
\fmf{phantom,tension=1.5}{i2,v2}
\fmf{phantom,tension=1.5}{o1,v2}
\fmf{gluon,label=$\phantom{x}$,tension=1.5,label.side=right}{i1,v1}
\fmf{photon,label=$\phantom{x}$,left,tension=0.05}{v1,v2}
\fmf{photon,label=$\phantom{x}$,left,tension=0.05}{v2,v1}
\fmf{gluon,label=$\phantom{x}$,tension=1.5}{v1,o1}
\fmfdotn{v}{1}
\fmfv{decor.shape=cross}{v2}
\fmfv{label=$V^\gamma$,label.angle=90,label.dist=0.2cm}{v2}
\end{fmfgraph*}
\end{gathered} 
= -i \frac{g^2\mu^{-2\epsilon}}{(4\pi)^2\epsilon} \delta^{ab} g^{\mu \nu} \frac{N}{2(N-1)}(3+\alpha^2)\left[\frac{3}{4}+\frac{1}{4\xi} \right].
\end{equation}

\begin{equation}
\begin{gathered}
\begin{fmfgraph*}(60,60)
\fmfleft{i1,i2}
\fmflabel{$V^j$}{i1}
\fmfright{o1,o2}
\fmflabel{$V^k$}{o1}
\fmf{phantom,tension=1.5}{i2,v2}
\fmf{phantom,tension=1.5}{o1,v2}
\fmf{photon,label=$\phantom{x}$,tension=1.5,label.side=right}{i1,v1}
\fmf{gluon,label=$\phantom{x}$,left,tension=0.05}{v1,v2}
\fmf{gluon,label=$\phantom{x}$,left,tension=0.05}{v2,v1}
\fmf{photon,label=$\phantom{x}$,tension=1.5}{v1,o1}
\fmfdotn{v}{1}
\fmfv{decor.shape=cross}{v2}
\fmfv{label=$\phantom{x}$,label.angle=90,label.dist=0.2cm}{v2}
\end{fmfgraph*}
\end{gathered} \ \ 
= -i \frac{g^2\mu^{-2\epsilon}}{(4\pi)^2\epsilon} \delta^{jk} g^{\mu \nu} \frac{3+\xi^2}{4}\left(3+\frac{1}{\xi}\right).
\end{equation}

\begin{equation}
\begin{gathered}
\begin{fmfgraph*}(60,60)
\fmfleft{i1,i2}
\fmflabel{$V^\gamma$}{i1}
\fmfright{o1,o2}
\fmflabel{$V^\gamma$}{o1}
\fmf{phantom,tension=1.5}{i2,v2}
\fmf{phantom,tension=1.5}{o1,v2}
\fmf{photon,label=$\phantom{x}$,tension=1.5,label.side=right}{i1,v1}
\fmf{gluon,label=$\phantom{x}$,left,tension=0.05}{v1,v2}
\fmf{gluon,label=$\phantom{x}$,left,tension=0.05}{v2,v1}
\fmf{photon,label=$\phantom{x}$,tension=1.5}{v1,o1}
\fmfdotn{v}{1}
\fmfv{decor.shape=cross}{v2}
\fmfv{label=$\phantom{x}$,label.angle=90,label.dist=0.2cm}{v2}
\end{fmfgraph*}
\end{gathered} \ \ 
= -i N \frac{g^2\mu^{-2\epsilon}}{(4\pi)^2\epsilon}  g^{\mu \nu} \frac{3+\xi^2}{4}\left(3+\frac{1}{\xi}\right).
\end{equation}

\begin{equation}
\begin{gathered}
\begin{fmfgraph*}(60,60)
\fmfleft{i1,i2}
\fmflabel{$V^j$}{i1}
\fmfright{o1,o2}
\fmflabel{$V^k$}{o1}
\fmf{phantom,tension=1.5}{i2,v2}
\fmf{phantom,tension=1.5}{o1,v2}
\fmf{photon,label=$\phantom{x}$,tension=1.5,label.side=right}{i1,v1}
\fmf{photon,label=$\phantom{x}$,left,tension=0.05}{v1,v2}
\fmf{photon,label=$\phantom{x}$,left,tension=0.05}{v2,v1}
\fmf{photon,label=$\phantom{x}$,tension=1.5}{v1,o1}
\fmfdotn{v}{1}
\fmfv{decor.shape=cross}{v2}
\fmfv{label=$V^l$,label.angle=90,label.dist=0.2cm}{v2}
\end{fmfgraph*}
\end{gathered} \ \ 
= -i \frac{g^2\mu^{-2\epsilon}}{(4\pi)^2\epsilon} \delta^{jk} g^{\mu \nu} (N-1)\frac{3}{4}(3+\lambda^2).
\end{equation}

\begin{equation}
\begin{gathered}
\begin{fmfgraph*}(60,60)
\fmfleft{i1,i2}
\fmflabel{$V^j$}{i1}
\fmfright{o1,o2}
\fmflabel{$V^k$}{o1}
\fmf{phantom,tension=1.5}{i2,v2}
\fmf{phantom,tension=1.5}{o1,v2}
\fmf{photon,label=$\phantom{x}$,tension=1.5,label.side=right}{i1,v1}
\fmf{ghost,label=$\phantom{x}$,left,tension=0.05}{v1,v2}
\fmf{ghost,label=$\phantom{x}$,left,tension=0.05}{v2,v1}
\fmf{photon,label=$\phantom{x}$,tension=1.5}{v1,o1}
\fmfdotn{v}{1}
\fmfv{decor.shape=cross}{v2}
\fmfv{label=$\omega$,label.angle=90,label.dist=0.2cm}{v2}
\end{fmfgraph*}
\end{gathered} \ \ 
= -2i \frac{g^2\mu^{-2\epsilon}}{(4\pi)^2\epsilon} \delta^{jk} g^{\mu \nu}.
\end{equation}

\begin{equation}
\begin{gathered}
\begin{fmfgraph*}(60,60)
\fmfleft{i1,i2}
\fmflabel{$V^\gamma$}{i1}
\fmfright{o1,o2}
\fmflabel{$V^\gamma$}{o1}
\fmf{phantom,tension=1.5}{i2,v2}
\fmf{phantom,tension=1.5}{o1,v2}
\fmf{photon,label=$\phantom{x}$,tension=1.5,label.side=right}{i1,v1}
\fmf{ghost,label=$\phantom{x}$,left,tension=0.05}{v1,v2}
\fmf{ghost,label=$\phantom{x}$,left,tension=0.05}{v2,v1}
\fmf{photon,label=$\phantom{x}$,tension=1.5}{v1,o1}
\fmfdotn{v}{1}
\fmfv{decor.shape=cross}{v2}
\fmfv{label=$\omega$,label.angle=90,label.dist=0.2cm}{v2}
\end{fmfgraph*}
\end{gathered} \ \ 
= -i2N \frac{g^2\mu^{-2\epsilon}}{(4\pi)^2\epsilon}  g^{\mu \nu}.
\end{equation}

\begin{equation}
\begin{gathered}
\begin{fmfgraph*}(60,60)
\fmfleft{i1,i2}
\fmflabel{$\omega$}{i1}
\fmfright{o1,o2}
\fmflabel{$\bar{\omega}$}{o1}
\fmf{phantom,tension=1.5}{i2,v2}
\fmf{phantom,tension=1.5}{o1,v2}
\fmf{ghost,label=$\phantom{x}$,tension=1.5,label.side=right}{i1,v1}
\fmf{ghost,label=$\phantom{x}$,left,tension=0.05}{v1,v2}
\fmf{ghost,label=$\phantom{x}$,left,tension=0.05}{v2,v1}
\fmf{ghost,label=$\phantom{x}$,tension=1.5}{v1,o1}
\fmfdotn{v}{1}
\fmfv{decor.shape=cross}{v2}
\fmfv{label=$\omega$,label.angle=90,label.dist=0.2cm}{v2}
\end{fmfgraph*}
\end{gathered} \ \ 
= -N \frac{g^2\mu^{-2\epsilon}}{(4\pi)^2\epsilon}  \delta^{ab} \frac{\xi}{2}.
\end{equation}

\begin{equation}
\begin{gathered}
\begin{fmfgraph*}(60,60)
\fmfleft{i1,i2}
\fmflabel{$\omega$}{i1}
\fmfright{o1,o2}
\fmflabel{$\bar{\omega}$}{o1}
\fmf{phantom,tension=1.5}{i2,v2}
\fmf{phantom,tension=1.5}{o1,v2}
\fmf{ghost,label=$\phantom{x}$,tension=1.5,label.side=right}{i1,v1}
\fmf{photon,label=$\phantom{x}$,left,tension=0.05}{v1,v2}
\fmf{photon,label=$\phantom{x}$,left,tension=0.05}{v2,v1}
\fmf{ghost,label=$\phantom{x}$,tension=1.5}{v1,o1}
\fmfdotn{v}{1}
\fmfv{decor.shape=cross}{v2}
\fmfv{label=$V^j$,label.angle=90,label.dist=0.2cm}{v2}
\end{fmfgraph*}
\end{gathered} \ \ 
= \frac{g^2\mu^{-2\epsilon}}{(4\pi)^2\epsilon} \delta^{ab} \frac{N(N-2)}{2(N-1)}(3+\lambda^2).
\end{equation}

\begin{equation}
\begin{gathered}
\begin{fmfgraph*}(60,60)
\fmfleft{i1,i2}
\fmflabel{$\omega$}{i1}
\fmfright{o1,o2}
\fmflabel{$\bar{\omega}$}{o1}
\fmf{phantom,tension=1.5}{i2,v2}
\fmf{phantom,tension=1.5}{o1,v2}
\fmf{ghost,label=$\phantom{x}$,tension=1.5,label.side=right}{i1,v1}
\fmf{photon,label=$\phantom{x}$,left,tension=0.05}{v1,v2}
\fmf{photon,label=$\phantom{x}$,left,tension=0.05}{v2,v1}
\fmf{ghost,label=$\phantom{x}$,tension=1.5}{v1,o1}
\fmfdotn{v}{1}
\fmfv{decor.shape=cross}{v2}
\fmfv{label=$V^\gamma$,label.angle=90,label.dist=0.2cm}{v2}
\end{fmfgraph*}
\end{gathered} \ \ 
=  \frac{g^2\mu^{-2\epsilon}}{(4\pi)^2\epsilon} \delta^{ab} \frac{N}{2(N-1)}(3+\alpha^2).
\end{equation}
These results for the Feynman diagrams imply that some of the 16 a priori non-vanishing renormalization factors actually become zero. The factors $Z_{10}^{(1)}$ and $Z_{22}^{(1)}$ given by equations \eqref{z10} and \eqref{z22} are vanishing because the involved diagrams are finite. Moreover, the factors $Z_2^{(1)}$, $Z_5^{(1)}$, $Z_{14}^{(1)}$ and $Z_{17}^{(1)}$ given by the equations \eqref{z2}, \eqref{z5}, \eqref{z14} and \eqref{z17}, respectively, are zero as well since in each case the involved diagrams cancel each other. This finally leaves us with 10 non-zero renormalization factors, yielding the equations \eqref{matrixentries} and \eqref{renmacalc} in the main text.
$\;$ \\
$\;$ \\

\end{fmffile}


\bibliography{mebib}



\end{document}